\documentclass[onecolumn]{mnras_edited}
\usepackage[utf8]{inputenc}
\usepackage{amsmath}
\usepackage{natbib}
\usepackage{makecell}
\usepackage{tabularx}
\usepackage{graphicx}
\usepackage{fancyhdr}
\usepackage{lipsum}
\usepackage{booktabs, makecell, tabularx}
\usepackage{rotating}
\usepackage[export]{adjustbox}
\usepackage[british]{babel}
\usepackage{hhline}
\usepackage{multirow}
\usepackage[most]{tcolorbox}
\usepackage[export]{adjustbox}
\usepackage[caption=false]{subfig}

\newcommand{\be}{\begin{equation}}
\newcommand{\ee}{\end{equation}}
\newcommand{\nn}{\mbox{} \nonumber \\ \mbox{} }
\newcommand{\ba}{\begin{eqnarray}}
\newcommand{\ea}{\end{eqnarray}}
\newcommand{\om}{\omega}
\newcommand{\Alfven}{Alfv\'{e}n }

\newcommand{\B}{{\bf B}}

\newcommand\eg{{\it{e.g.\ }}}
\newcommand\cf{{\it{cf.\ }}}

\newcommand{\Bf}{{magnetic field}}
\newcommand{\Bfs}{{magnetic fields}}

\newcommand{\NS}{neutron star}
\newcommand{\NSs}{{neutron stars}}
\newcommand{\EM}{electromagnetic}

\newcommand{\Sc}{Schwarzschild}
\newcommand{\ms}{magnetosphere}
\newcommand{\mss}{magnetospheres}

\newcommand{\LC}{light cylinder}
\newcommand{\Lf}{Lorentz factor}

\begin{document}

\title[Relativistic coronal mass  ejections]{Relativistic coronal mass  ejections  from magnetars}

\author[Sharma et al.] {Praveen Sharma $^{1}$, Maxim V.~Barkov$^{2}$, Maxim Lyutikov$^{1}$ \\
$^{1}$ Department of Physics, Purdue University, 
 525 Northwestern Avenue,
West Lafayette, IN, USA\\
$^2$ Institute of Astronomy, Russian Academy of Sciences, Moscow, 119017, Russian Federation}

\maketitle

\begin{abstract}
We study dynamics of relativistic Coronal Mass Ejections (CMEs), from launching by shearing of foot-points (either slowly - the ``Solar flare'' paradigm, or suddenly - the ``star quake" paradigm), to propagation in the preceding magnetar wind. For slow shear, most of the energy injected into the CME is first spent on the work done on breaking through the over-laying magnetic field. At later stages, sufficiently powerful CMEs may experience  ``detonation"  and lead to   opening of the magnetosphere beyond some equipartition radius $r_{eq}$, where the energy of the CME becomes larger than the decreasing external magnetospheric energy. Post-CME magnetosphere relaxes via formation of a plasmoid-mediated current sheet, initially at $\sim r_{eq}$ and slowly reaching the light cylinder (this transient stage has much higher spindown rate and may produce an  ``anti-glitch''). Both the location of the foot-point shear and the global magnetospheric configuration affect the frequent-and-weak versus rare-and-powerful CME dichotomy - to produce powerful flares the slow shear should be limited to field lines that close near the star. After the creation of a topologically disconnected flux tube, the tube quickly (at $\sim$ the light cylinder) comes into force-balance with the preceding wind, and is passively advected/frozen in the wind afterward. For fast shear (a local rotational glitch), the resulting large amplitude Alfven waves lead to opening of the magnetosphere (which later recovers similarly to the slow shear case). At distances much larger than the light cylinder, the resulting shear Alfven waves propagate through the wind non-dissipatively. Implications to  Fast Radio Bursts are discussed.
\end{abstract}


\section{Introduction}

Magnetars, a class of highly magnetized \NSs, produce  X-ray and $\gamma$-ray bursts  \citep{TD95,2007MNRAS.382.1029K,2008A&ARv..15..225M,2017ARA&A..55..261K,1992Natur.357..472U}, and occasional giant flares  \citep[][]{2005Natur.434.1107P,2005Natur.434.1098H}.
 Discoveries related to Fast Radio Bursts   \citep{2019A&ARv..27....4P,2019ARA&A..57..417C}, especially   simultaneous observations of radio and X-ray bursts from a magnetar  \citep{2020Natur.587...54C,2021NatAs...5..372R,2020Natur.587...59B,2020ApJ...898L..29M} renewed interest in the  dynamics of magnetar's explosions. 

To set-up the stage, we first qualitatively divide FRB  models into two types - magnetospheric and wind models. Also  qualitatively we divide  magnetar flares' models into   \emph{Solar flare paradigm}  and \emph{Starquake paradigm}, with a clear understanding that the actual separation of models is/may not be as clearly defined. 

In the case of FRBs, one set of theories, advocates that FRBs are magnetospheric events  \citep[\eg][]{2003MNRAS.346..540L,2013arXiv1307.4924P,2016MNRAS.462..941L,2020arXiv200505093L}. Alternative suggestion is generation of FRBs  in the wind  or in the wind termination shock \citep[\eg][]{2014MNRAS.442L...9L,2017ApJ...843L..26B,2019MNRAS.485.4091M,2022arXiv220911136T, 2022ApJ...927....2K,2022MNRAS.515.4217B}. 
Observations of contemporaneous magnetar X-ray flares and FRB strengthened  the evidence for  magnetospheric {\it loci} 
\citep[as argued by][]{2020arXiv200505093L}; the detection of sub-second periodicity  \citep{2022Natur.607..256C}  leaves little doubt 
in our view. Recent detection of anti-glitch in FRB-associated magnetar \cite{2022arXiv221011518Y}  is also consistent with the magnetospheric model, see \cite{2013arXiv1306.2264L}.

The wind models of FRBs appeal to the generation of strong shock, or magnetic shell that propagates through the wind. As we demonstrate in the present paper the assumption of strong shock/magnetic shell propagating  {\it  through} the wind is incorrect: the magnetic shells can naturally come into force balance near the  \LC, and  are then passively advected with the wind.
We model quite a small magnetosphere radius $R_{LC}= 5 R_{NS}$. It allows to keep pressure balance due to small Lorentz factors of the flow. In the case of larger dynamical range (1$0^4$) it can be not so.
acceleration of the blob inside magnetosphere up to $\Gamma > 10 $ will leads to lose of the casual connection and blob can escape in strongly unbalanced conditions and form explosion like solution in the wind zone.

  As for magnetars' flares,  the   \emph{Solar flare paradigm}  for magnetar explosions  \citep{2006MNRAS.367.1594L,2015MNRAS.447.1407L} argues that the underlying mechanism that causes magnetars' flares may be similar to those operating in the  solar corona. According to the model, GFs are magnetospheric events. Alternative view is that magnetar flares are    crustal events \citep{TD95}.
  
In the  \emph{Solar flare paradigm} the energy that will eventually power  magnetar flares is first  stored  inside  the neutron star right following the core-collapse of the progenitor star.   Slowly over time, hundreds to thousands of years, the internal magnetic twist  is pushed into the \ms\ via Hall (electron-MHD) drift  \citep{RG,2013MNRAS.434.2480G,2014PhPl...21e2110W},  gated by {slow, plastic deformations} of the neutron star crust  \citep{2015MNRAS.447.1407L}.
This leads to gradual twisting of the external  magnetospheric field lines, on time scales much longer than the magnetar's GF, and creates active magnetospheric regions similar to  the Sun's spots. As more and more current is pushed into the magnetosphere, it eventually reaches a point of dynamical instability. The loss of stability leads to a rapid restructuring of magnetic configuration, on the Alfven crossing time scale, to the formation of narrow current sheets, and to the onset of magnetic dissipation.
As a result, a large amount of magnetic energy is converted into the kinetic and  bulk motion  and radiation  \citep{2003MNRAS.346..540L,2007MNRAS.374..415K,2019MNRAS.485..299R,2020ApJ...900L..21Y}. The coherent emission may be produced due to some kind of plasma instability, \eg\ via the Free Electron Laser mechanism \citep{2021ApJ...922..166L}. 
Perhaps the best argument in favor of the ``Solar flare paradigm'' is that the observed sharp rise of $\gamma$-ray flux during GF, on a time scale similar to the \Alfven crossing time of the inner magnetosphere, which takes $\sim 0.25 $ msec \citep[][]{2005Natur.434.1107P}. This unambiguously points to the magnetospheric origin of GFs  \citep{2006MNRAS.367.1594L}. Since in the  \emph{Solar flare paradigm}, GFs are magnetospheric events, no large baryonic loading is expected in the ensuing outflows.

Another model of magnetars' flares, which we call the  \emph{Starquake model} of \cite{TD95,2001ApJ...561..980T} (though the  ``starquake'' is not used in these papers -  we thank Chis  Thompson for pointing this out -  we use this terms as a classification marker; the models do appeal to crustal faults) , whereas sudden fraction of the crust leads to fast motion of the magnetic foot-points. \citep[][criticized this set-up: even if the elastic properties of the crust allow the creation of a shear crack, the strongly sheared magnetic field around the crack leads to a back-reaction from the Lorentz force which does not allow large relative displacement of the crack surfaces.]{2012MNRAS.427.1574L}

For the present purposes, the difference between slow shear of the  \emph{Solar flare}  and fast shear of the  \emph{Starquake} models is that for the slow shear the whole \ms\ remains in the  causal contact, while the fast shear corresponds to a packet of \Alfven waves generated by the foot-point motions. We emphasize that our separation  of models into \emph{Solar flare} -  \emph{Starquake} clearly misses many details and is introduced here to highlight the two different dynamics regimes in the ensuing discussion.

In the present paper we seek answers to the two sets of questions, one related to launching the CMEs from the \ms, and the second related to the propagation of the resulting structures through the wind:  (i) how  the  model of sheared/inflated magnetic flux tubes  in the Sun  \citep{1999ApJ...510..485A}  transports into relativistic highly magnetized regime; (ii) what is the role of the \LC\ in generating the CMEs; (iii) what underlying physical parameters distinguish  magnetars' giant flares, from  the less energetic   bursts;  (iv)  what  is the dynamics of  magnetospheric perturbation as they enter the wind. By  the FRB-magnetar association, these question may carry the answer to why FRBs are different? Investigations are done with the code PHAEDRA  \citep{2012MNRAS.423.1416P}, Appendix  \ref{NumericalMethod}. The code invokes force-free electrodynamics, an appropriate limit for the study of \NS\ magnetosphere, considering their extremely high magnetic field. In this limit, hydrodynamic forces can be safely neglected and therefore the electromagnetic Lorentz force can be approximated as zero.

The plan of the paper is as follows. In \S  \ref{theor}  we describe theoretical expectations that would guide us through the following research. In  \S \ref{PHAEDRA1} we describe the code. In \S  \ref{generation-inside}
we concentrate on the inner-most dynamics of the CMEs, neglecting rotation/presence of the \LC.
 In \S \ref{generation} (slow shear) we adapt a model of generation of  Solar CMEs by  \cite{1999ApJ...510..485A,2007ApJ...671..936A} to relativistic {\it rotating}  \mss\  of \NSs. In \S \ref{Jerked}  (fast shear)   we consider dynamics of a ``
glitched \ms'' -  when a part of the \NS's crust experience sudden change in the rotational angular velocity. 
 In \S \ref{Fluxtube} we consider dynamics, from the \mss\ to  the wind, of an ejected magnetic flux tube.

\section{Magnetar's CMEs}
\label{theor}

\subsection{The  \emph{Solar Flare} paradigm}

Coronal mass ejections (CMEs) are  the most explosive events in our solar system and have been long studied in solar physics  \citep{vourlidas2002solar, 2000JGR...10523153F}. 
According to  the model of  Solar flares by \cite{1999ApJ...510..485A,2007ApJ...671..936A}, the underlying cause of the manifestations of solar activity - CMEs, eruptive flares and filament ejections - is the disruption of a force balance between the upward pressure
of the strongly sheared field of a filament channel and the downward tension of a potential (non-current carrying) overlying field.  Thus, an eruption is driven solely by  the magnetic free energy stored in a closed, sheared magnetic field that opens toward infinity during a CME. Initially, the magnetic field has a complicated multipolar topology while reconnection between a sheared arcade and neighboring flux systems triggers the eruption.  We also mention an important Aly's theorem, that open topologies have the largest energy given poloidal magnetic field distribution on the surface \citep{1991ApJ...375L..61A}. The presence of the light cylinder change this picture:  if an arc reaches the light cylinder it will  become open.

 We  first explore models of magnetar giant flares based on the same paradigm as Solar Flares and  Coronal Mass Ejections  (CME)  \citep{1999ApJ...510..485A,2007ApJ...671..936A}, that they are driven by slow surface shear leading to catastrophic rearrangement of  the neutron star's magnetospheric fields. 

The principal difference between Solar  and magnetar  CMEs is that  the  magnetar plasma is
relativistic and strongly magnetized, with  Alfven velocity of the order of the speed of light. Perhaps it is more correct to them Coronal Flux Ejections (Jens Mahlmann, priv. comm.), but we keep the more familiar notation of a CME.
In addition, presence of a light cylinder play the most important part in the generation of CMEs in magnetars, if compared with non-rotating calculations of    \citep[][the light cylinder  is the analogue of the  Alfven surface in rotating stars]{1999ApJ...510..485A,2007ApJ...671..936A}.

Through numerical experiments we found that several complementary ingredient control  the overall dynamics of  the generation of CMEs in magnetars:  global magnetospheric structure,  rotation,  and the location of  foot-point shear. 
To make the following discussion clear the term ``shearing''  refers to the dynamical motion of magnetic foot-points.

We first study step-by-step different global configurations and different shearing prescriptions.
 In Appendix \ref{NumericalMethod} we study separately/reproduce analytical results for separate ``ingredients'' of the model:  (ii) Sheared non-rotating \mss, Appendix \ref{nonrotatingsystem};  (ii)  Rotating stars with no foot-point shearing, Appendix 
 \ref{noshear1} and in particular Michel's solution, Fig. \ref{pr_analytical_numerical}.

\subsection{Theoretical expectations}

\subsubsection{The set-up}


Let us  first discuss  dynamics of a topologically isolated  flux tubes/magnetic blobs (called CME below)  injected deep within a \ms, so that the presence of a  \LC\ is not important.
Consider an injected  isolated magnetic structure - two possible geometries include  a magnetic flux tube and magnetic ball. Let the injection occur near the stellar surface with 
  typical size $R_{CME,0} \leq R_{NS}$ and associated energy $E_{CME,0}$, Table \ref{scales11}. The \Bf\ inside the CME is of the order of the surface \Bf\ $B_0$, so that initially the CME  is just slightly unbalanced - internal \Bf\ matches approximately the magnetospheric field. The gradient of the external field pushes the CME out.

 \begin{table}
\begin{center}
\begin{tabular}{ |c|c|c|c| } 
\hline
Model & flux tube & small CME &  large  CME $(r=R_{CME})$\\
\hline
Initial volume of CME
& $2\pi  R_{NS}  \times \pi R_{CME,0}^2 $, flux tube  & $ (4 \pi/3) R_{CME,0}^3$, sphere&  $(4 \pi/3) R_{CME,0}^3$, sphere \\
\hline
injected energy $E_{CME,0}$ &  $2\pi  R_{NS}   R_{CME,0}^2 {B_0^2}/{(8\pi)} $& $ (4 \pi/3) R_{CME,0}^3 {B_0^2}/{(8\pi)} $ & $ (4 \pi/3) R_{CME,0}^3 {B_0^2}/{(8\pi)} $ \\
\hline
CME's linear size  at $r$ $R_{CME}/R_{CME,0}$ & $ \left(\frac{r}{R_{NS}}\right)^{3/2} $ &  $ \left(\frac{r}{R_{NS}}\right)^{3/2} $  & $ \left(\frac{r}{R_{NS}}\right)^{3} $ \\
\hline
energy  $E_{CME}/E_{CME,0}$ at  $r$ & $ \left( { R_{NS} }/{r}  \right) ^2$ & $ \left( { R_{NS} }/{r}  \right) ^{3/2}$ & $ \left( { R_{NS} }/{r}  \right) $\\
\hline
Equipartition radius $r_{eq}/R_{NS} $ & $  \eta_{CME}^{-1} $ & $   \eta_{CME}^{-2/3} $ & $ \eta_{CME}^{-2/3}  $  \\
\hline
Energy 
remaining at $r_{eq}$ & $ \eta_{CME}^2$ & $\eta_{CME} $ & $\eta_{CME}$ \\
\hline
\end{tabular}
\end{center}
\caption{Typical scales for dynamics of CME injected within the \ms. Only small fraction of the injected energy remains within the CME after it escapes from the \ms.}
\label{scales11}
 \end{table}

An important parameter is the total magnetic energy of the \ms,
\be
E_{B, NS} \sim B_0^2 R_{NS}^3
\ee
Naturally, the injected energy is much smaller than the total energy,
\be
\eta_{CME} = \frac{ E_{CME,0}}{E_{B, NS}}.
\ee

Conservation of the magnetic flux  within CME plays the most important role. The injected flux is 
\be
\Phi_B \sim B_0 R_{CME,0}^2= {\rm Const} = B_{CME} R_{CME}^2
\ee
It is conserved during evolution. Thus, \Bf\ inside is
\be
B_{CME} = B_0  \left ( \frac{R_{CME,0}}{R_{CME}}  \right) ^2
\label{BCME} 
\ee

We can then identify three different geometrical case: (i)  flux tube (a toroidally-symmetric configuration), (ii)  small magnetic ball (spherical ball displaced from the center); (iii)  large magnetic ball ($R_{CME,0} \sim R_{NS}$) (centered ball). In the ``large magnetic ball'' case the quasi-spherical injected structure is of order of the \NS\ from the beginning. 

\begin{figure}
 \begin{center}
\includegraphics[width=0.85\linewidth]{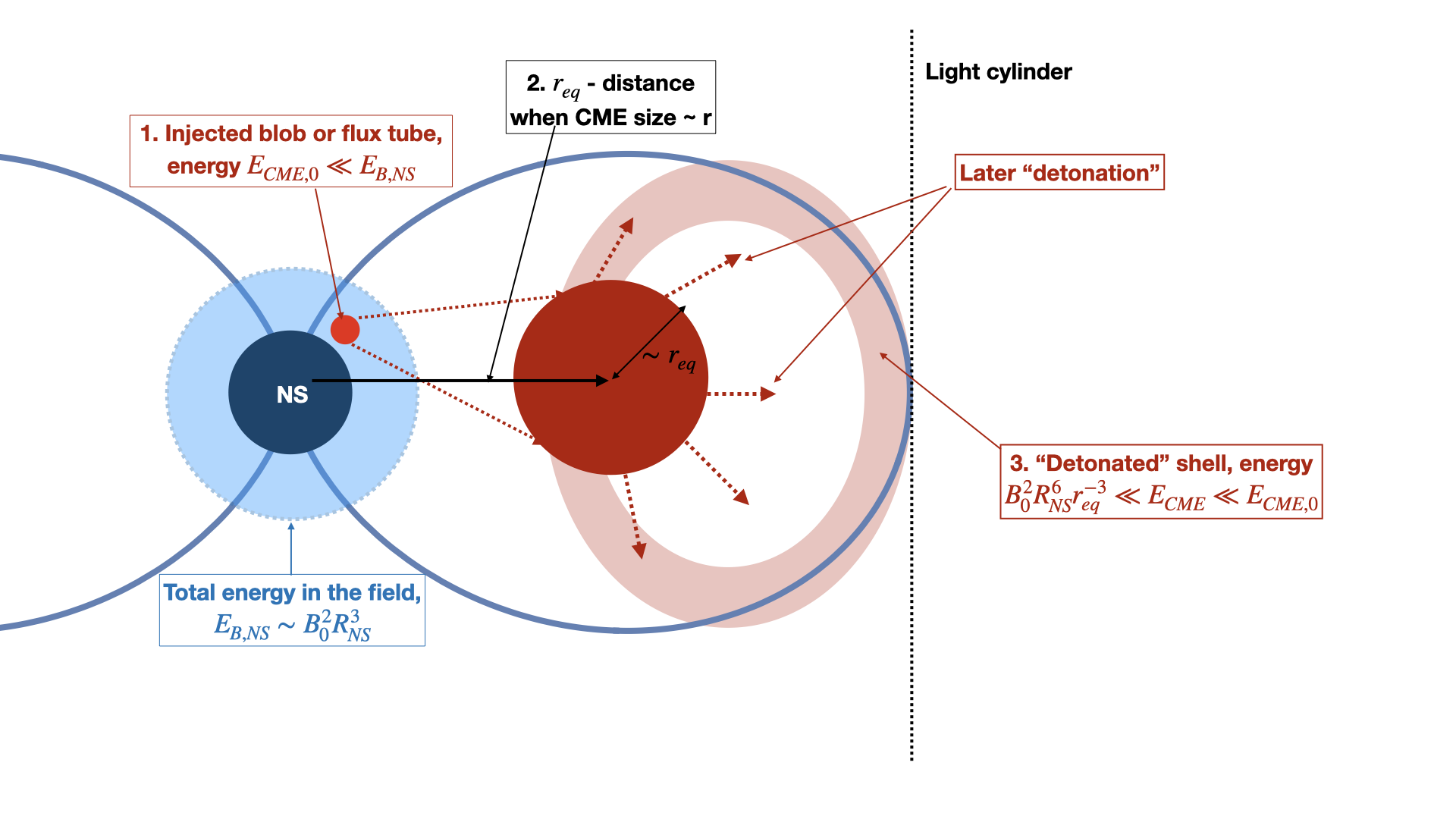}
 \vskip -40pt
\includegraphics[width=0.85\linewidth]{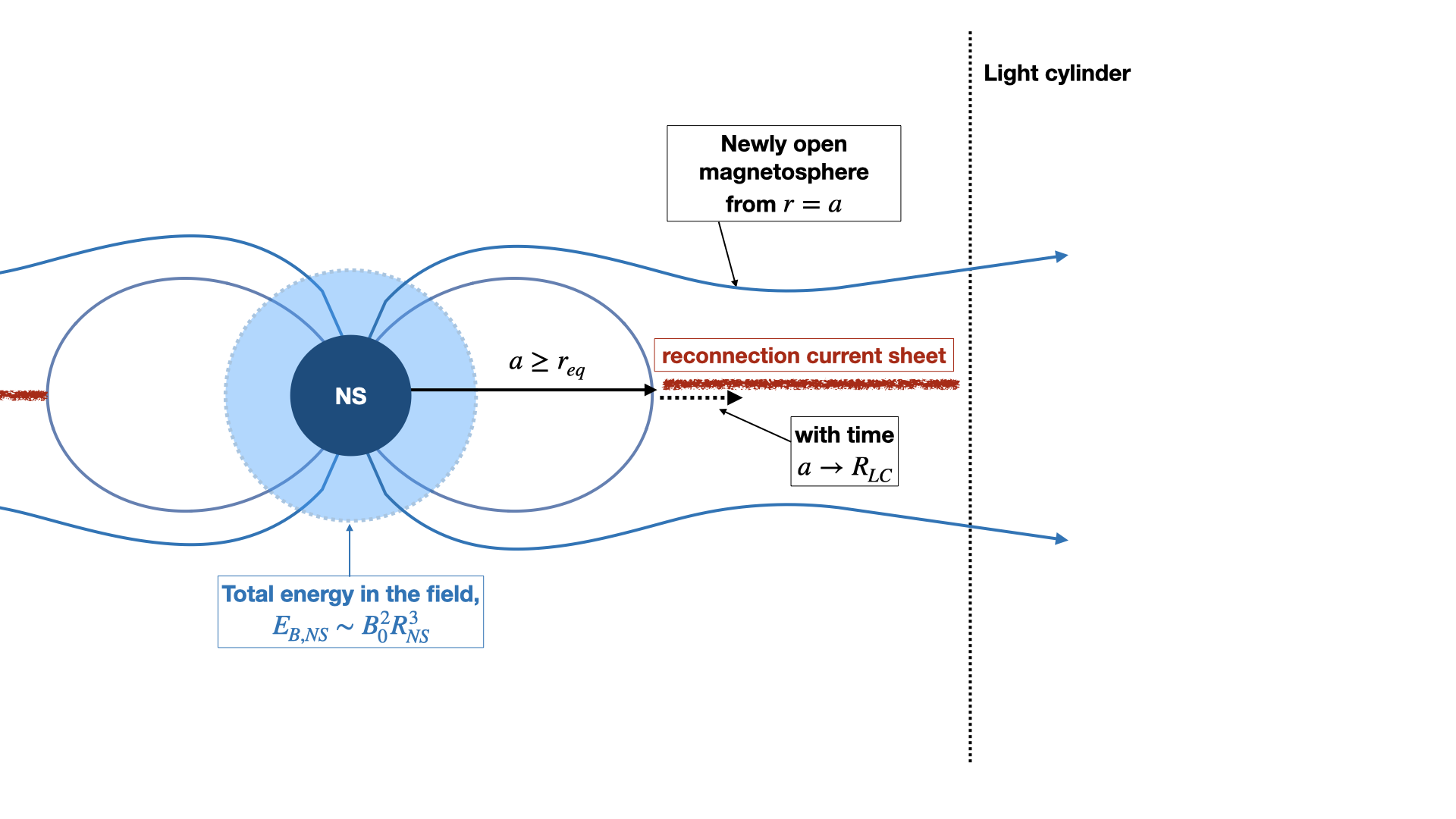}
 \vskip -40pt
      \includegraphics[width=0.65\textwidth]{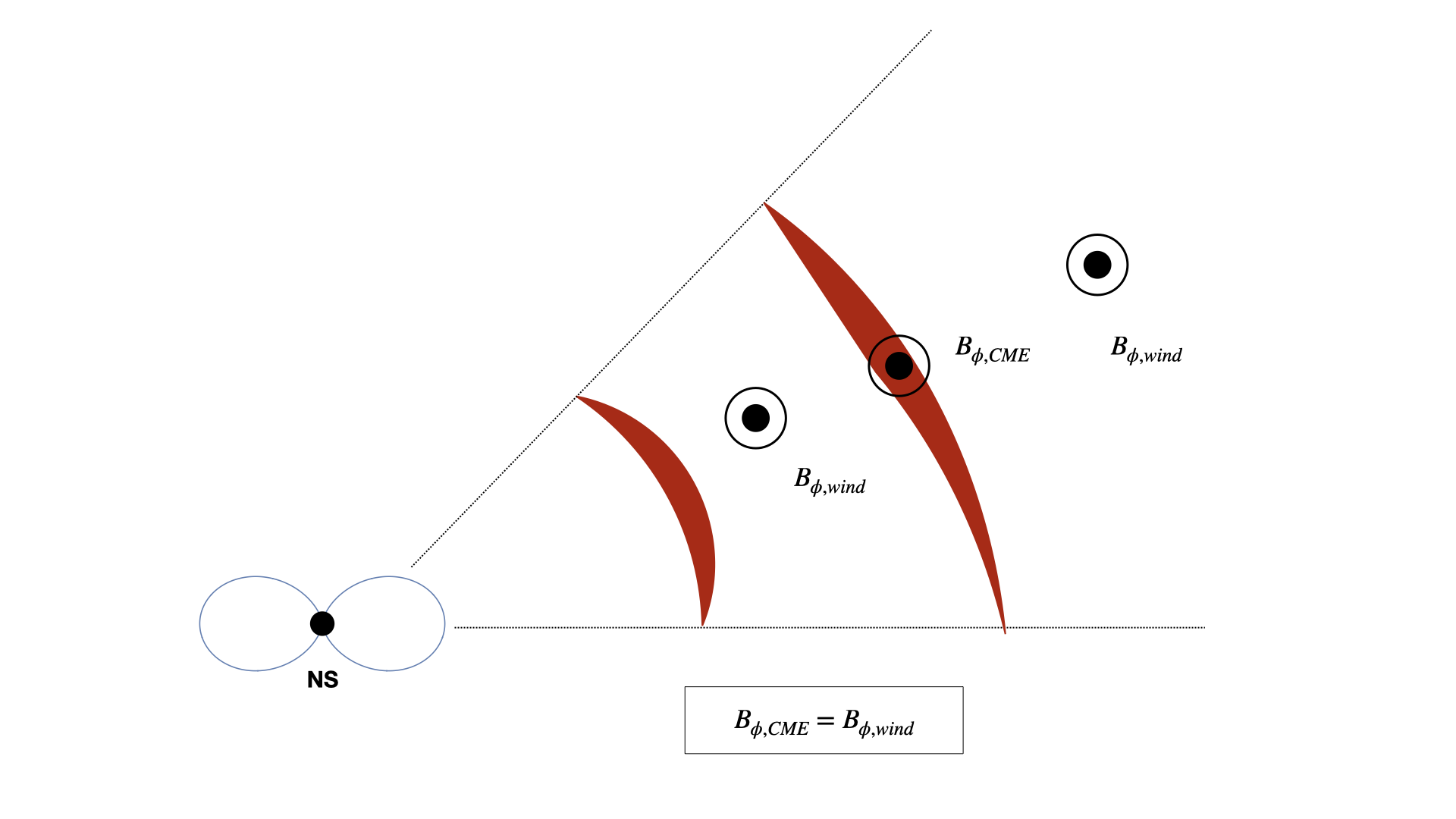}
    \caption{CME dynamics inside the \ms\ and in the wind. Top panel:  A CME, a flux tube or magnetic blob, carrying initial  energy $E_{CME,0}$   is released near the surface. The energy of the CME is much smaller than the total magnetic energy of the \ms, $ E_{CME,0} \ll E_{B,NS}$.
  As the CME   expands it is doing work on the over-laying \Bf, and loses energy.  Sufficiently powerful CME  may still  reach a size comparable to the local distance to the star $r_{eq}$ (while still within the \LC). After that the injected structure would expand quasi-spherically, opening field lines beyond  $r_{eq}$. Central panel: post explosion relaxation. Generation of CME lead to opening of the \ms\ at $r_{eq} \leq R_{LC}$. The post-CME \ms\ recovers by forming a current sheet from  $r_{eq} $ to $ R_{LC}$.
  Bottom  panel: CME in the wind.  A flux tube is injected  within the \ms, CME first expands with the \ms, losing energy doing  work  on the magnetospheric fields to break out,  coming to a force balance near the \LC, and then is advected passively within the wind as a shell of constant radial and lateral extension. }
\label{flux-tube-inside-magnsph1}
\end{center}
    \end{figure}

Importantly, we can then identify three  regimes for the dynamics of the CME: (i)  ``breaking-out''; (ii)  ``detonation''; (iii) magnetospheric recovery; (iii)  CME's expansion in the  wind, Figs. \ref{flux-tube-inside-magnsph1}.  During the  early ``breaking-out''  phase the CME expands while doing work on the overlaying \Bf. As a result, the energy of the CME reduces dramatically,  Table \ref{scales11}. During  ``detonation'' stage the CME expands nearly freely, opening the \ms. After the CME's break-out, the \ms recovers by forming a current sheet, while the CME is mostly passively advected with the wind. 


\subsubsection{The ``breaking-out'' stage}

At the ``breaking-out'' stage the internal \Bf\ (\ref{BCME}) matches the magnetospheric field  at the location of the CME, 
\be
B_{CME} = B_{NS}(r) = B_0  \left ( \frac{r}{R_{NS}}  \right) ^{-3} 
\label{balance}
\ee
(assuming $R_{CME}\ll r$; this is  not applicable for the ``large  CME'' case, right column in  Table \ref{scales11}). 
 Combining  (\ref{BCME})  and (\ref{balance}),
 \be
 \frac{ R_{CME}}{R_{CME,0}} =  \left ( \frac{r}{R_{NS}}  \right) ^{3/2}  
 \ee
 Thus, the cross-sectional area $\propto R_{CME}^2 \propto r^3$. This scaling is true for both the flux tube case and small blob case.
 Importantly, the CME expands laterally  
 \be
  \Delta \theta = \frac{ R_{CME}}{r}=  \frac{ R_{CME,0}}{R_{NS}} \left ( \frac{r}{R_{NS}}  \right) ^{1/2}  
  \ee

As the CME is  breaking-out through the overlaying \Bf, it does work on the magnetospheric \Bf. As a result, its internal energy sharply  decreases:  {\it at least} as the ratio of the CME energy to the total energy of the \ms, $ \eta_E$,  Table \ref{scales11}:
\be
\frac{r_{eq} } {R_{NS}}  \sim
\left\{
\begin{array}{cc}
\frac{E_{B, NS}} {E_{CME,0}} = \eta_{CME}^{-1}, & \mbox {spherical CME}\\
\left( \frac{E_{B, NS}} {E_{CME,0}}\right)^2 = \eta_{CME}^{-2}, & \mbox {flux tube}
\end{array}
\right.
\label{req}
\ee

We arrive at an  important conclusion: only a small fraction of the injected CME's energy affects the wind, at $r \geq R_{LC}$ - most energy is spent on work against the over-laying \Bf. Later-on, when the \ms\ recovers, the energy deposited into the \ms\ during CME break-out is dissipated in the newly created current sheet, see Fig \ref{rot_nonrot_comaparison}, on times scales much longer than the dynamic times scale of the injection.
 
 \subsubsection{The ``detonation'' stage} 
The dynamics changes from ``breaking-out'' to ``detonation'' when the total energy contained in the confining \Bf\ exterior to the position of the CME ($\sim B_0^2 R_{NS} ^6 r^{-3}$)  becomes smaller than the CME's internal energy (equivalently, when the size of the CME becomes comparable to the distance to the star).  This occurs at some equipartition radius $r_{eq}$, possibly  within the \LC, see Table \ref{scales11} and Eq. (\ref{req}).

 Beyond the $r_{eq}$ the dynamics changes: the CME has much more energy than the confining dipolar \Bf\ (from  $r_{eq}$ to infinity) - as a result the expansion enters ``detonation stage'' - nearly vacuum-like expansion \citep{2022ApJ...934..140B}. At this stage most of the \Bf\ is concentrated near the surface of exploding structure. Most importantly, the whole structure becomes causally disconnected. 

To enter  ``detonation'' stage the radius  $r_{eq}$ should be (much) smaller that the \LC\ radius. This requires sufficiently high injection energy, \eg  for ``small CMR''  column in Table \ref{scales11}, 
\be
\frac{ E_{CME,0} }{E_{B, NS}} \gg   \frac{R_{NS}} {R_{LC}} = 2 \times 10^{-4}  \, P^{-1} 
\ee
where $P$ is the spin period in seconds.
Thus, for spin period of one second, only CMEs that carry energy much  larger few thousandths of the total magnetospheric energy reach the ``detonation'' stage. 

 For example, for a magnetar with surface field $B_{NS} = 10^{15}$, the total magnetospheric  energy $E_{B, NS} \sim 10^{48} $ erg. Then to enter the detonation stage, the CME should have energy $\geq 10^{44} P^{-1} $ ergs. Only very powerful events experience detonation stage. Even if the CME's energy exceeds the critical, only small fraction, at  most $\sim \eta_E$ is transferred to the wind in the form of EM pulse.

For very energetic explosions, when equipartition radius $r_{eq}$ is  smaller than the \LC, the resulting CME ``detonates'': creates a causally disconnected shell of thickness $\sim r_{eq}$ that expands freely within the \ms. Locally, the dynamics is governed by the solutions of \cite{2010PhRvE..82e6305L,2012PhRvE..85b6401L} describing 1D expansion of magnetized fluid into vacuum. Most of the magnetic energy is concentrated near the surface of the expanding ball (see also Fig. \ref{zoomed_explodingring}).


 \subsubsection{Magnetospheric recovery} 
 
 The ``detonation'' stage, if it occurs, leads to temporary  opening of the 
\ms\ beyond the radius $r_{eq}$. The  post-CME \ms\ recovers by forming  a current sheet from  $r_{eq} $ to $ R_{LC}$, Fig. \ref{flux-tube-inside-magnsph1} middle panel.  Recovery proceeds slow - the rate of recover is controlled by dissipative processes in the current sheet. 

For $r_{eq}\ll R_{LC}$ the overall magnetic structure can be approximated as a \ms\ plus diamagnetic disk  \citep{1980A&A....86..192A,2023MNRAS.tmp..317L}. In this configuration the structure of the \ms\ beyond $r_{eq}$  is approximately monopolar  \citep[the case of balanced magnetic dipole in notation of][]{2023MNRAS.tmp..317L}. The location of the inner edge of the reconnection current sheet  $a$ slowly  approaches the \LC. At each moment the spindown power 
\ba&&
L_{sd} \approx  B_{NS}^2  \frac{R_{NS} ^6 \Omega^2}{ c a^2} \approx \left( \frac{R_{LC}}{a} \right)^2 \times   L_{sd,dipole} \gg   L_{sd,dipole}
\nn &&
 L_{sd,dipole} \approx  B_{NS}^2  \frac{R_{NS} ^6 \Omega^4}{c^3}
\ea
Thus, if detonation occurs, the  post-explosion spindown is much higher than the average. As argued by \cite{2013arXiv1306.2264L}, magnetospheric modifications naturally  explain the "anti-glitches" seen in some magnetars \citep{2013Natur.497..591A}.

\subsubsection{Beyond  \LC}

Dynamics beyond the   \LC\  depends on whether the CME reached the detonation stage or not. In the more likely scenario when the  detonation stage is not reached, the CME is just frozen into the wind, with the 
the lateral and radial extensions remaining nearly constant, see Fig. \ref{flux-tube-inside-magnsph1}, bottom panel,  so that it's cross-section $S$ and internal \Bf\ evolve according to
    \ba && B_0
    S \sim \Delta r \times r \Delta \theta \propto r
    \nn &&
    B_{in} = \Phi_0 /S \propto r^{-1}
    \label{Bin} 
    \ea
($\Phi_0$ is the value of the injected flux.) Scaling of $  B_{in} $ (\ref{Bin}) matches the scaling of the external wind \Bf.  {\it  Thus, after reaching a force balance close to the \LC\ the ejected flux tube remains in force-balance with the wind, and is passively advected.}  The flux tube expands along conical trajectory, with constant radial thickness. The energy contained in the flux tube remains constant:   the   expanding magnetic flux tube does not do any work on the surrounding wind.

If the   flare energy is sufficiently  large and the detonation stage is achieved, the \ms\ will open up at $r_{eq}$. As a  result an  \EM\ pulse  will be launched in the wind.  The energy of the pulse will  much smaller than the initial injection energy, $E_{CME}( r_eq) \ll 
E_{CME,0}$.

\section{Simulations with PHAEDRA code}
\label{PHAEDRA1}

\subsection{Global magnetospheric structure} 
\label{shearing_expressions}

 Investigations are done with the code PHAEDRA  \citep{2012MNRAS.423.1416P}, Appendix  \ref{NumericalMethod}. We have verified that  for non-sheared configurations our procedure reproduces the analytical solution and key known results (\eg formation of plasmoids at the Y-point), see Appendix \ref{noshear1}.


The first important ingredient that affects the  generation of  flares is the global structure of the  \ms. To investigate the influence of global magnetic structure on the generation of CMEs we first
consider several  initial magnetospheric configurations: purely dipole, twisted dipole-like configurations, dipole+quadrupole and dipole+octupole fields.

The expressions for magnetic fields of dipole, quadrupole and octupole, normalized with respect the field at the pole $B_p$  is given by
\ba &&
\B_{d}= \left\{  \cos \theta, \frac{\sin \theta}{2}, 0 \right\} B_p  \frac{R^3}{r^3} 
\nn &&
 \B_q = \left\{\frac{1}{4} (3 \cos (2 \theta )+1),\sin (\theta ) \cos (\theta ),0\right\} B_p \frac{R^4}{r^4}
\nn &&
\B_{o}=  \left\{\frac{1}{2} (5 \cos ( \theta )^2 -3),\frac{3}{4}\sin (\theta ) (5 \cos (\theta )^2 -1),0\right\}B_p \frac{R^5}{r^5}
\nn &&
\B_{tot} = \B_d + \mu_{q,o} \B_{q,o}
\label{multipole_field}
\ea
See Appendix \ref{multipole} for more  detailed description of analytically tractable case of dipole+quadrupole configuration. 

\begin{figure}
  \includegraphics[width=0.3\textwidth]{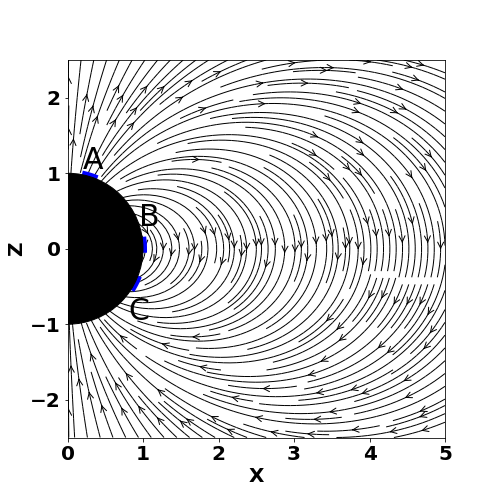}
  \includegraphics[width=0.3\textwidth]{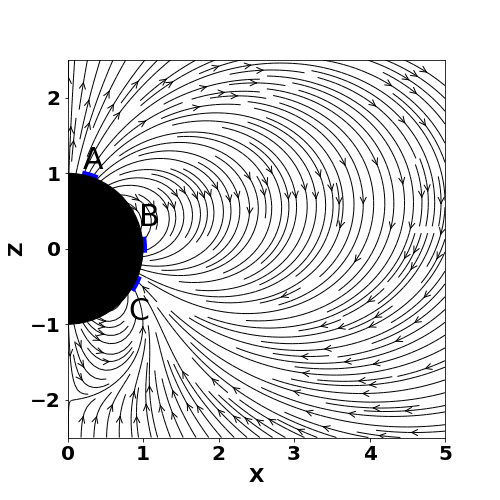}
  \includegraphics[width=0.3\textwidth]{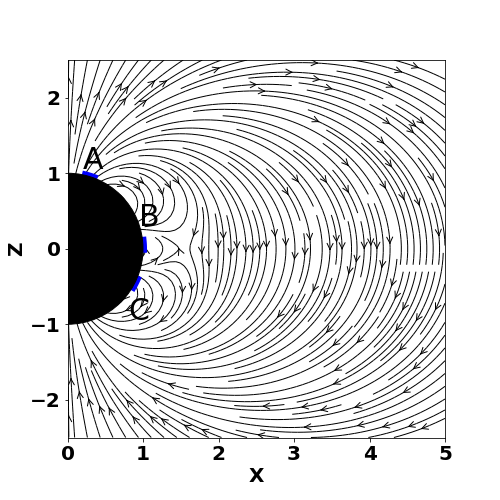}
    \caption{Initial poloidal field lines for superposition of dipole, dipole plus quadrupole, and dipole plus octupole (left to right) for  $\mu_{q}= 2$ and $\mu_{o}=3$, see Eq. (\ref{multipole_field}). 
     Shearing region A,B, and C are at $20 ^{\circ}$, $90 ^{\circ}$, and $120 ^{\circ}$ respectively from the z-axis. Here and below scales are normalized to \NS\ radius.}
    \label{intialfieldlines}
\end{figure}

We then study three different configurations:  (i) dipole; (ii) mixed dipole-quadrupole; (iii) mixed dipole-octuple. The relative strength of the higher multipoles is parameterized by $\mu_q$ and $\mu_o$.  In what follows we use $\mu_q =2$ and $\mu_o =3$ - in these cases  the higher order multipoles introduce non-trivial corrections to the surface fields, if compared with dipolar (\eg, in case of   dipole-quadrupole configuration, a ``dome'' appears near the south pole), see Figs. \ref{intialfieldlines}  and \ref{multipoles001}.

\subsection{Prescriptions for foot-point shear} 

The second important ingredient is the location of the shear.  The imposed shear on magnetic foot-points is confined to a small band, to be located at various latitudes and of latitudinal extent. The shearing is applied at the inner boundary of our simulation, the radius of the star.

When the two foot-points of the sheared arcade are well separated, we employ symmetric shear, moving azimuthally only one set  of footprints (This is nearly  equivalent to anti-symmetric shear, when the two footprints are moved in the opposite direction, given the overall spin of the star). The symmetric shear fails to create an expanding flux tube for the case of equatorial shear -  in that case  symmetric shear moves both footprints in the same direction, so that  the global magnetosphere can remain stationary \citep{2022MNRAS.513.1947L}. To induce explosion for the equatorial shear we apply antisymmetric prescription, Eq. \ref{asymmetric_shear_expression}.

\begin{figure}
 \centering
  \includegraphics[width=.75\linewidth]{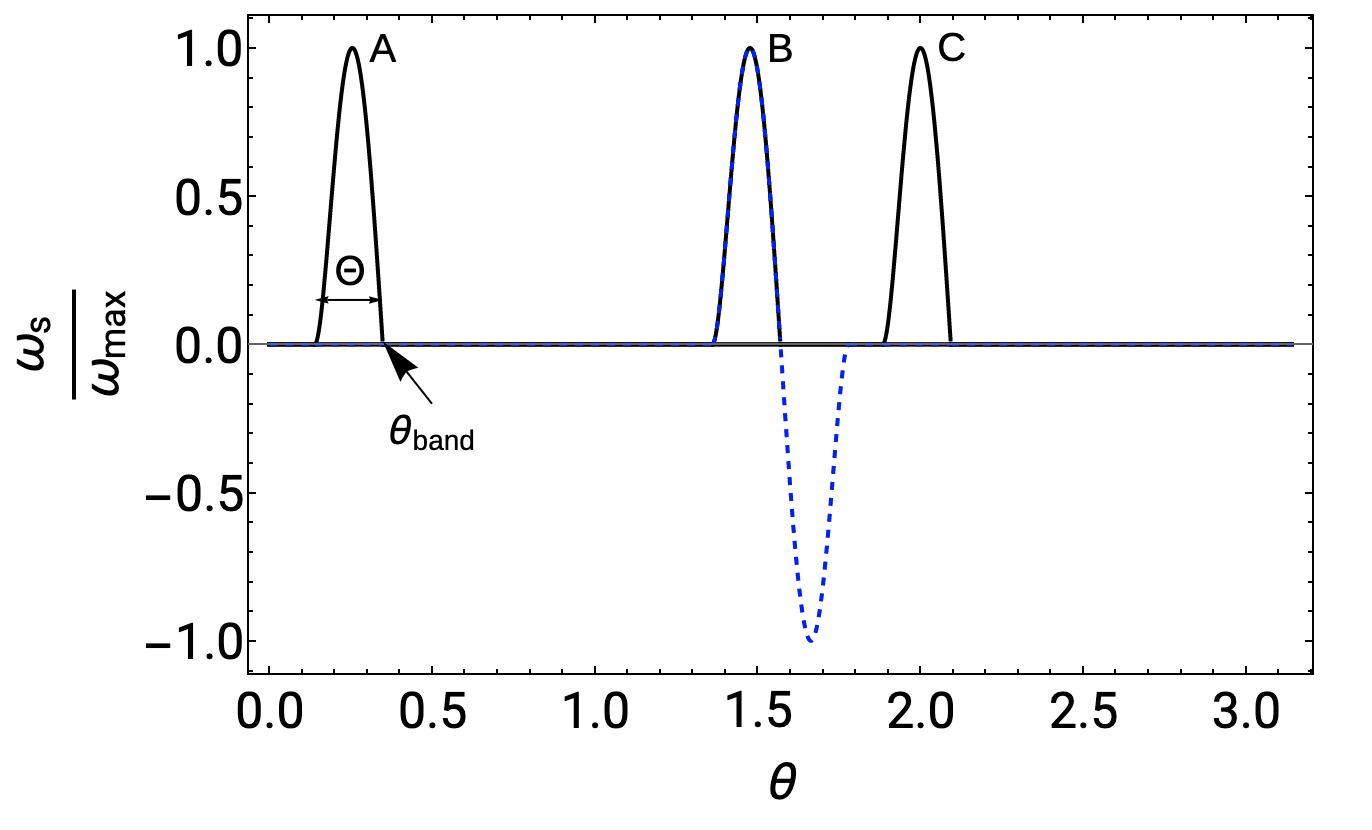}
     \caption{Normalized shear velocity ($\omega_{s}/\omega_{max}$) as a function of co-altitude $\theta$ (see Eq. \ref{symmetric_shear_expression},\ref{asymmetric_shear_expression}). Solid black curves shows shear profile for symmetric mode at three different regions (A,B,C) whereas blue dashed line represent the shear profile for anti-symmetric mode near the equator (region B). }
\label{shearingband}
    \end{figure}

We employ several prescriptions for symmetric shear. First, 
we follow the discussion in section 3  of \cite{1999ApJ...510..485A}. In that case the  angular velocity of the foot-points  $\omega_{s}$ is (see Fig. \ref{shearingband})
\begin{equation}
\omega_{s}(\theta)=\left\{\begin{array}{ll} \omega_{max}g(\theta), & \text { for } \theta_{band} - \Theta \le \theta \le  \theta_{band}\\
0, & \text { otherwise }
\end{array}
\right.
\label{symmetric_shear_expression}
\end{equation}
Here,   $ \omega_{max}$ is the maximum value of applied shear,  and function $g(\theta)$,
\ba &&
g(\theta) = C \left(\psi^{2}-\Theta^{2}\right)^{2} \sin \psi 
\nn &&
 \psi = \theta_{band} - \theta,
 \ea 
defines the latitudinal extent
of the shear region, and $\theta_{band} $ is the polar angle around which shearing is applied. $C$ is a normalization constant introduced to ensure that max$|g(\theta)| = 1$ and $\Theta = \pi/15$ is the assumed latitudinal extent of the shear layer.

We can also construct expression for anti-symmetric shearing as below: 
\begin{equation}
\omega_{s}(\theta)=\left\{\begin{array}{ll} \omega_{max}g(\theta), & \text { for } \theta_{band} - \Theta \le \theta \le  \theta_{band} +  \Theta \\
0, & \text { otherwise }
\end{array}
\right.
\label{asymmetric_shear_expression}
\end{equation}

Since for the rotating case, we are mostly interested in ratio of shear velocity to the stars rotation velocity, we rearrange Eq. \ref{symmetric_shear_expression}, to get

\begin{equation}
\frac{\omega_{s}}{\Omega_{\star}}=\left\{\begin{array}{ll}
\xi g(\theta), & \text { for } \theta_{band} - \Theta \le \theta \le  \theta_{band}\\
0, & \text { otherwise }
\end{array}\right.
\end{equation}
Where $\xi = {\omega_{max}}/{\Omega_{\star}}$.

The shear profile for both symmetric and anti-symmetric case, as the function of $\theta$ is plotted in Fig. \ref{shearingband} for three different location of shearing band. 

\section{Coronal mass ejections deep inside magnetospheres.}
\label{generation-inside}

 \subsection{Sheared non-rotating \mss}
 \label{nonrotatingsystem}

We start this work by probing static non-rotating configurations with shearing introduced at different locations.  The main justification is the limited dynamic range of simulations of the rotating \mss, \S \ref{generation}.  Our typical \LC\ radius is only 5 stellar radii, while in case of magnetar the expected ratio is in the tens of  thousands (for $\sim 1$ second period). 

As a key new ingredient, we  probe the effect of various magnetic field topologies by adding  contributions from other multipoles. We achieved this by superimposing quadrupole and octupole field on star's dipolar field (see  \S \ref{shearing_expressions}). Since we are mostly interested in the plasmoid ejections, we chose relatively high shearing rate to ensure the magnetosphere enters into a non-equilibrium dynamic states \citep{2013ApJ...774...92P,1994ApJ...430..898M}. For this and the subsequent simulations, the maximum shearing rate $\omega_{max}$ was chosen as $0.1$ (so that the \LC\ corresponding to the shearing motion is at 10 stellar radii.)

\begin{figure*}
 \centering
 \rotatebox[origin=c]{0}{$\textbf{(a)Region A: $20^{\circ}$}$}\hspace*{5em}
    \rotatebox[origin=c]{0}{$\textbf{(b)Region B: $90^{\circ}$}$}\hspace*{5em}
    \rotatebox[origin=c]{0}{$\textbf{(c)Region C: $120^{\circ}$}$}\hspace*{-1.5em}\\
    \rotatebox{90}{\hspace{3em}  Dipole}
        \includegraphics[width=0.3\textwidth]{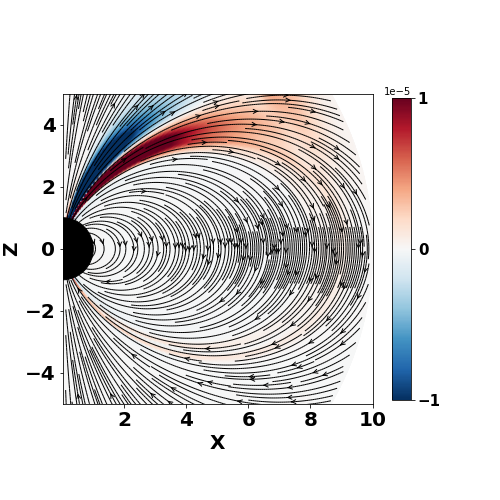}
        \includegraphics[width=0.3\textwidth]{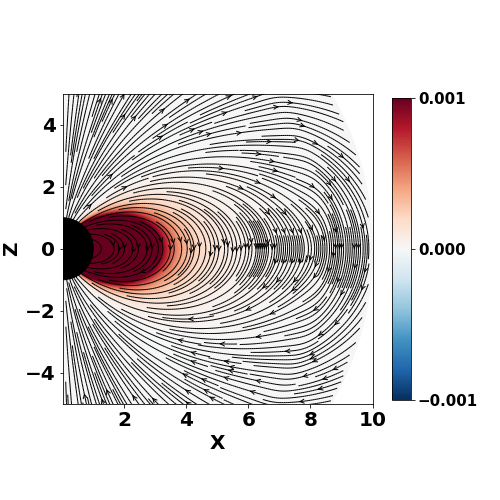}
        \includegraphics[width=0.3\textwidth,cfbox=red 1pt 1pt]{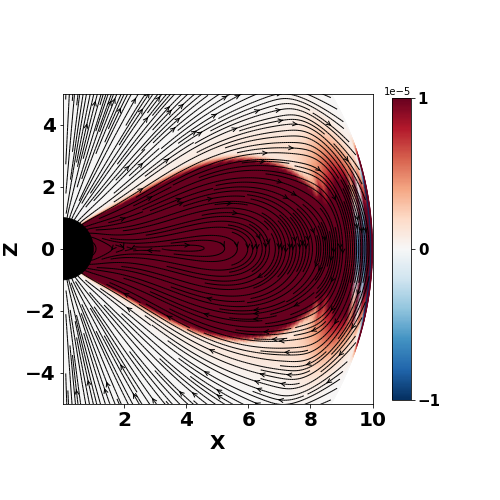}\\
\rotatebox{90}{\hspace{2em}  Dipole+Quadrupole}
        \includegraphics[width=0.3\textwidth]{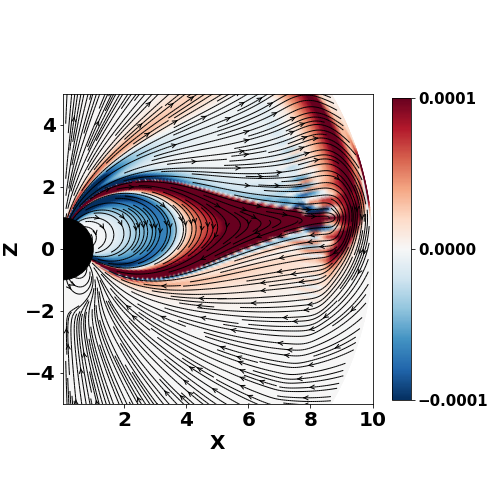}
        \includegraphics[width=0.3\textwidth,cfbox=red 1pt 1pt]{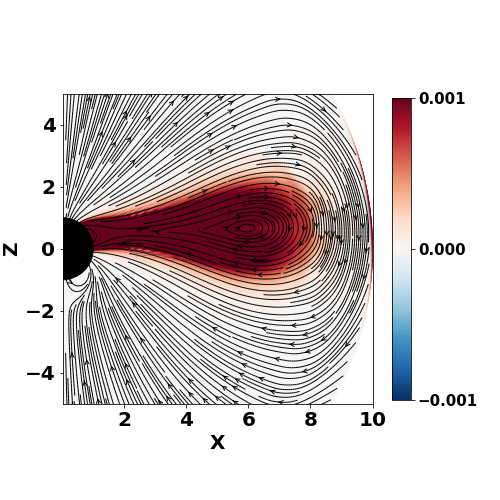}
        \includegraphics[width=0.3\textwidth]{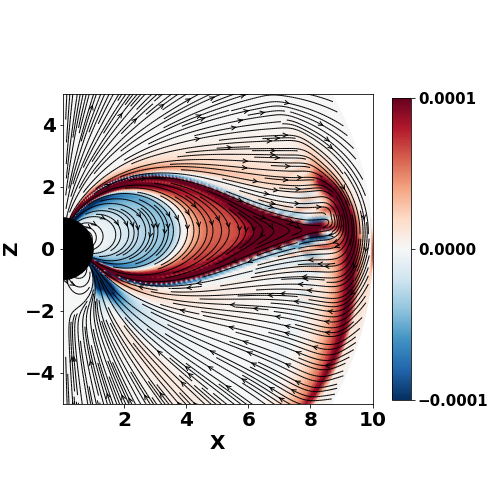}\\
\rotatebox{90}{\hspace{3em}  Dipole+Octupole}
        \includegraphics[width=0.3\textwidth]{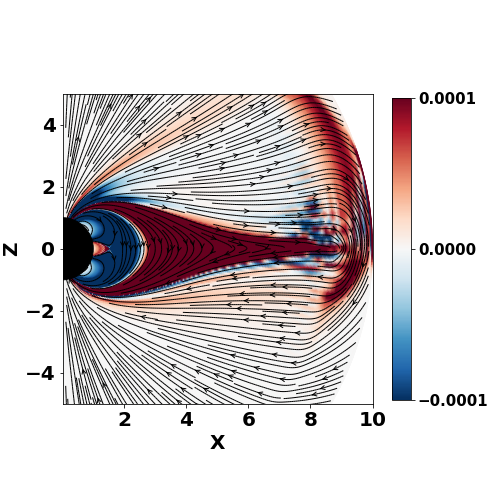}
        \includegraphics[width=0.3\textwidth,cfbox=red 1pt 1pt]{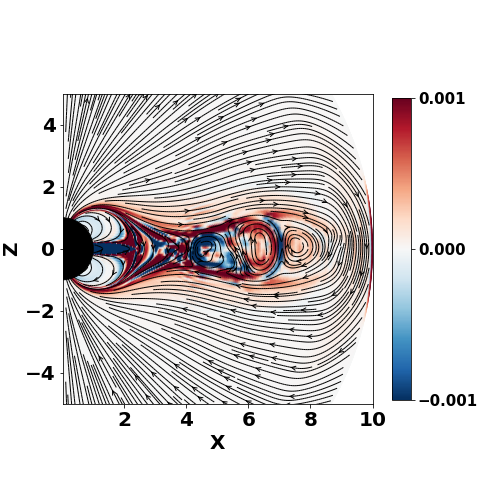}
        \includegraphics[width=0.3\textwidth,cfbox=red 1pt 1pt]{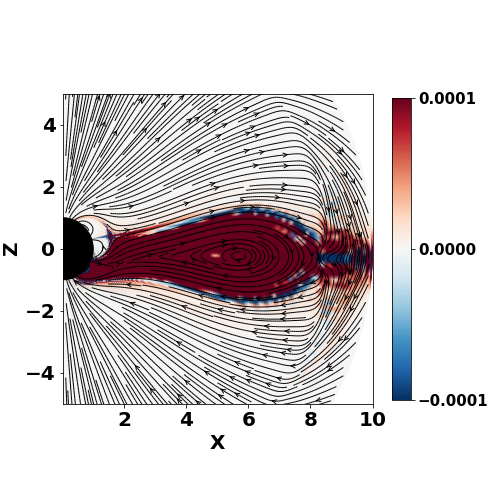}
    \caption{ Toroidal current density$( J_{\phi}) $ for non-rotating configurations shear at various  altitudes ($20^{\circ}$, $90^{\circ}$, and $120^{\circ}$ from left to right). Shearing rate is    $\omega_{max}$ =0.1. In all cases, when shearing  is done close to the north pole (region A, left column), no major ejection events are observed. When shearing near the equator (region B, middle column), in all cases we observe powerful ejections. When shearing  is done at region C (right column) whether or not ejections are observed depend on the magnetic field topology. Red boxes are drawn around configurations where a clear expulsion of plasmoids is observed.} 
\label{Jphi_nonrotatingstar_chi=01} 
\end {figure*}

The shearing of foot-points starts immediately at beginning of simulations $t=0$, causing the field lines to twist. The subsequent evolution of the system is visualized in Fig. (\ref{Jphi_nonrotatingstar_chi=01}), where we plot toroidal current density $J_{\phi}$ for combination  of different shearing altitude and magnetic field topology. To demonstrate the importance of location of foot-point shearing, we consider three different shearing regions: near the poles (region A), at equator (region B), and at $\sim 120^{\circ}$  from the poles (region C).

\begin{figure*}
     \centering
     \subfloat[\label{0a}]{\includegraphics[width=.32\linewidth]{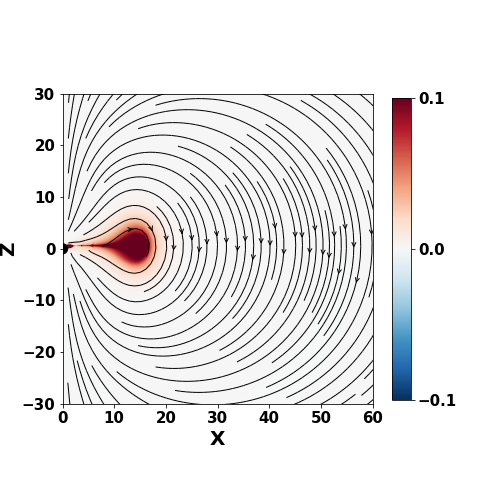}}
      \subfloat[\label{0b}]{\includegraphics[width=.32\linewidth]{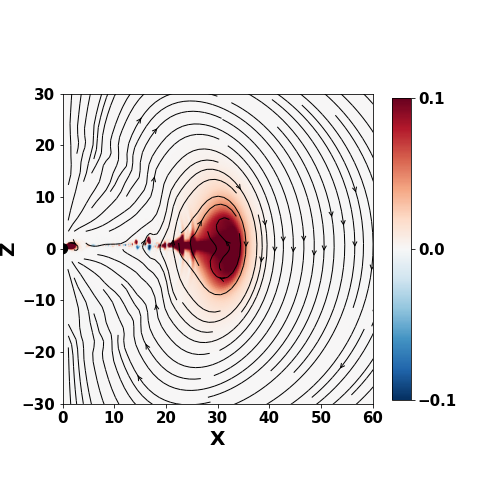}}
      \subfloat[\label{0c}]{\includegraphics[width=.32\linewidth]{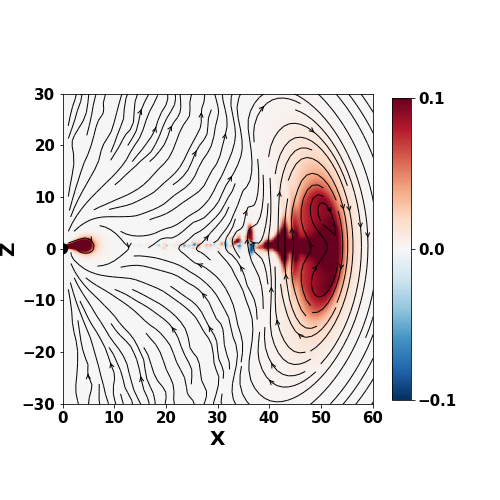}}\\
       \subfloat[\label{0d}]{ \includegraphics[width=0.32\textwidth]{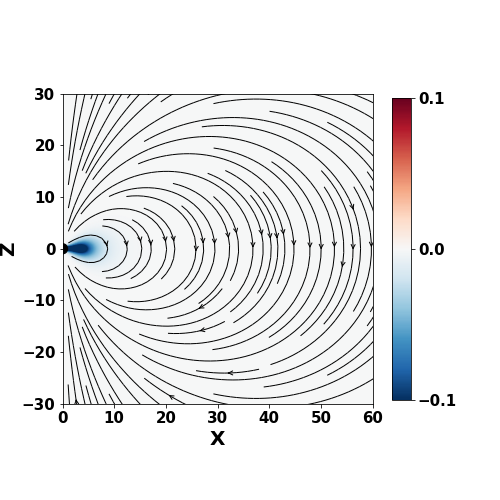}}
        \subfloat[\label{0e}]{ \includegraphics[width=0.32\textwidth]{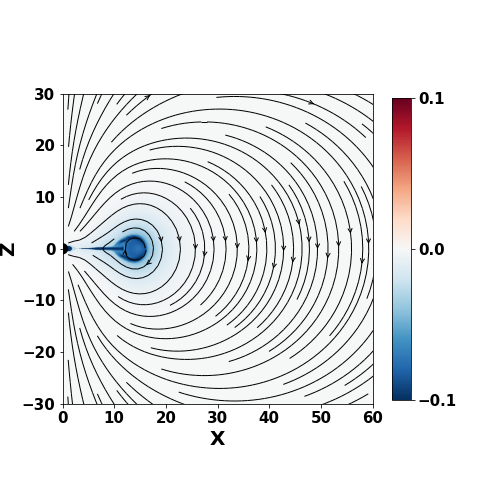}}
         \subfloat[\label{0f}]{\includegraphics[width=0.32\textwidth]{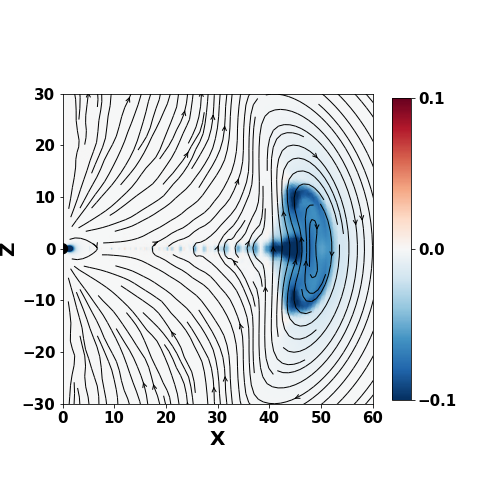}}
   \caption{Time evolution for non-rotating dipole+ quadrupole sheared at point B (top panel), and dipole sheared anti-symmetrically at point B (bottom panel).  Color is toroidal magnetic field $r \sin(\theta) B_{\phi} $ (this applies  other figures unless stated otherwise), streamlines are poloidal fields. Shearing starts at beginning of the simulation. }
 \label{nonrotating_largescale}
\end{figure*}

\begin{figure*}
     \centering
     \includegraphics[width=.5\linewidth]{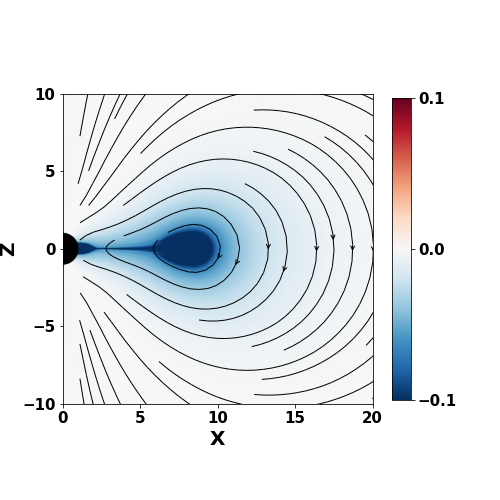}
     	\includegraphics[width=.45\linewidth]{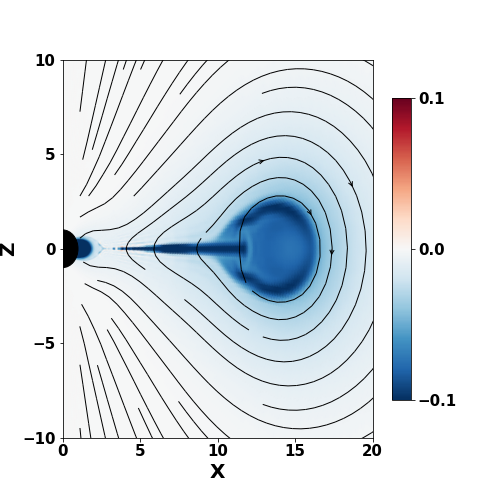}
   \caption{Detonating CME within the \ms.  Right panel: zoomed-in version of Fig. \ref{0e}: notice  the  edge-enhances  structure of the detonating CME, \cf\ Fig \protect\ref{flux-tube-inside-magnsph1}. As discussed by \cite{2022ApJ...934..140B}, detonating magnetic structures are concentrated near the surface. Left panel: we show the system  approximately a  third  of  a period before the detonation. } 
 \label{zoomed_explodingring}
\end{figure*}

No ejections were seen when shearing region is located near the poles (Region A) for all three initial magnetic field topology. As observed in Fig. \ref{Jphi_nonrotatingstar_chi=01} left panel, there isn't substantial poloidal expansion and the  system attains a quasi-equilibrium state where most of the field lines remain closed.

For the case equatorial shearing (region B), we see major explosive events for superposition of dipole and quadrupole/octupole topologies. We attribute this to the significant opening of the closed lines. While the ejection is evident for the case of dipole and octupole superposition, we show the final inflated state for remaining two configurations: the structure breaks away and exits the simulation box at the next time step. This equatorial expansion is consistent with previous simulations by other authors \citep{2013ApJ...774...92P,1994ApJ...430..898M}. As argued in above mentioned works, field line opening causes the formation of a current sheet and the subsequent reconnection of field lines triggers ejection of magnetic energy in the form of plasmoids.

The ejections profile while shearing region C, however depends on the magnetic topology, as depicted in Fig. \ref{Jphi_nonrotatingstar_chi=01} right panel. Unlike in the case of dipole and dipole+octupole magnetosphere, the field lines for dipole+quadrupole topology are only partially open and the system achieves a quasi-equilibrium state. In Fig. \ref{Jphi_nonrotatingstar_chi=01} we highlight those cases where the system explodes and ejects plasmoids with a red box.

We further show the large scale time evolution of two selected configuration : dipole+quadrupole, and dipole sheared anti-symmetrically in Fig. \ref{nonrotating_largescale}. We see ejection events in both scenarios, albeit at different time.  In Fig. \ref{zoomed_explodingring}, we zoom in on Fig. \ref{0e} to highlight  the structure of the exploded shell - most of the energy/\Bf\  is concentrated near the surface of  a detonating flux tube as discussed/ simulated by \cite{2022ApJ...934..140B}.

\section{Coronal mass ejections by rotating  magnetospheres (slow shear)}
\label{generation}


 

\subsection{Results: magnetospheric dynamics for slow shear}
\label{Magnetospheric}

Next  we proceed to the main topic of this paper: dynamics of  Coronal Mass Ejections (CMEs) in relativistic rotating \mss\ with sheared foot points.   
 For slow shear we  set $\omega_{max}=0.1$, which corresponds to $\xi =0.5$  (maximal shearing rate is half the spin). 
  The shearing of the stellar surface begins after one rotational time period, to ensure that our initial un-sheared system is in an equilibrium state.

\begin{figure*}
     \centering
     	\includegraphics[width=.32\linewidth]{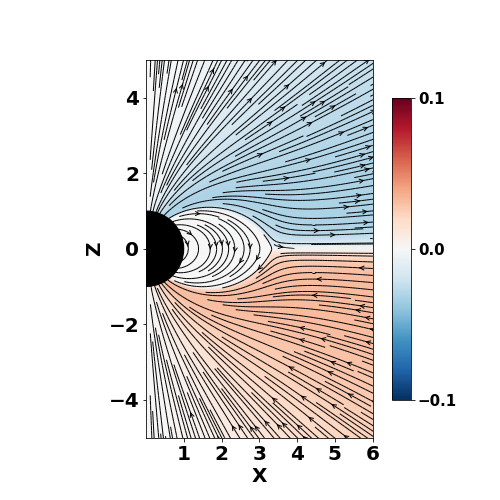}
	\includegraphics[width=.32\linewidth]{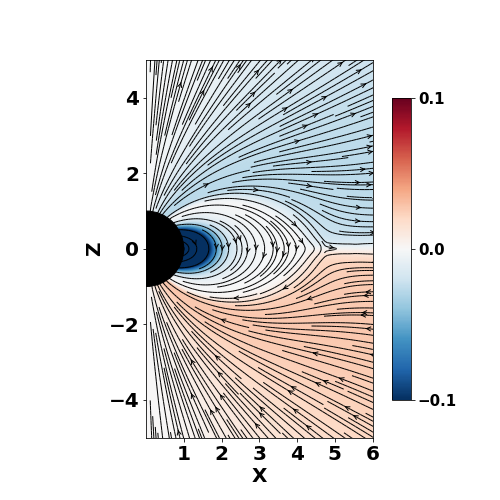}
	\includegraphics[width=.32\linewidth]{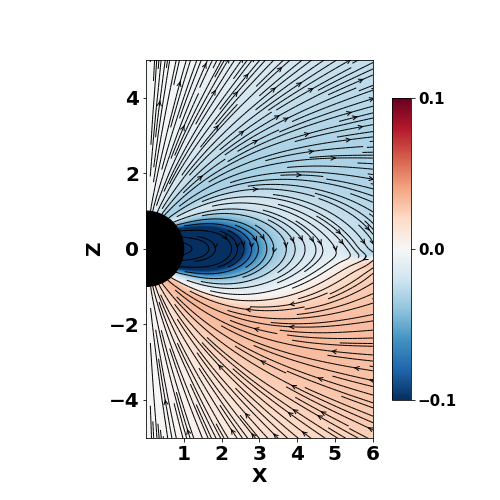}\\
       \includegraphics[width=0.32\textwidth]{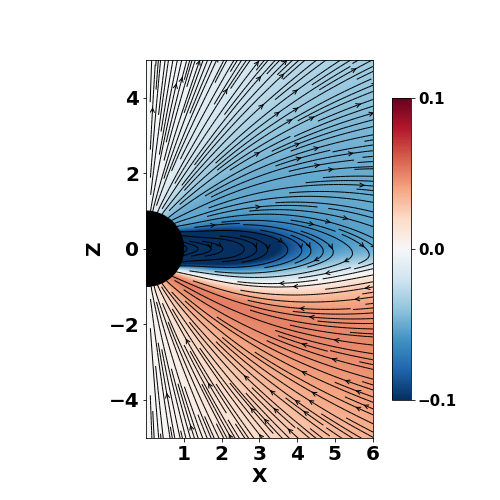}
        \includegraphics[width=0.32\textwidth]{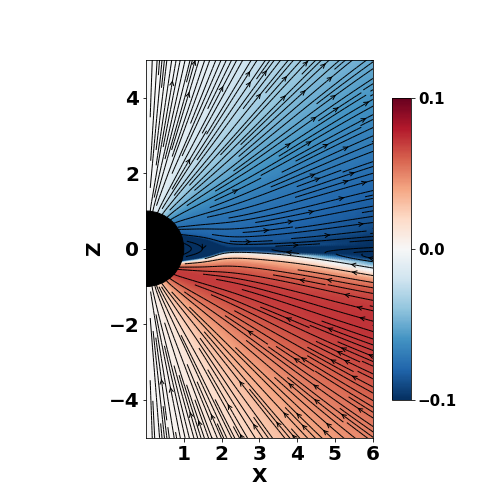}
        \includegraphics[width=0.32\textwidth]{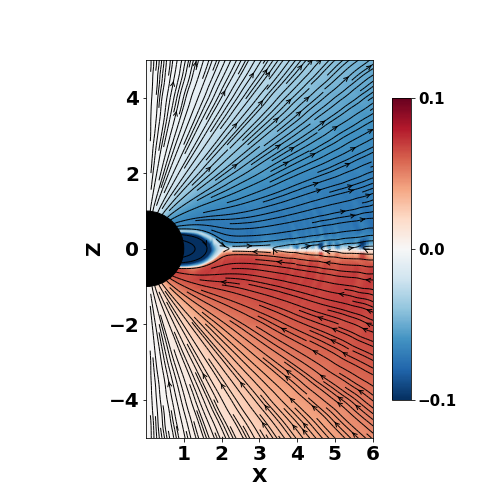}
   \caption{Time evolution  for rotating  dipole configuration sheared anti-symmetrically at point B (equatorial), slow shear. Snapshots are taken at $t=\{0.95,3.2,5.4,6.4,7.0,7.3\}$.  Shearing starts at $t=1$  (after one rotation period). One  can clearly observe the opening of field line and ejection of a CME.  After the  CME is ejected  the closed part of the \ms\ is smaller, with the current sheet showing plasmoid instability.  The final configuration has non-zero twist on closed field lines - compare first and last panels. See also Fig. \protect\ref{dipoleantisymmetric_gamma} for the large scale dynamics.}
 \label{dipole_anitsymmetric}
\end{figure*}

 We start with basic case of dipolar \ms\ sheared at the equator with anti-symmetric shearing profile (\ref{asymmetric_shear_expression}), Fig. \ref{dipole_anitsymmetric}.  We introduce magnetic foot-point shearing to rotating \NSs.  
 One  can clearly observe the opening of field line and ejection of a CME.  After the  CME the closed part of the \ms\ is smaller, with the current sheet showing plasmoid instability.  The final configuration has non-zero twist on closed field lines.
 
 The behavior matches the expectations:  the closed field lines  become partially or even fully opened in response to finite foot-point shearing due to additional magnetic pressure from the toroidal component of the magnetic field \citep{1992ApJ...391..353W}.  The opened field lines  subsequently causes the expulsion of magnetic energy in the form of Coronal Mass Ejections. 
 The post-CME relaxation is a new effect: formation of smaller closed \ms, with  plasmoid-mediated  current sheet deep inside the \LC, and a slow, reconnection-mediated relaxation to a new equilibrium with twisted field lines.

\begin{figure*}
     \centering
	\includegraphics[width=.40\linewidth,height=.22\textheight]{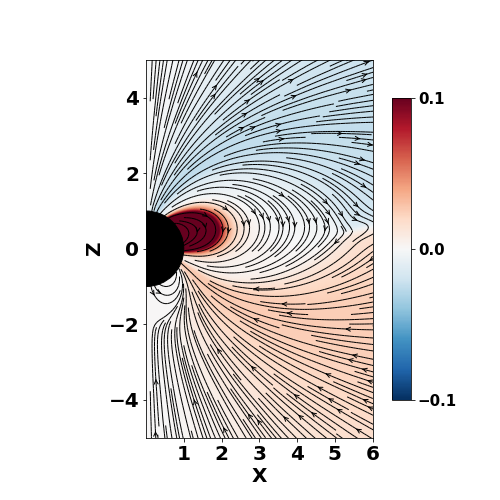}
	\includegraphics[width=.40\linewidth,height=.22\textheight]{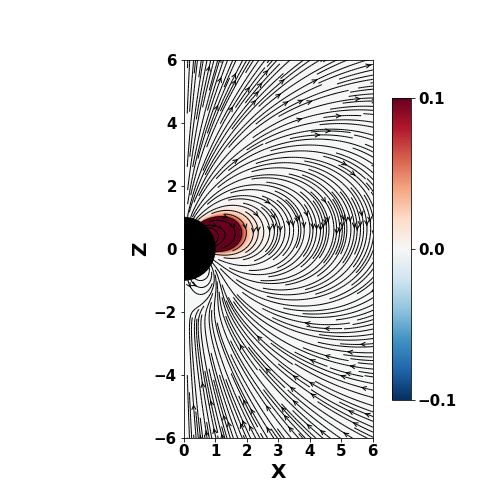}\\
       \includegraphics[width=0.40\textwidth,height=.22\textheight]{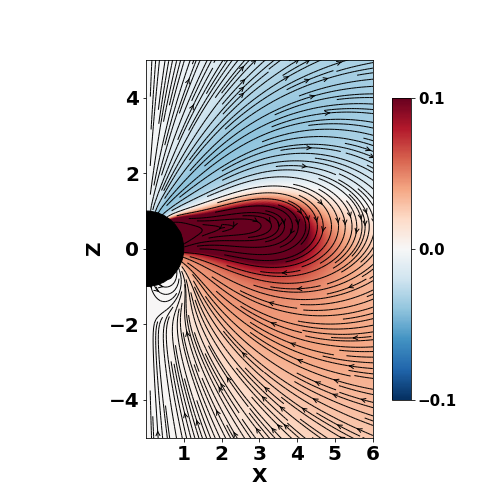}
        \includegraphics[width=0.40\textwidth,height=.22\textheight]{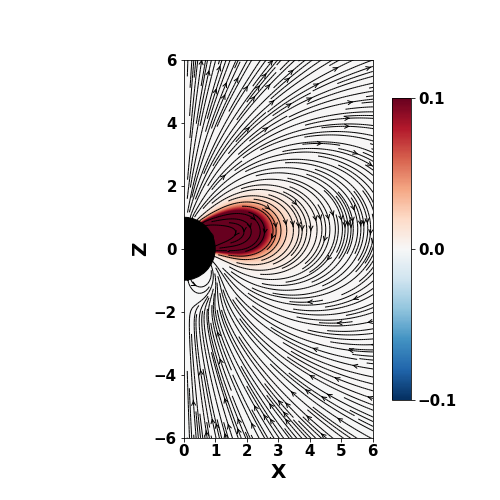}\\
         \includegraphics[width=0.40\textwidth,height=.22\textheight]{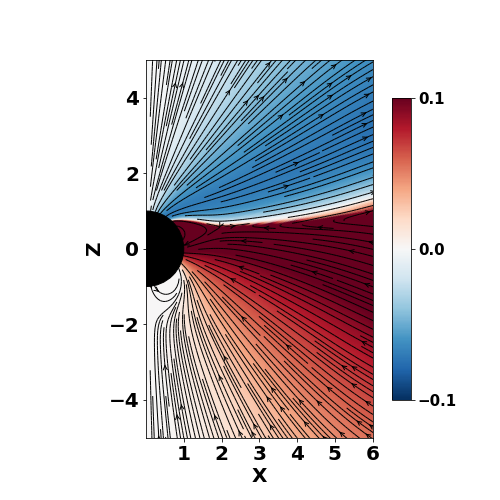}
        \includegraphics[width=0.40\textwidth,height=.22\textheight]{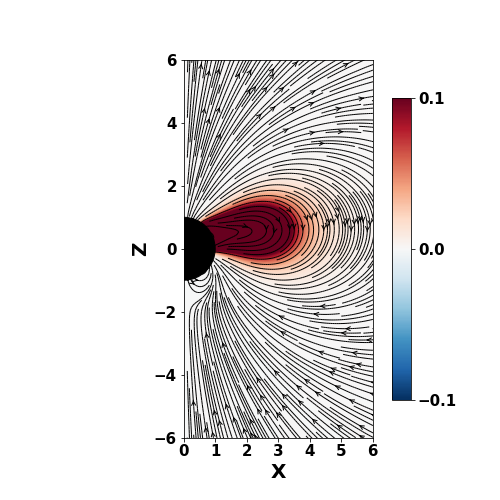}\\
        \includegraphics[width=.40\linewidth,height=.22\textheight]{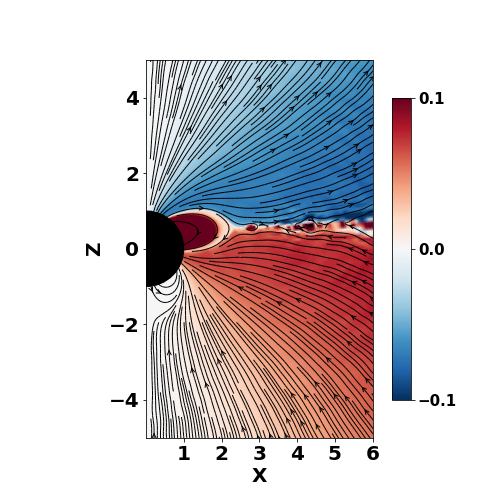}
	\includegraphics[width=.40\linewidth,height=.22\textheight]{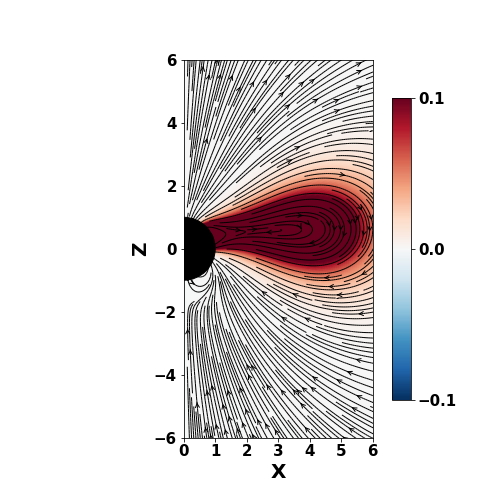}\\
   \caption{ Dipole + quadrupole configuration sheared at point B (equatorial), slow shear.
   Left column: rotating case, right column: non-rotating case (snapshots are taken at same time, as measured from the start of the shearing.). One clearly sees the 
   influence of the \LC\ on the development of the CME.  This demonstrates that the \LC\ makes it ``easier'' for the CME to break out.
 }
\label{rot_nonrot_comaparison}
\end{figure*}

Next we show  time evolution of sheared dipole and quadrupole system for equatorial shearing  in Fig. \ref{rot_nonrot_comaparison}, left column. The opening of field lines and subsequent ejection of CME is clearly evident.  In Fig. \ref{rot_nonrot_comaparison} we compare the dynamics  the same configuration (dipole + quadrupole configuration sheared at point B), between rotating and non-rotating cases.
We clearly see that it is much easier to break out from the rotating \ms. This is expected, since in the rotating case the breakout occurs when the top of the inflated loop reaches the \LC. Fig. \ref{rot_nonrot_comaparison}  also demonstrates that though our dynamic range is not very large (\LC\ at only five stellar radii), we do  correctly capture the dynamics of the inflated flux tube within the \ms.

 In Fig. \ref{rot_nonrot_comaparison} we compare the dynamics  the same configuration (dipole + quadrupole configuration sheared at point B), between rotating and non-rotating cases.
We clearly see that it is much easier to break out from the rotating \ms. This is expected, since in the rotating case the breakout occurs when the top of the inflated loop reaches the \LC. Fig. \ref{rot_nonrot_comaparison}  also demonstrates that though our dynamic range is not very large (\LC\ at only five stellar radii), we do  correctly capture the dynamics of the inflated flux tube within the \ms.

\begin{figure*}
 \centering
 \rotatebox[origin=c]{0}{$\textbf{(a)Region A: $20^{\circ}$}$}\hspace*{5em}
\rotatebox[origin=c]{0}{$\textbf{(b)Region B: $90^{\circ}$}$}\hspace*{5em}
\rotatebox[origin=c]{0}{$\textbf{(c)Region C: $120^{\circ}$}$}\hspace*{-1.5em}
\\
\rotatebox{90}{\hspace{3em}  Dipole}
\includegraphics[width=0.3\textwidth]{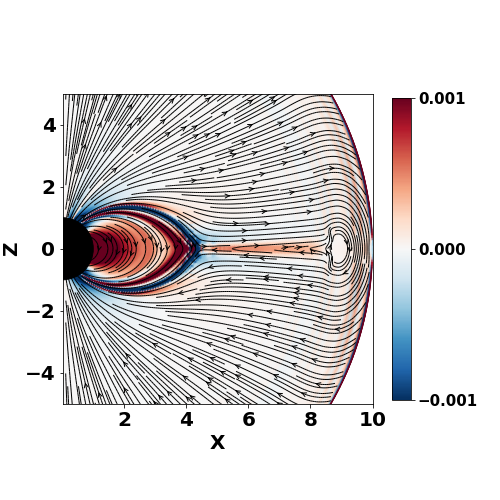}
\includegraphics[width=0.3\textwidth]{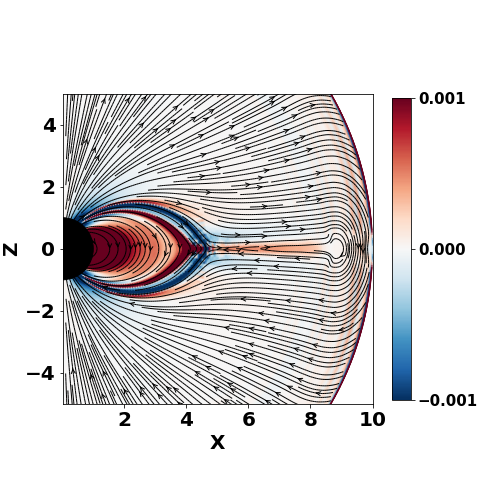}
\includegraphics[width=0.3\textwidth,cfbox=red 1pt 1pt]{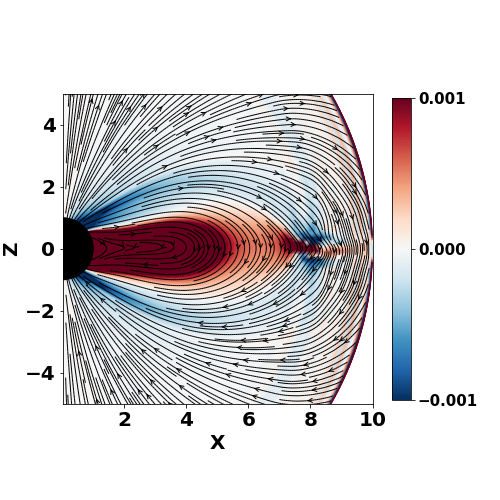}\\
\rotatebox{90}{\hspace{2em}  Dipole+Quadrupole}
\includegraphics[width=0.3\textwidth]{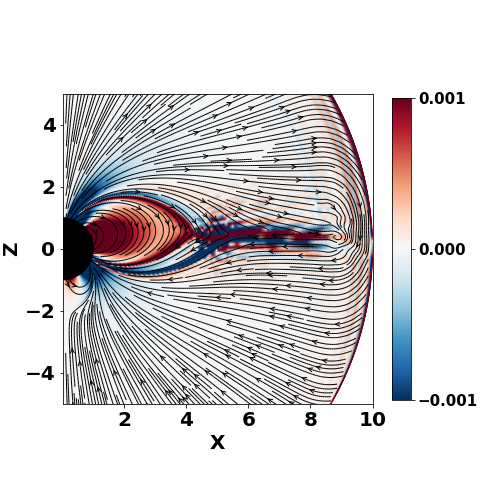}
\includegraphics[width=0.3\textwidth,cfbox=red 1pt 1pt]{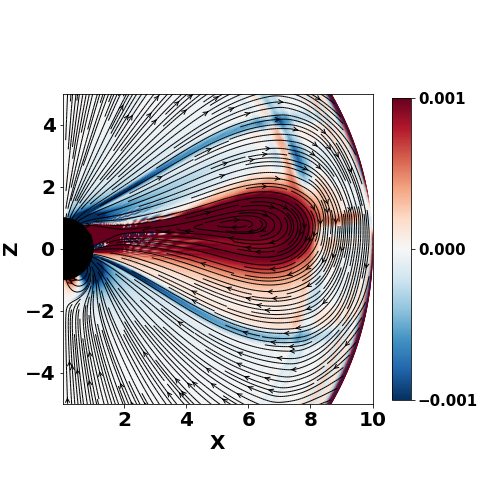}
\includegraphics[width=0.3\textwidth]{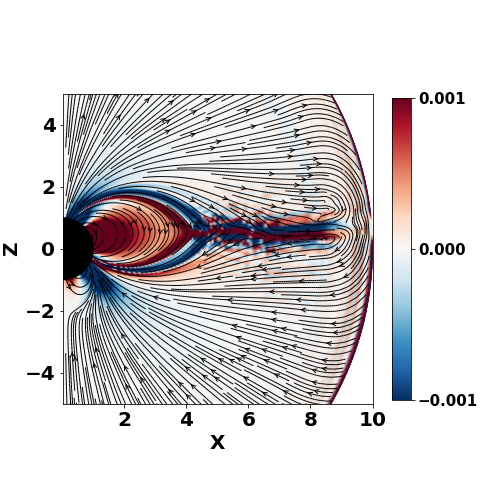}\\
\rotatebox{90}{\hspace{3em}  Dipole+Octupole}
\includegraphics[width=0.3\textwidth]{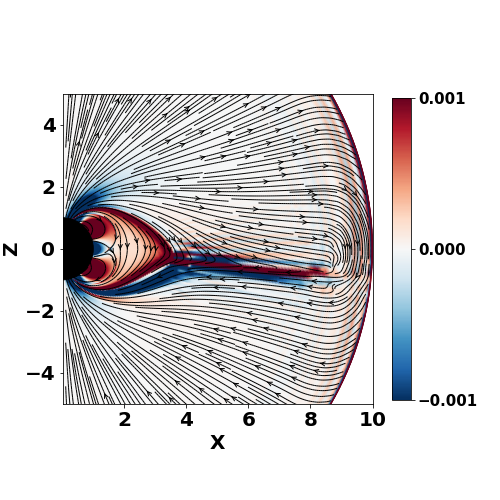}
\includegraphics[width=0.3\textwidth,cfbox=red 1pt 1pt]{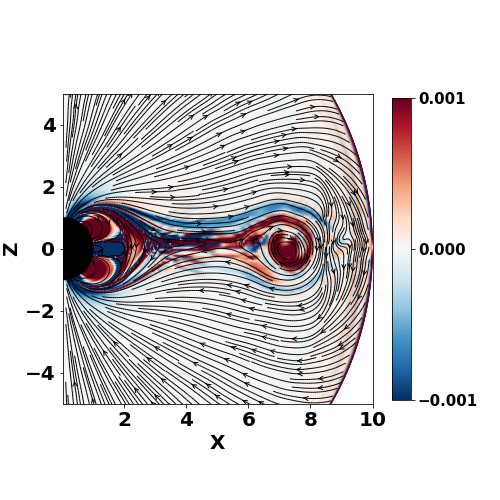}
\includegraphics[width=0.3\textwidth,cfbox=red 1pt 1pt]{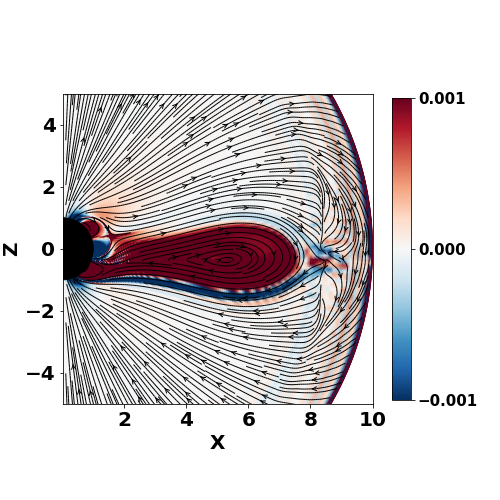}\\
\caption{ Toroidal current density $ J_{\phi} $ for rotating and sheared  configurations. Overall angular velocity of the star is $\Omega = 0.2$ (so that the \LC\ is at $x=5$), 
 normalized shear parameter $\xi =0.5$ (so that maximal shearing angular velocity is $\om_{max} =0.1$), symmetric shear (so that all footpoints are moved in the same azimuthal direction)  In all cases, when shearing is done close to the north pole (region A, left column), no major ejection events are observed. When shearing near the equator (region B, middle column), in all cases we observe powerful ejections. When shearing is done at region C (right column) whether or not ejections are observed depend on the magnetic field topology: in the Dipole+Quadrupole case (second row, right column) no powerful ejections are observed.  Red boxes are drawn around configurations where a clear expulsion of plasmoids is observed.}
\label{Jphi_rotatingstar_chi=01} 
\end {figure*}

Finally,  in Fig. \ref{Jphi_rotatingstar_chi=01} we discuss  all three magnetic configuration sheared at various locations.
The top panel shows  systems with only dipole field. No ejections were observed when shear is applied near the polar area (region A) and the system remained in quasi-equilibrium state. Similar to what we observed in previous section for non-rotating system, this observation will hold true even for more complicated magnetic topologies. 
We also don't observe ejections while shearing near equator (region B). This is consequence of the fact that our shearing profile is symmetrical i.e. for the equatorial case, the shearing is confined to one hemisphere.


In middle panel we consider rotating star system with superposition of dipole and quadrupole field, we find that powerful ejection events are observed when shearing between $\sim 30^{\circ}$ and $100^{\circ}$. This is consistent with our hypothesis that strong ejections are observed while shearing region with closed field lines (bigger loop in Fig. \ref{cartoon1}). We demonstrate our results by plotting the toroidal current at three different locations : near the poles (region A), at equator (region B), and at $\sim 120^{\circ}$  from the poles (region C). 
In the bottom panel we consider simulations with superposition of dipole and octupole field.  Powerful plasmoid ejection events are observed once the shearing region is away from the polar region . 

Based on above results, we can safely conclude that the effects of shearing highly dependent on how far the field lines extent. Shearing closed field lines leads to powerful ejections whereas if the shearing region is located in an area where field lines have started opening out, no or weak pulsating ejections are observed. Following the discussion in \S\ \ref{nonrotatingsystem}, we highlight those cases where the explosion can be observed with a red box.  In Table \ref{result_table_rot} we summarize our results for rotating sheared configurations (see also \S \ref{Conclusionfast} for a related case of fast shear).


 \begin{table*}
  \centering
  \renewcommand{\arraystretch}{1.2}
  \begin{tabular}{|p{0.21\linewidth}|p{0.18\linewidth}|p{0.18\linewidth}|p{0.18\linewidth}|}
    \hline
    \multirow{2}{*}{\textbf{Field Topology}} & \multicolumn{3}{c|}{\textbf{Shearing Region}}\\
    \cline{2-4}
    & \textbf{Region A} & \textbf{Region B} & \textbf{Region C}\\
    \hline
    Dipole & Weak pulsating eruptions [Fig. \ref{Jphi_rotatingstar_chi=01}(a),Top Panel] & 
    Weak pulsating eruptions [Fig.\ref{Jphi_rotatingstar_chi=01}(b),Top Panel)] &
    Powerful ejections [Fig.\ref{Jphi_rotatingstar_chi=01}(c),Top Panel)]\\
    \hline
    Dipole+Quadrupole & Weak pulsating eruptions [Fig. \ref{Jphi_rotatingstar_chi=01}(a), Middle Panel]  &
    Few but powerful [Fig. \ref{Jphi_rotatingstar_chi=01}(b),Middle Panel] &
    Weak pulsating eruptions [Fig. \ref{Jphi_rotatingstar_chi=01}(c),Middle Panel] \\
        \hline
    Dipole+octupole &Weak pulsating eruptions  [Fig. \ref{Jphi_rotatingstar_chi=01}(a),Bottom Panel] &
    Frequent and powerful [Fig. \ref{Jphi_rotatingstar_chi=01}(b),Bottom Panel] &
    Frequent and powerful [Fig. \ref{Jphi_rotatingstar_chi=01}(c),Bottom Panel] \\
         \hline
      \end{tabular}
  \caption{Table summarizing results from simulations when the magnetic field lines are sheared for rotating stars, following the prescription in  \cite{1999ApJ...510..485A}.}
  \label{result_table_rot}
\end{table*}

\subsection{Conclusion 1: three  important  ingredients for generation of CME:  global magnetospheric structure,   location of foot-point shear, and rotation}

Different magnetic field configurations are required for CME initiations by different models  \citep{2005AGUSMSH51C..08L}. The \emph{Break-out} model proposed by  \citep{1999ApJ...510..485A} has a multi-flux topology with four distinct flux systems. The shearing of the central arcade (which straddles the equator) and the subsequent reconnection of the sheared  magnetic arcade with the overlying un-sheared field leads to build up of large energy excess in closed sheared field lines, to power CME. 

Another similar model was proposed by \cite{1994ApJ...430..898M}, where the trigger for plasmoid ejections is the introduction of resistivity in the plasma when the shearing is turned off. The introduction of resistivity causes the magnetic field lines to reconnect in the current and the subsequent formation and ejection of plasmoid islands. In absence of plasma resistivity no eruption occurs, field lines becomes fully opened  and the system remains in equilibrium.

The major difference between this work and model by  \cite{1999ApJ...510..485A} is the inclusion of the rotation of the star. The rotation of star, which in turn leads to formation of light cylinder, removes the need for magnetic reconnection. The flux tube opens up to infinity approximately when the top  point reaches the \LC.

\begin{figure}
\includegraphics[width=.99\linewidth]{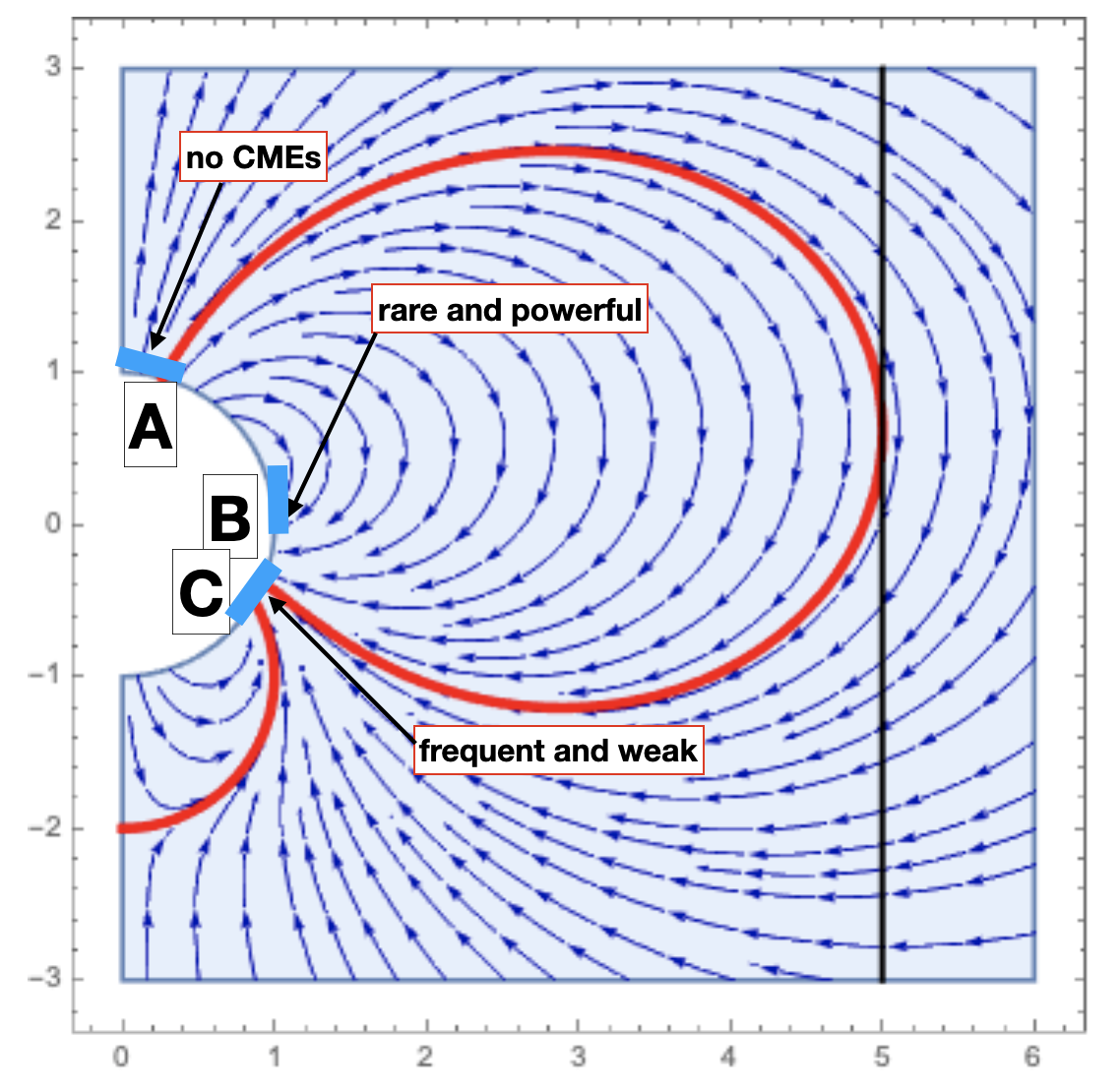}
\caption{Qualitative description of flare dynamics depending on the location of the shear (dipole-quadrupole case).  Vertical line at $5$ is a \LC. }
\label{cartoon1}
\end{figure}

We find a relatively simple picture of shear-generated explosion:  the location of the shear for a given global magnetospheric structure determines the presence or absence of strong ejection events. 
Qualitatively, we show the results for dipole+quadrupole  configuration  in Fig. \ref{cartoon1}.




\subsection{Large scale dynamics of ejected CMEs  in the wind, and conclusion 2}

\begin{figure*}
     \centering
	\includegraphics[width=.32\linewidth]{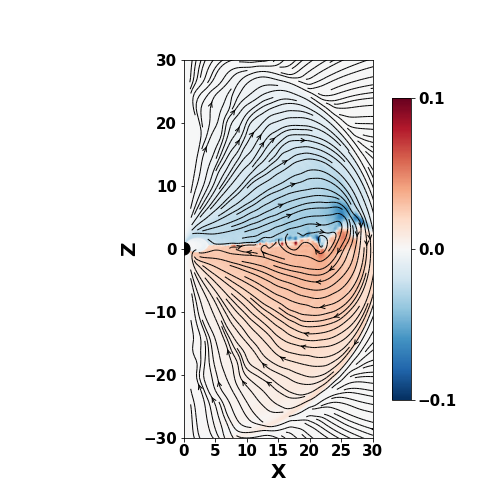}
	\includegraphics[width=.32\linewidth]{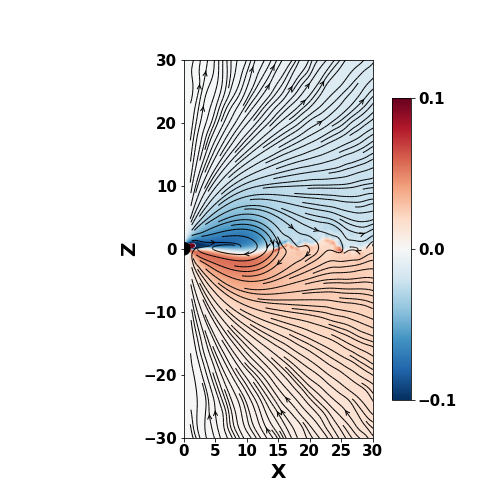}
       \includegraphics[width=0.32\textwidth]{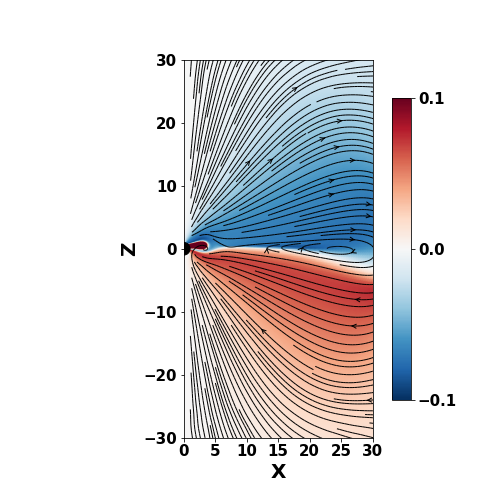}\\
        \includegraphics[width=0.32\textwidth]{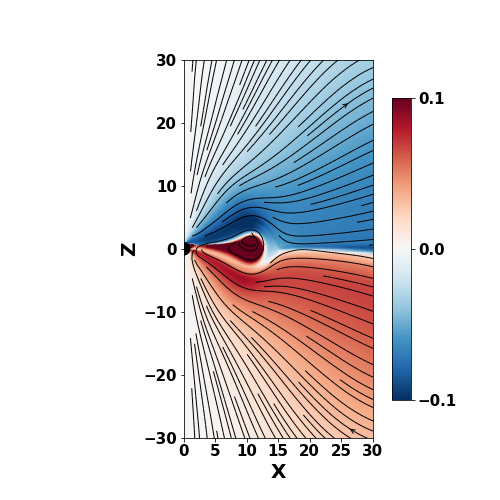}
         \includegraphics[width=0.32\textwidth]{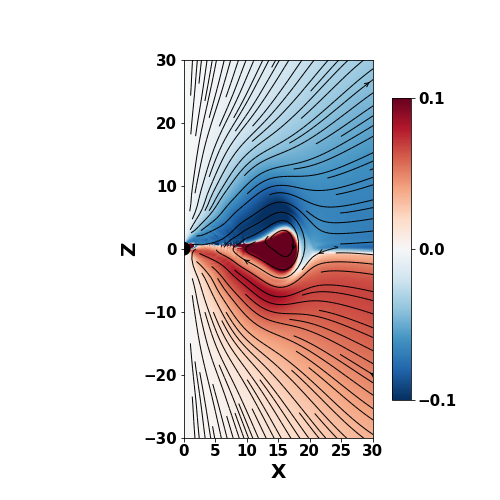}
        \includegraphics[width=0.32\textwidth]{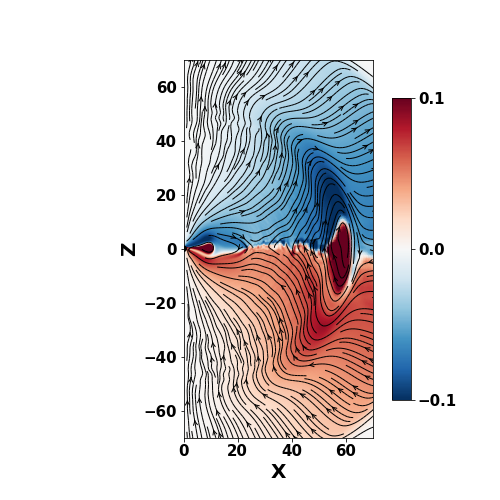}
   \caption{Rotating  dipole + quadrupole configuration sheared at point B (equatorial), slow shear on large scale. The snapshots are taken at $ t={0.99,2.4,3.2,3.6,3.8,5.3}$  of the rotational period of the star. Shearing is introduced after one rotation. One can clearly observe the opening of field line and ejection of a CME.  After the CME the closed part of the \ms\ is smaller, with the current sheet showing plasmoid instability.  The final configuration has non-zero twist on closed field lines.}
\label{dipolequadrupole_timeevol_largescale}
\end{figure*}

Previously, in \S \ref{Magnetospheric} we discussed generation of a CME within the magnetar's \mss. Next we study the large scale dynamics of the resulting CME. We stat with large scale simulation showing time evolution of a sheared dipole+quadrupole configuration, see Fig. \ref{dipolequadrupole_timeevol_largescale}. Here we set the outer boundary far away from the light cylinder.
 
\begin{figure*}
     \centering
	\includegraphics[width=.31\linewidth,height=0.22\textheight]{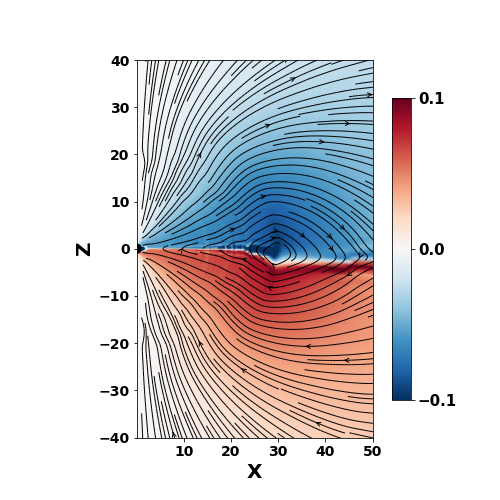}
	\includegraphics[width=.31\linewidth,height=0.22\textheight]{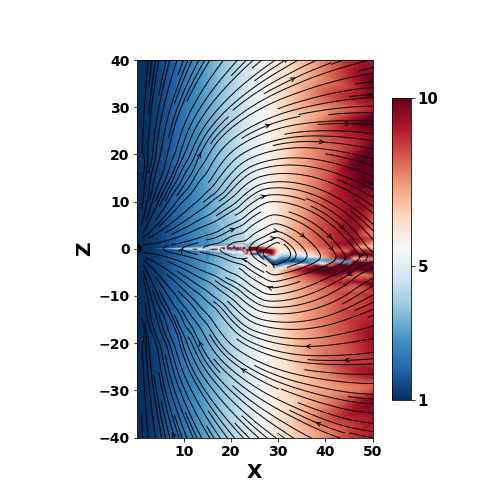}
	 \includegraphics[width=0.31\textwidth,height=0.22\textheight]{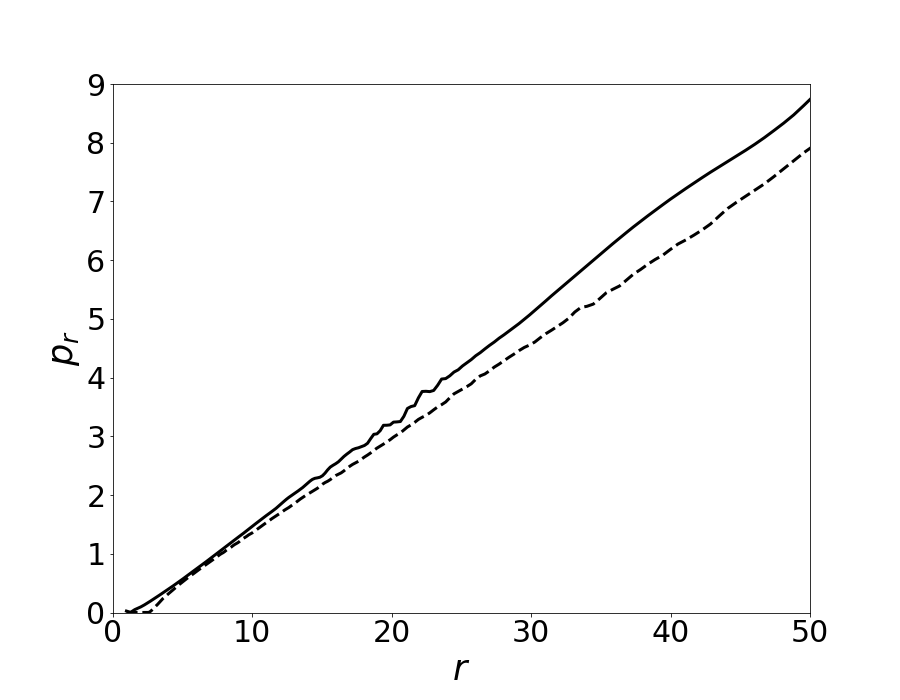}
\caption{ Similar to Fig. \protect\ref{dipole_anitsymmetric} (dipolar field sheared anti-symmetrically at region B) but on large scales. Plotted are values of $r \sin \theta B_\phi$ (left panel), \Lf\ (middle panel) and 
  radial momentum $p_(r) $  as a function of radial distance r at a zenith angle of $60^{\circ}$ from the pole for the above configuration (solid) and for unsheared rotating dipolar configuration (dashed). The snapshots are taken at t= {7.8}. 
 As observed from middle and left plot there is minimal effect of the ejection on the Lorentz factor $\Gamma$. }
\label{dipoleantisymmetric_gamma}
\end{figure*}

\begin{figure*}
     \centering
	\includegraphics[width=.31\linewidth,height=0.22\textheight]{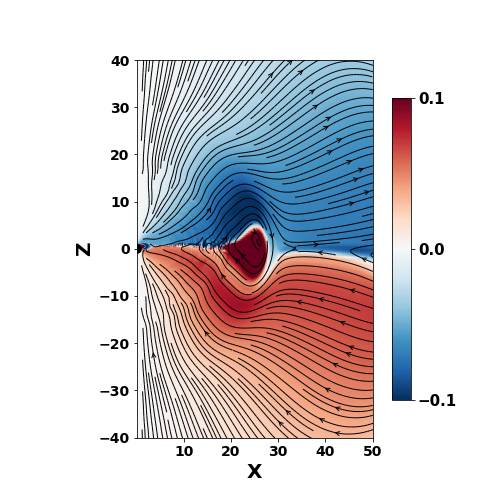}
	\includegraphics[width=.31\linewidth,height=0.22\textheight]{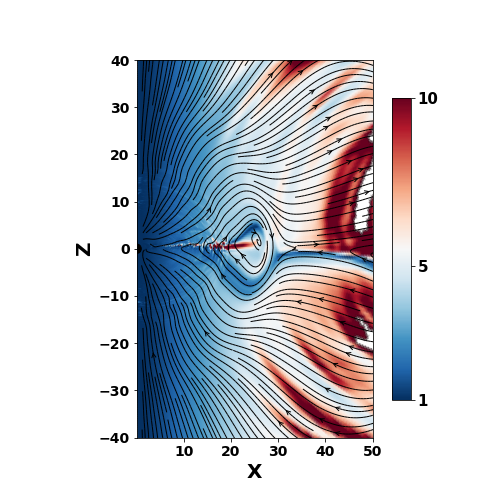}
	 \includegraphics[width=0.31\textwidth,height=0.22\textheight]{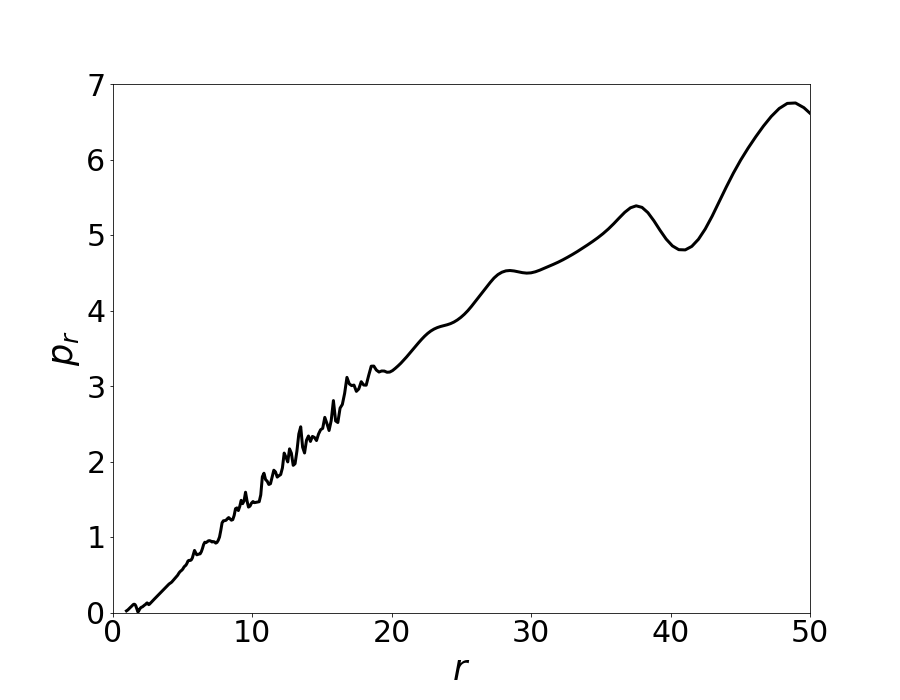}
\caption{ Similar to Fig. \protect\ref{dipolequadrupole_timeevol_largescale} (dipole + quadrupole configuration sheared at point B)  but here we show Lorentz factor $\Gamma$ and radial momentum $p_{r}$ as a function of radial distance r at a zenith angle of $60^{\circ}$ from the pole along with scaled toroidal magnetic field at t= 4.1. As observed from middle and left plot there is minimal effect of the ejection on the Lorentz factor $\Gamma$. }
\label{dipolequadrupole_gamma}
\end{figure*}

In Figs. \ref{dipoleantisymmetric_gamma} and \ref{dipolequadrupole_gamma}  we plot a large scale snapshot for the two cases of dipolar fields sheared anti-symmetrically  and dipole plus  quadrupole configuration sheared at point B.

Recall that shearing results in the generation of topologically disconnected flux tube, a CME.  In Figs.  \ref{dipoleantisymmetric_gamma} and \ref{dipolequadrupole_gamma}  an ejected CME is clearly identified  in the left panels around $x\approx 25$. At the same time, the CME are barely seen in the \Lf/ radial momentum plots (center and left panels): topologically disconnected CME is frozen into the wind and propagates with the local \Lf\ of the wind.

Overall, our numerical results are in excellent agreement with analytics \citep{2022MNRAS.509.2689L}, see Fig. \ref{flux-tube-inside-magnsph1}.  
     
\section{Glitched Magnetosphere (fast shear)}  
\label{Jerked}

\subsection{Magnetar's CMEs in the  \emph{Star Quake} paradigm}

In a complementary approach, which can be supported by  a fully analytical model, we consider  a model of propagation of the force-free \EM\ pulse generated by a sudden {\it local}  spin-up of a \NS, a ``Glitched Magnetosphere".  This type of dynamics mimics the starquake model of \cite{TD95,2020ApJ...900L..21Y}

\subsection{Locally glitched Michel's \ms:  analytical approach}
 
\begin{figure*}
\includegraphics[width=.32\linewidth]{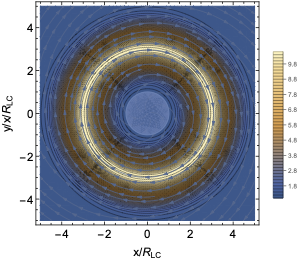}
\includegraphics[width=.32\linewidth]{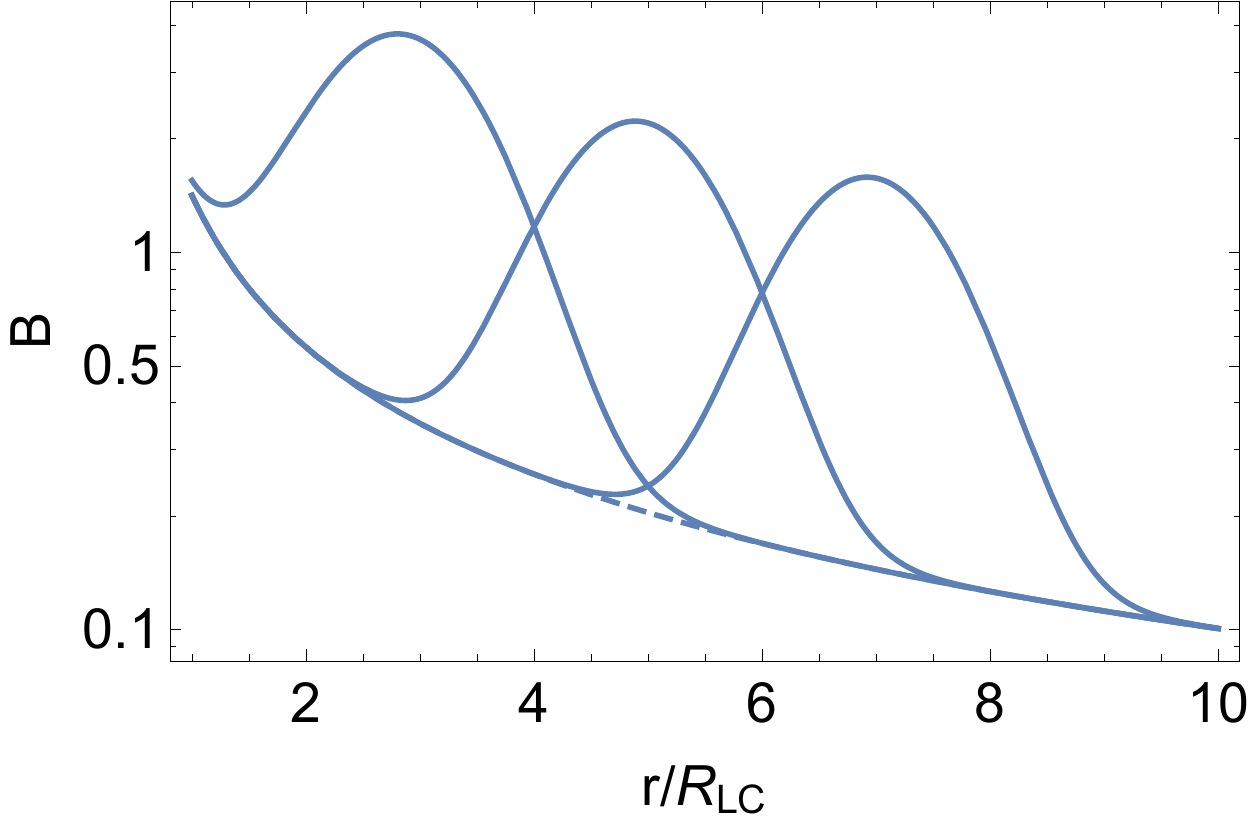}
\includegraphics[width=.32\linewidth]{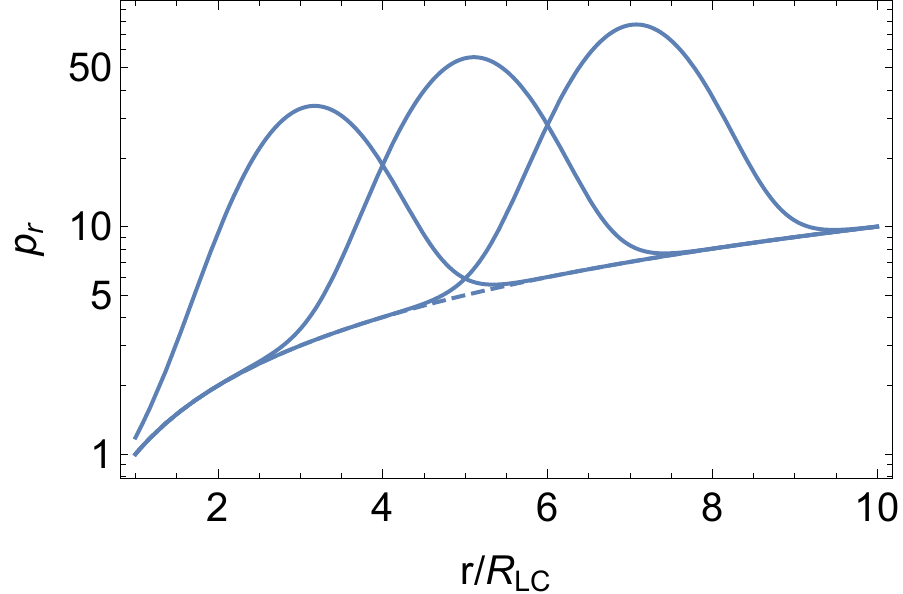}\\
\includegraphics[width=.32\linewidth,height=0.20\textheight]{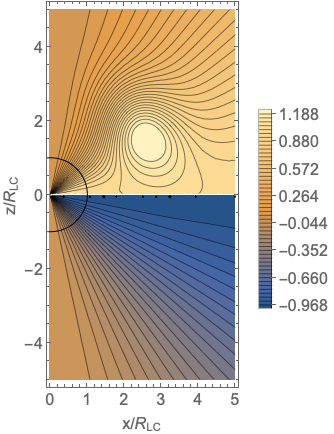}
\includegraphics[width=.32\linewidth,height=0.20\textheight]{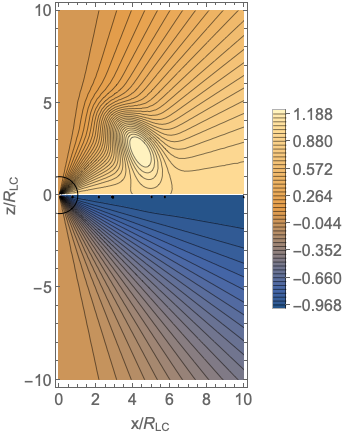}
\includegraphics[width=.32\linewidth,height=0.20\textheight]{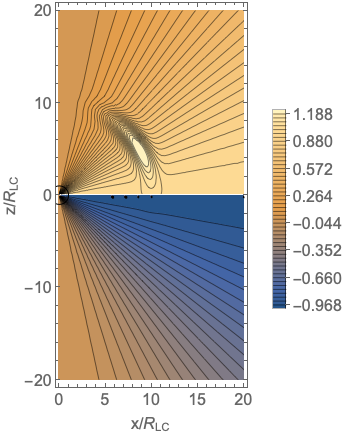}
\caption{Top row. Left Panel: Structure of the \Bf\ in the equatorial plane of an \EM\ Gaussian  pulse with amplitude $10$ times the average propagating through Michel's wind. The peak of the pulse is at $r/R_{LC}=3$. The \LC\  is at $\sqrt{x^2+y^2}=1$. Color scheme corresponds to $\ln B/B_M$, where $B_M$ is the local value of the magnetic  field for Michel's solution. Center Panel: plot of $ B$ showing EM pulse propagating with the wind for times $t=3,5,7$ (in units of $R_{LC}/c$); dashed line is the Michels' solution.  Right panel: plot of $ p_r(r)$. The pulse propagates with the flow with constant relative amplitude, without experiencing any distortions.
Bottom row: value of $r \sin \theta B_{\phi}$
}
\label{pulseff1}
\end {figure*}

Effects of ``glitch in spin'' on the structure of the wind can actually be considered analytically and non-perturbatively for the case of \cite{1973ApJ...180L.133M}  \mss\  and the preceding wind:
\ba &&
\mathbf{B}= B_{0} \left\{\frac{r_{0}^2}{r^{2}}, 0,-\frac{r_{0}^{2} \Omega \sin (\theta)}{r}\right\} 
\nn &&
\mathbf{v}= \left\{\frac{r^{2} \Omega^{2} \sin(\theta)^{2}}{1+r^{2} \Omega^{2} \sin(\theta)^{2}}, 0, \frac{r \Omega \sin (\theta)}{1+r^{2} \Omega^{2} \sin (\theta)^{2}}\right\} 
\nn  &&
\mathbf{\Gamma}=\sqrt{1+r^{2} \Omega^{2} \sin(\theta)^{2}} 
\nn  &&
\mathbf{p}=\left\{\frac{r^{2} \Omega^{2} \sin(\theta)^{2}}{\sqrt{1+r^{2} \Omega^{2} \sin(\theta)^{2}}}, 0, \frac{r \Omega \sin (\theta)}{\sqrt{1+r^{2} \Omega^{2} \sin(\theta)^{2}}}\right\}
\label{michels_expressions}
\ea
$B_{0}$ is the the fiducial magnetic field magnitude at the light cylinder ($r_{0}$) and we set $c=1$. 
  Realistic dipolar \mss\ do evolve asymptotically to the  \cite{1973ApJ...180L.133M} solution \citep{1999A&A...349.1017B,1999ApJ...511..351C,2006MNRAS.367...19K}.

One can generalize Michel's solution  for any arbitrary time- and angle dependent rotation $\Omega=\Omega  [r-t] g( \theta)$   \cite{2011PhRvD..83l4035L} \citep[see also][]{2014MNRAS.445.2500G}.  \citep[The solution can also be generalized to \Sc\ metric using the  Eddington-Finkelstein coordinates][]{2011PhRvD..83l4035L}. This \emph{glitch in spin}  time-dependent nonlinear solution (nonlinear both in a sense that  the current is a  non-linear  function of the magnetic flux function, and that the perturbation can be of large amplitude) preserves both the radial and $\theta$ force balance. 
Qualitatively, "a 
glitch" in the angular rotation velocity $\Omega$  mimics a symmetric  shearing motion of a patch of field lines (we remind: ``symmetric'' means overall motion in one direction along $\phi$). Approximation of  \cite{1973ApJ...180L.133M}  \mss\  misses the magnetospheric dynamics, but it capture the wind dynamics.

In Fig. \ref{pulseff1}  we show the  evolution of single \Alfven pulse  using  $g(\theta)= \sin^{10}(  \theta + \pi/4) $ and $\Omega = 1+ e^{-(r-t)^2}$.  To complement the analytical work, we show the complete evolution of a pulse via numerical simulation in \S\ \ref{glitch_simulations}

A pulse  of shearing \Alfven waves with $\Omega [r-t] g(\theta) $  propagates  with  radial 4-momentum
\be
p_r =   \frac{ r^2  \sin^2 \theta  \Omega^2  [r-t]  g(\theta)^2}{ \sqrt{ 1+  r^2  \sin^2 \theta  \Omega ^2 [r-t]  g(\theta)^2}}
\ee
which is larger than that of the wind for $ \Omega  [r-t]   g(\theta) \geq \Omega_0$, the constant value.
Higher radial  momentum  (than  that of the background flow) {\it does not mean} that plasma is swept-up: it's just an EM pulse propagating through the accelerating  wind.

\subsection{Locally glitched  \ms:  simulations with PHAEDRA}
\label{glitch_simulations}

In a numerical implementation we limit ourselves to just dipolar \mss.
We
 use glitch parametrization as
\be
\Omega = \Omega_0 \left( 1+ g(\theta) f(t) \frac{\delta \Omega}{ \Omega_0}\right) 
\ee

Several types of shearing were implemented:
(i)  overall glitch  $g(\theta) =$ constant (so in this case the glitch is actually global); (ii)
symmetric 
$
g(\theta)= \sin^{10}(  \theta + \pi/4)
$; (iii)
and anti-symmetric near equator
$
  g(\theta)= \sin^{3}(  \theta )\cos(  \theta)
$.
The extra rotation within the shearing band is fast ${\delta \Omega}/{ \Omega_0} =5$. 

Time dependence of the glitch is  
\ba && 
 f(t) =  \sin [\pi (t-t_{on})/\tau] 
 \nn &&
 t_{on} =  P\,  \mbox{(one period)}
 \nn &&
 \tau = P/10
 \nn &&
 t_{off}= t_{on} +  \tau
 \ea
 so that  the glitch is implemented after one rotation for one tenth of the  with  maximum rate  reached at $t= t_{on} + \tau/2$. Thus a total shearing angle is $\Delta \phi = \pi$.
Note  that the shearing expression $g(\theta)$ used in this section is somewhat  different from the one used for slow shearing. 

\subsubsection{Overall glitch}

\begin{figure*}
  \centering
 \subfloat[\label{1a}]{\includegraphics[width=0.33\textwidth]{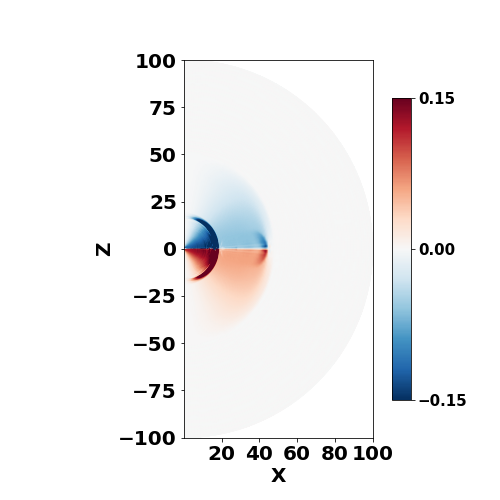}}\hfill
  \subfloat[\label{1b}]{\includegraphics[width=0.33\textwidth]{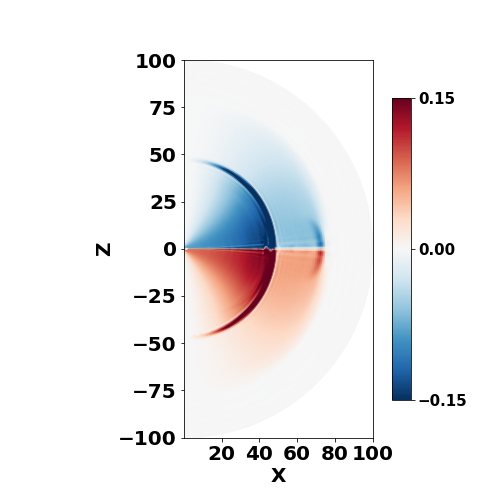}}\hfill
   \subfloat[\label{1c}]{\includegraphics[width=0.33\textwidth]{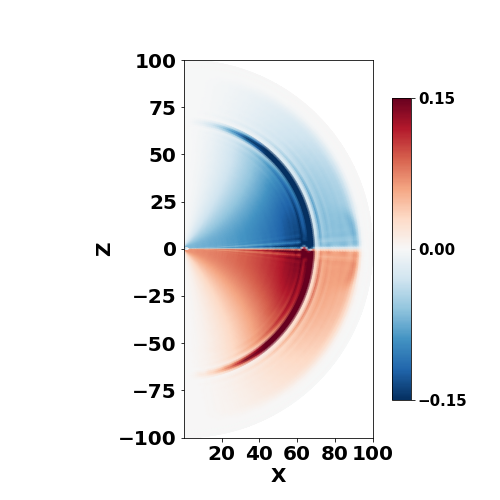}}\hfill
 \caption{Global  glitch  with $g(\theta)= const $, large scale view. 
 Snapshots  of  $r\sin(\theta)B_{\phi}$ are taken at $ t=1.6,2.5, 3.2$ rotation periods. A glitch produced a global \Alfven wave propagating through the wind.
 }
\label{xBphi_GlobalShearing}
\end{figure*}

We first consider the case of constant $g(\theta)$ i.e. the entire magnetosphere is glitched instead of a narrow band. We demonstrate our findings in Fig. \ref{xBphi_GlobalShearing} where we plot $r \sin(\theta)B_{\phi}$ at different time steps. 

\subsubsection{Narrow symmetric  glitch  at $\theta =\pi/4$}

\begin{figure*}
  \centering
 \subfloat[\label{2a}]{\includegraphics[width=0.33\textwidth]{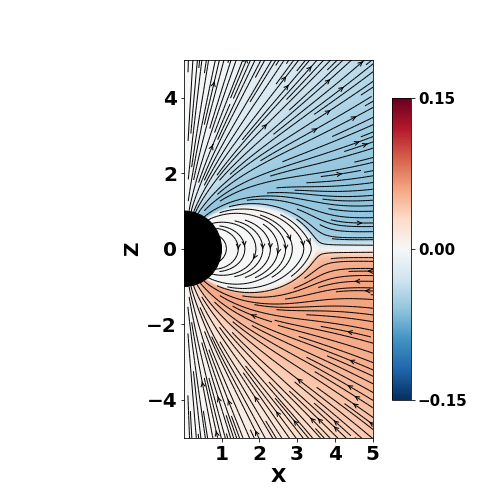}}\hfill
  \subfloat[\label{2b}]{\includegraphics[width=0.33\textwidth]{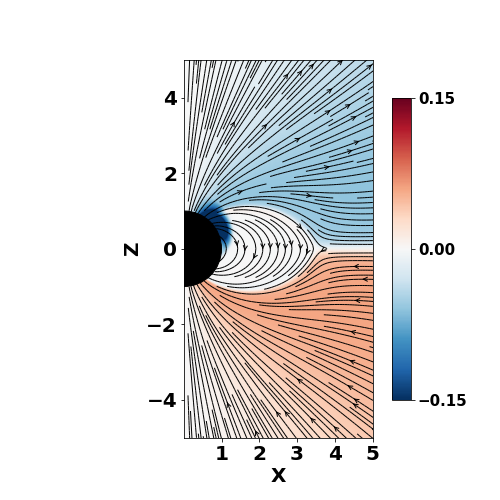}}\hfill
   \subfloat[\label{2c}]{\includegraphics[width=0.33\textwidth]{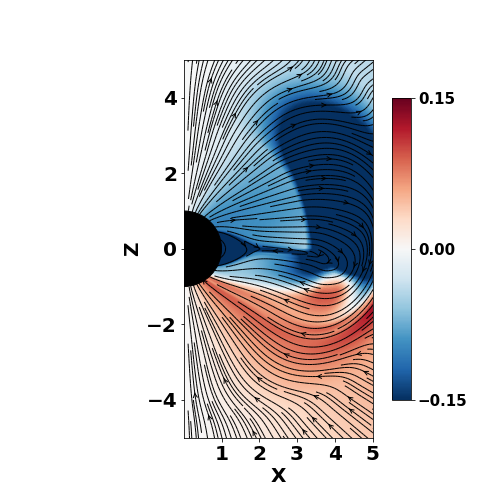}}\hfill\\
    \subfloat[\label{2d}]{\includegraphics[width=0.33\textwidth]{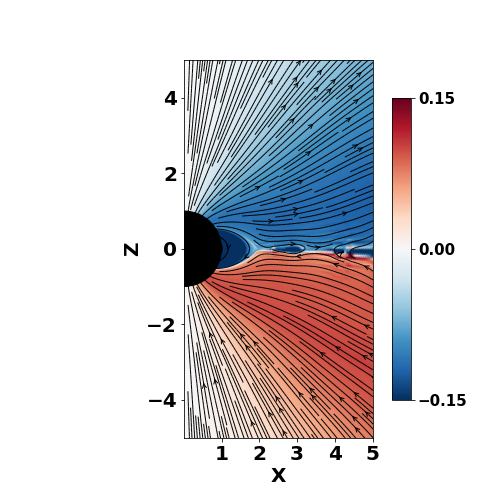}}\hfill
     \subfloat[\label{2e}]{\includegraphics[width=0.33\textwidth]{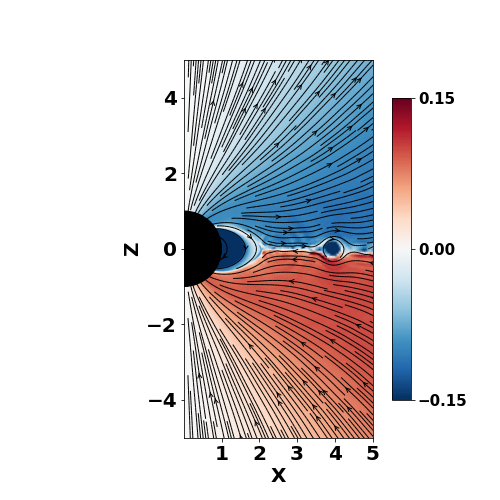}}\hfill
       \subfloat[\label{2f}]{\includegraphics[width=0.33\textwidth]{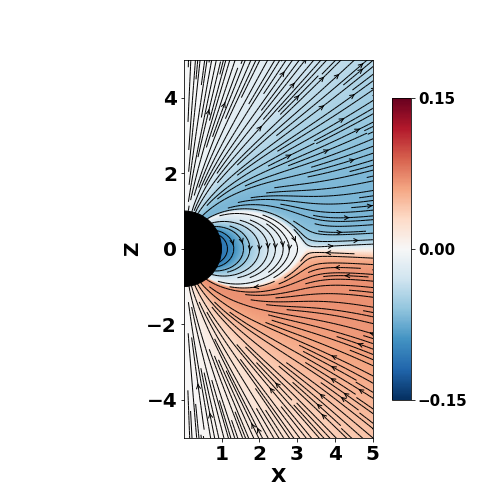}}\hfill
 \caption{Fast glitch near $\theta = \pi/4$,  $g(\theta)= \sin^{10}(  \theta + \pi/4) $, dipolar \mss, symmetric shear Fig.  \ref{2a} is at time step just before the 
glitch is applied whereas Fig.  \ref{2b}  is just after. We observe plasmoid island formation in Fig.  \ref{2e}.  Fig.  \ref{2f} show final equilibrium state.  See Fig. \ref{pulseff11} for large-scale view.}
\label{StreamlinexBphi_LocalShearing_Symmetric_Pi4}
\end{figure*}

\begin{figure*}
     \centering
       \includegraphics[width=0.3\textwidth]{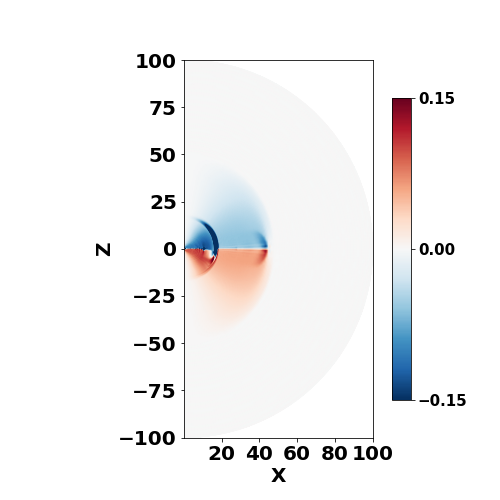}
       \includegraphics[width=0.3\textwidth]{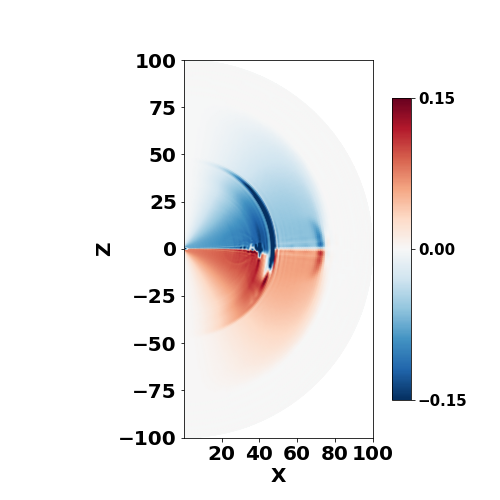}
       \includegraphics[width=0.3\textwidth]{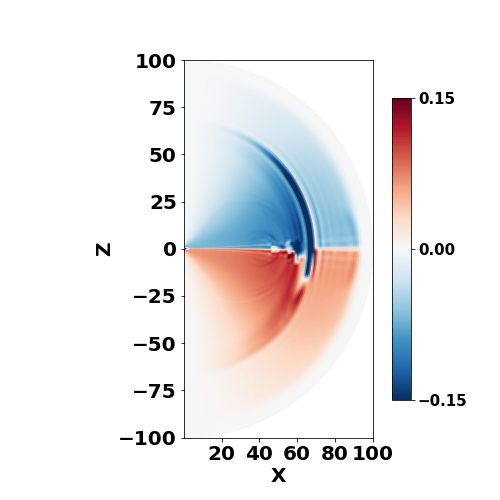}
 \caption{
 Same as Fig. \ref{StreamlinexBphi_LocalShearing_Symmetric_Pi4}, large scale view. 
 Snapshots are taken at $ t=1.6,2.5, 3.2$, see Fig. \ref{StreamlinexBphi_LocalShearing_Symmetric_Pi4} for zoomed-in view of the simulations. One clearly sees an \EM\ pulse propagating through the wind.}
\label{pulseff11}
\end{figure*}

Results of simulation are presented in 
Fig. \ref{StreamlinexBphi_LocalShearing_Symmetric_Pi4} (zooming in close to the star), and bottom row of Fig. \ref{pulseff11} (long time scale evolution). In  Fig. \ref{StreamlinexBphi_LocalShearing_Symmetric_Pi4}  we show zoomed-in plots for  $r \sin(\theta) B_{\phi}$  superimposed on poloidal field lines for symmetric shear: narrow band near $\theta = \pi/4$ is suddenly moved with angular velocity 5 times the spin. We start with unperturbed magnetosphere (Fig. \ref{2a}), one period after the star of overall rotation.  Then, shear is introduced, Fig. \ref{2b} - blue region near the star at $\theta \approx \pi/4$. The resulting shear \Alfven  wave breaks out from the \ms, Fig. \ref{2c}. The \ms\ recovers: bottom row. A new Y-point is formed close to the star (compare locations of the Y-points before the shear is introduced in Fig.  \ref{2a} and right after break-away, Fig.  \ref{2d}).  Outside of the newly formed Y-point reconnection layer forms. It is subjected to plasmoid instability Fig.  \ref{2d}-\ref{2e}.  Eventually, the \ms\ recovers,  to approximately the same location of the Y-point,  Fig.  \ref{2f}.  Notice that the newly formed \ms\ is twisted:  there is non-zero toroidal \Bf\ on closed field lines.

\subsubsection{Narrow anti-symmetric glitch  at $\theta =\pi/2$}

Here anti-symmetric shear is needed to produce a CME (otherwise the flux surfaces are just rotated as a whole, see \citep{2022MNRAS.513.1947L}. In order to generate anti-symmetric we chose $g(\theta)= \sin^{3}( \theta )\cos(\theta) $. 

\begin{figure*}
  \centering
 \subfloat[\label{3a}]{\includegraphics[width=0.33\textwidth]{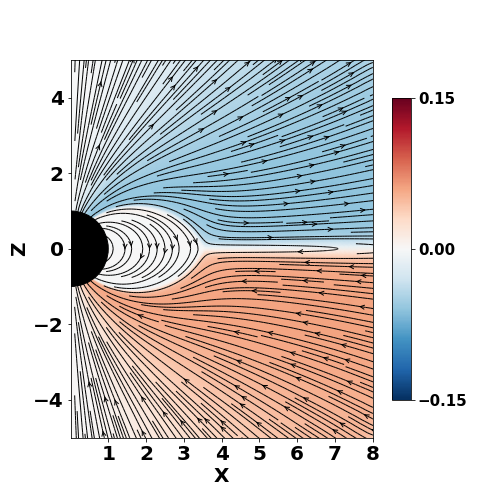}}\hfill
  \subfloat[\label{3b}]{\includegraphics[width=0.33\textwidth]{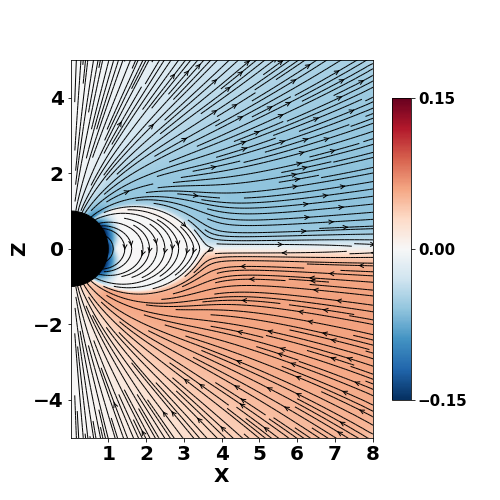}}\hfill
   \subfloat[\label{3c}]{\includegraphics[width=0.33\textwidth]{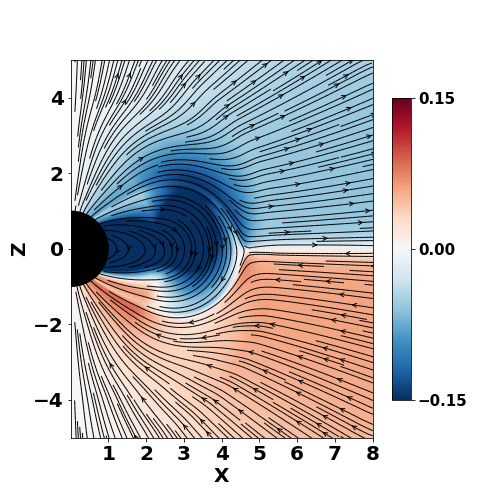}}\hfill\\
    \subfloat[\label{3d}]{\includegraphics[width=0.33\textwidth]{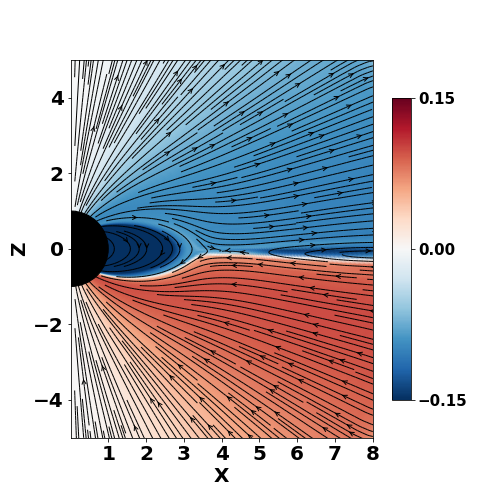}}\hfill
     \subfloat[\label{3e}]{\includegraphics[width=0.33\textwidth]{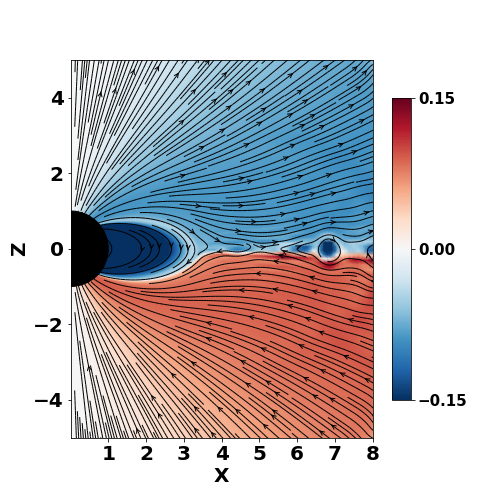}}\hfill
       \subfloat[\label{3f}]{\includegraphics[width=0.33\textwidth]{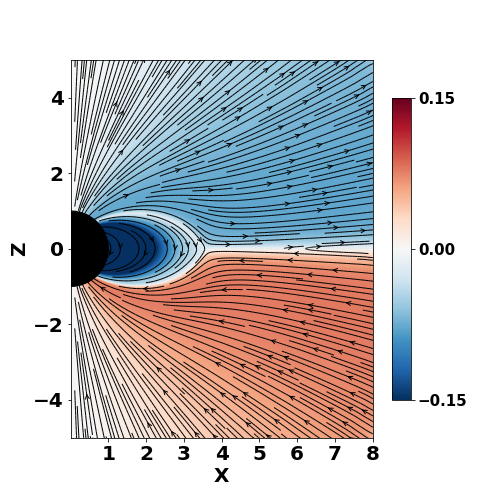}}\hfill
 \caption{Anti-symmetric equatorial glitch with  $g(\theta)= \sin^{3}(  \theta )\cos(\theta) $, zoomed-in view.
   Fig.  \ref{3a} is at time step just before the glitch is applied,  Fig.  \ref{3b}  is just one time step  after the initiation of the glitch. Fig.  \ref{3c}  shows shearing  \Alfven  wave breaking away from the magnetosphere. In lower row we observe formation of a new current sheet, subject  to plasmoid instability.  Fig.  \ref{3f} is the final equilibrium state, with twisted closed field lines.}
\label{StreamlinexBphi_LocalShearing_Antisymmetric_Pi2}
\end{figure*}

\begin{figure*}
  \centering
 \subfloat[\label{4a}]{\includegraphics[width=0.33\textwidth]{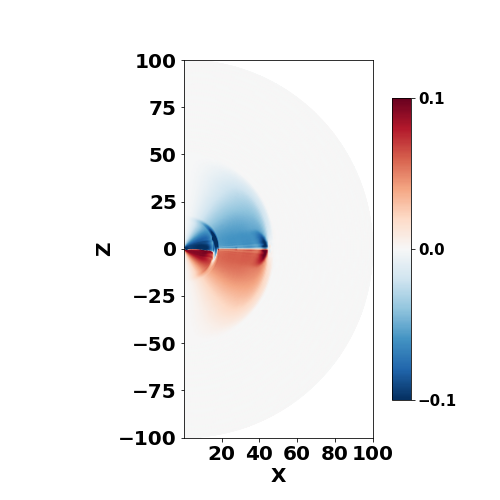}}\hfill
  \subfloat[\label{4b}]{\includegraphics[width=0.33\textwidth]{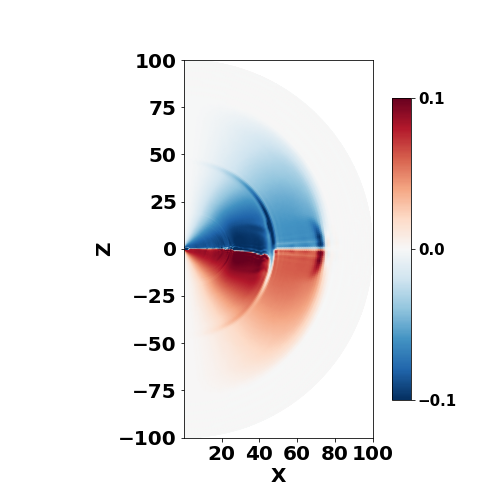}}\hfill
   \subfloat[\label{4c}]{\includegraphics[width=0.33\textwidth]{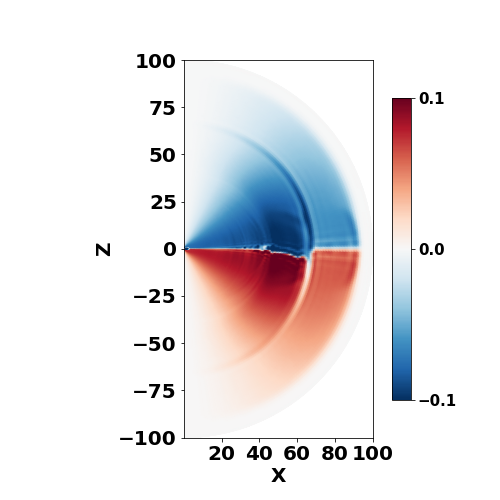}}\hfill
 \caption{Anti-symmetric equatorial glitch, global view at  times $ t=1.6,2.5, 3.2$. }  
\label{xBphi_LocalShearing_Antisymmetric_Pi2}
\end{figure*}

As in the previous subsection, we present our results by  zooming in close to the star (Fig. \ref{StreamlinexBphi_LocalShearing_Antisymmetric_Pi2}) and showing the long time scale evolution (Fig. \ref{xBphi_LocalShearing_Antisymmetric_Pi2}). The results are similar to what we observe for fast symmetric shearing at a band around $\pi/4$: once the shearing is introduced resulting shear \Alfven  wave breaks out, reconnection is observed outside the new Y-point is formed and plasmoid instability is detected, Fig.  \ref{3d}-Fig.  \ref{3e}. Equilibrium state is shown in Fig.  \ref{3f}.

%


 \subsection{Comparison of slow and fast shear, and discussion of previous results}
 \label{Conclusionfast}

Previously, in \S \ref{generation} we considered CME dynamics for slow shear. Let us compare slow and fast shear cases. Concisely: slow shear generates topologically disconnected CME that is frozen into the wind, while fast shear generates an \Alfven wave propagating through the wind.
 In both cases  opening of the \ms\ is  followed by formation of a reconnection sheet.  In the case of slow shear this opening is  achieved by the  inflation of the field lines followed by the break-out near the $r_{eq}$ (or near the \LC\ for weaker injections),  like in the classical Solar flare models. In the case of fast shear the opening is achieved by the \Alfven packet itself, exerting a ram pressure on the closed field lines, and breaking them open.

 Thus, large amplitude \Alfven waves \emph{do not break-down} within the \ms, as suggested by \cite{TD95}.  Instead, they open-up the \ms\ and  form propagating \EM\ pulses. The \ms\ recovers 
 by forming a current sheet deep inside the \LC, subject to  plasmoid instability. Thus the \Alfven packet in the wind 
eventually become causally disconnected.
    
Our case of fast shear resembles simulations of  \cite{2020ApJ...900L..21Y} who considered  the dynamics of shear \Alfven waves within the \ms. In that simulation an \Alfven wave was added at the initial moment (see our  \S \ref{Fluxtube} for similar approach) when we perform similar injection.
  Since the relative amplitude of the waves increases as 
$\delta B/B = \left(\delta B/B\right)_\star   (r/r_\star)^{3/2} $  for sufficiently large $ \left(\delta B/B\right)_\star $ the waves would break with in the \ms\ \citep[such wave breaking is the key ingredient of][model of magnetar flares]{TD95}. To avoid wave breaking  \cite{2020ApJ...900L..21Y} fine-tuned the initial amplitude of the \Alfven wave to $ \left(\delta B/B\right)_\star$, that $\delta B/B \sim 1 $ near the \LC. 

 In contrast, we generate the  \Alfven wave self-consistently by shearing the foot-points. (Our simulation code use pseudo-spectral method which can't capture breaking  of the waves.) Our wave amplitude is large: the initial twist is 180 degrees. Thus, the fields in the wave quickly become  much larger than the background \Bf. The \Alfven pulse breaks out from the \ms. 
During break-out the pre-explosion  closed \Bf\ lines  are first stretched out, opening the \ms, then reconnection ``behind'' the wave pulse sets in. 

Our 2D  fast shear  simulations are  generally consistent with 3D force-free simulations of  \cite{2022ApJ...933..174Y}. The similarity includes that the resulting  \Alfven pulse opens the magnetosphere. The differences are as follows. First, for the non-rotating case we impose slow shear, so that the expanding structure is in an approximate force balance, while fast shear results in launching of \Alfven waves; the slow shear case also possibly  leads to the detonation stage. Second, we demonstrate that the resulting \Alfven pulse within the wind leads to an EM pulse, or even anti-pulse, not a strong shock wave.

The opening of the \ms\ in the fast shear case is somewhat different form the slow shear case. In the case of  fast shear, let's assume that the initially generated wave near  the  \NS\ surface has amplitude $\delta B = (\Delta \phi) B_0 $, where $(\Delta \phi) $ is a typical angle that the fields lines are sheared. The amplitude of the wave decreases as $\propto R_{NS}/r$ (both \Alfven and X-modes are excited), while the magnetospheric field decreases as   $\propto (R_{NS}/r)^3$. 
The amplitude of the wave becomes lager than the guiding field for 
\be
\frac{r_{eq, EM}} {R_{NS} }  \geq (\Delta \phi) ^{-1/2}
\ee
This is the estimate of the opening scale of the \ms\ for fast  shear, and of the ensuing  initial  size of the current sheet. 

Opening of the \ms\ requires energy  to be spent by the \EM\ pulse, of the order of
\be
B_0^2 \left(\frac{r_{eq, EM}} {R_{NS} }  \right) ^{-6}r_{eq, EM}^3
\ee
Thus, more powerful pulses open the \ms\ earlier, and experience larger energy losses.


\section{Large scale dynamics of  injected shear}
\label{Fluxtube}

To complement our analysis of slow and fast shear we also performed a series of experiments when a packet of  shear \Alfven waves is injected into the \ms\ (instead of the foot point motion).  We inject a packet of shear \Alfven waves carrying toroidal \Bf. In this work we focus on the scenario where the injection is performed  within the \ms.
 In what follows we  conduct a thorough  investigation of the system:  how does the location of the ejection influences the dynamics (ejection on open versus closed field lines), and  how does the flow reacts to the value of the injected flux (strong and weak ejections), and how do multiple ejections interact.
 
 In what follows we call the injected \Alfven wave packed as \emph{flux tube}, with clear understanding that the resulting structure is not topologically isolated - it is an \Alfven wave packet resembling the flux tube.


\subsection{Injection procedure} 
\label{Injectionprocedure} 

The set-up in this section is as follows. We start with a dipole configuration and let the system evolve unperturbed for two time period. We then introduce a flux tube in the magnetosphere of a rotating \NS\  in approximate  force equilibrium, just slightly out of force-balance. We  do this by introducing external $\mathbf{B_{\phi}}$ given by Eq.  (\ref{flux_tube_Bphi}) for a small but finite time interval $\Delta t$. 

The flux tube is taken to be a torus like structure and embedded in a force-free magnetic environment with magnetic field along the azimuthal direction. The magnitude of  the toroidal field inside the tube is equal to the total poloidal field of a dipole at  $r= a r_\star$, where $a$ can be considered as the location of the center of the flux tube. The flux tube is introduced over a small radial interval at a fixed zenith angle ($\theta_{ft}$)from the z-axis.

In this subsection, since we are interested in tubes inserted inside the light cylinder we set  $a = 2$. 
\ba &&
B_{\phi,tube} = \sqrt{B_{r}^2 + B_{\theta}^2} 
\nn && 
= B_{0} \sqrt{\frac{4 \cos[\theta_{ft}]^{2}}{a^{6}}+\frac{\sin [\theta_{ft}]^{2}}{a^{6}}}
\label{flux_tube_Bphi}
\ea
The strength and direction of the toroidal magnetic field inside the tube is controlled by the parameter $B_{0}$. For flux inserted along the magnetic wind $B_{0}$ is positive while negative when the tube is inserted against the wind.
Thus, the initial configuration is just slightly unbalanced: the dipolar field at the inner edge is somewhat larger than at the outer edge of the flux tube. But as the  tube is pushed radially away from the star the flux conservation quickly leads to the creation of highly over-pressurized  tube. The tube then both inflates and is pushed out. 

Construction of a self-confined flux tube implies that there are surface currents. Surface current $\mathbf{K} $  can be calculated via interface conditions of magnetic field \citep[][]{JacksonEM}.
\ba &&
\mathbf{K} = \frac{c}{4 \pi} [\mathbf{n} \times\left(\mathbf{B}_{2}-\mathbf{B}_{1}\right)] 
\nn  &&
\mathbf{n} \cdot \left(\mathbf{B}_{2}-\mathbf{B}_{1}\right) =0
\label{boundary}
\ea
Here, $B_{2} =0,B_{1}= B_{\phi,tube}$ and $\mathbf{n}$ is the unit vector from region 1 (inside the flux tube) to region 2 (outside the tube).

Since the code is sensitive to sudden changes on the \Bf, the flux tube insertion is  inserted over a finite period of time:
\begin{equation}
     B_{\phi}= 
\begin{cases}
B_{\phi,tube},& \text{if }  t_{insertion} \le  t \le t_{insertion} + \Delta t\\
0,& \text{otherwise }  
\end{cases}
\end{equation}

One important quantity  is the toroidal flux added within the flux tube. 
\be
\Phi=\int_{S_{t}} \vec{B} \cdot \vec{n} d S
\ee

In the case of magnetar-generated CME, 
the magnetic flux carried  by the flux tube originating near the surface can be estimated   as 
\begin{equation}
\Phi_{f} \sim B_{\star} R_{f}^{2} =  B_{\star} \eta_{f}^2 R_{\star}^{2} 
\label{injected_flux_analytics}
\end{equation}
where $R_f$ is a typical size of the active region and in the latter equality we scaled 
 flare's  size to the radius of the \NS,  $R_{f} = \eta_{f}R_{\star} $,

The value of the added toroidal flux can be compared with the total toroidal flux within the \LC\ (in one hemisphere) generated by the rotating dipole. The  model of  \cite{1969ApJ...157..869G}  gives an estimate
\begin{equation}
\Phi_{i} \sim B_{\star} R_{\star}^{2} \left(\frac{\Omega_{\star} R_{\star}}{c}\right)
\label{GJ_flux_analytics}
\end{equation}

The injected magnetic flux therefore can be calculated via,
\begin{equation}
\Delta \Phi = \Phi_{f} - \Phi_{i},
\label{flux_added}
\end{equation}
where  $\Phi_{f}$ is the toroidal flux at $t= t_{f}= t_{insertion} + \Delta t$ and $\Phi_{i}$ is the flux at $t= t_{i}= t_{insertion} $. Here $t_{insertion}$ is the simulation time step at which flux tube was introduced in the system for a duration of $\Delta t$. In this work we display our results for $t_{insertion}$ = 2 and $\Delta t = 0.06$, with time expressed in term of the unperturbed rotational period of the star.

For  $ \Delta \Phi  \sim \Phi_{f}$   we expect 
\begin{equation}
|\frac{\Delta \Phi}{\Phi_{i} }| \sim \eta_{f}^2   \frac{c }{\Omega_{\star} R_{\star}} \ge 1 
\label{injected_GJ_flux_comparison}
\end{equation}
Thus, the toroidal magnetic flux injected by the flare is expected to be of the order of the total toroidal magnetic flux of unperturbed \ms. Our simulations' parameters, Table \ref{injectedfluxtable}, use similar values.

\begin{table}
\centering
 \begin{tabular}{||c c c c c||} 
 \hline
 $ \Delta t$  & $\Phi_{i}$ & $\Phi_{f}$ &  $ \Delta \Phi $ & $|\frac{\Delta \Phi}{\Phi_{i}}|$\\ [0.5ex] 
 \hline\hline
0.03 & -0.24 & -0.54 & -0.30 &1.25\\ 
0.06 & -0.24 & -0.85 & -0.61 & 2.5 \\ 
 0.16 & -0.24 & -1.43 & -1.19 & 4.9 \\ 
 \hline
 \end{tabular}
  \caption{Table showing toroidal flux injected to the system $ \Delta \Phi$ for different values of $\Delta t$}
   \label{injectedfluxtable}
\end{table}


\subsection{Dynamics of  shear  \Alfven waves in the \ms\ and the preceding wind}

For our first sets of experiment, we add a single toroidal flux tube following the procedure described in  \S\ \ref{Injectionprocedure}. The tube is launched at $a=2$ (we remind that for our basic set-up the \LC\ is at $x=5$). We explored two injection sites: at $\theta_{injection}= 60^{\circ}$ (so that the injection is on closed field lines) and at $\theta_{injection}= 30^{\circ}$ (so that the injection is on closed field lines). We also explored two polarizations of the injected  waves which we call symmetric (so that the toroidal field in the wave is of the same sign as the toroidal field in the corresponding hemisphere of the  wind), and  antisymmetric (so that the toroidal field in the wave is of the same sign as the toroidal field in the corresponding hemisphere of the  wind).

Though the waves are injected with similar procedures, addition of the toroidal component, this, in fact, corresponds to  somewhat different modes. On the open field lines there is already $B_\phi$ present. This toroidal field determines the spin-down: addition of extra toroidal field modifies the spindown, see \S \ref{Jerked} and \S \ref{multiple_flux_tube}.  Addition of the toroidal component on the close field lines generates both \Alfven waves (propagating mostly along the \Bf), and compressional X-mode (propagating approximately radially). 

\begin{figure*}
     \centering
     \includegraphics[width=0.30\textwidth]{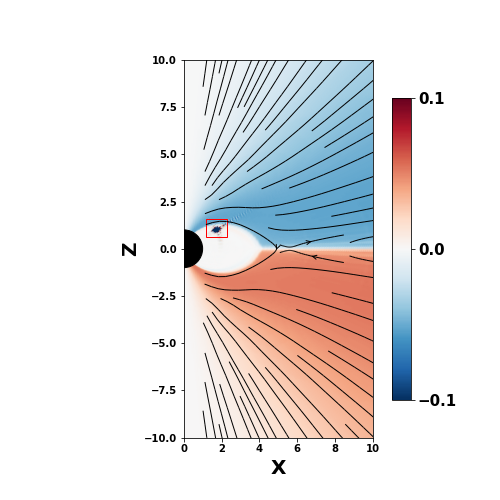}
     \includegraphics[width=0.30\textwidth]{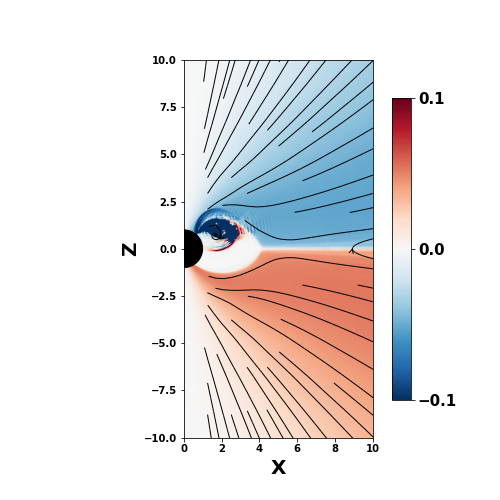}
       \includegraphics[width=0.30\textwidth]{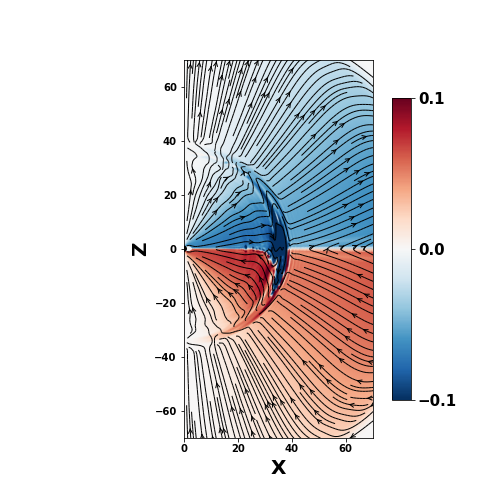}\\
       \includegraphics[width=0.30\textwidth]{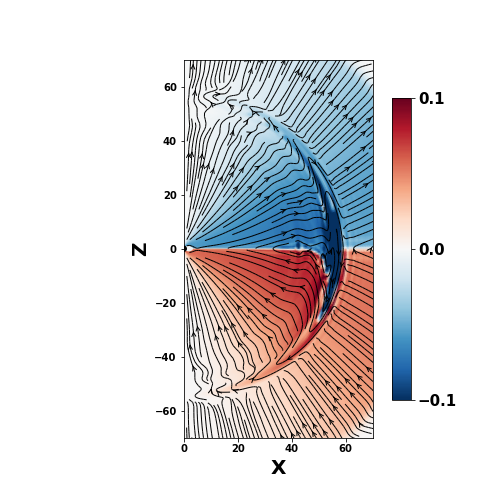}
       \includegraphics[width=0.30\textwidth]{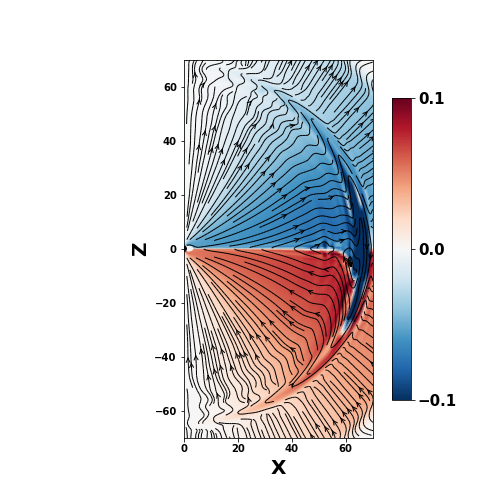}
       \includegraphics[width=0.30\textwidth]{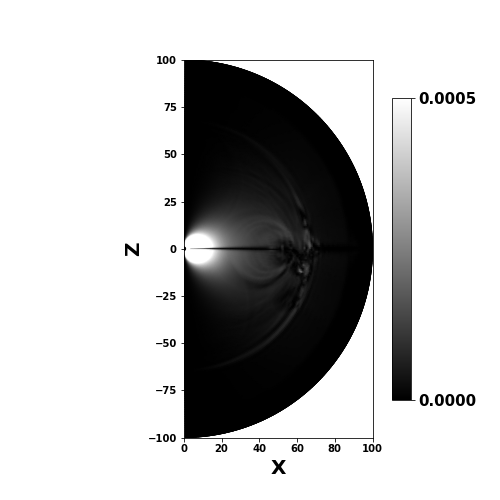}\\
        \includegraphics[width=0.30\textwidth]{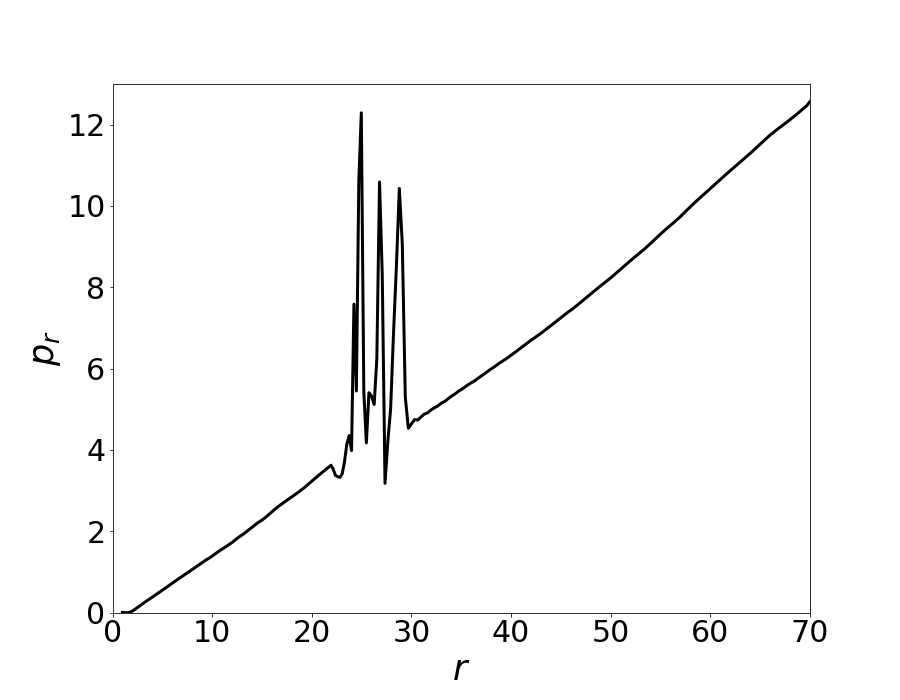}
        \includegraphics[width=0.30\textwidth]{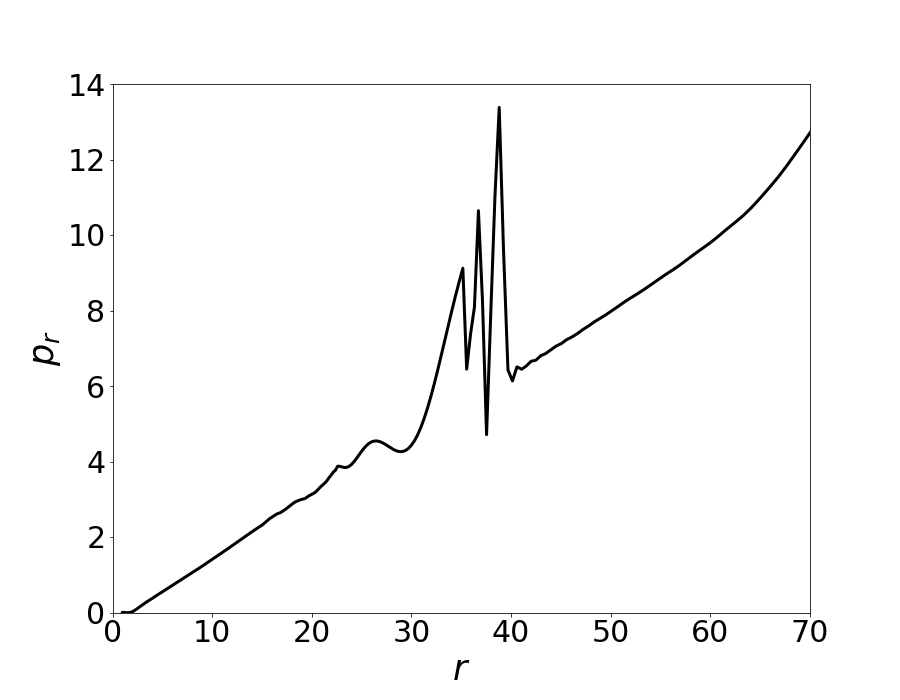}
         \includegraphics[width=0.30\textwidth]{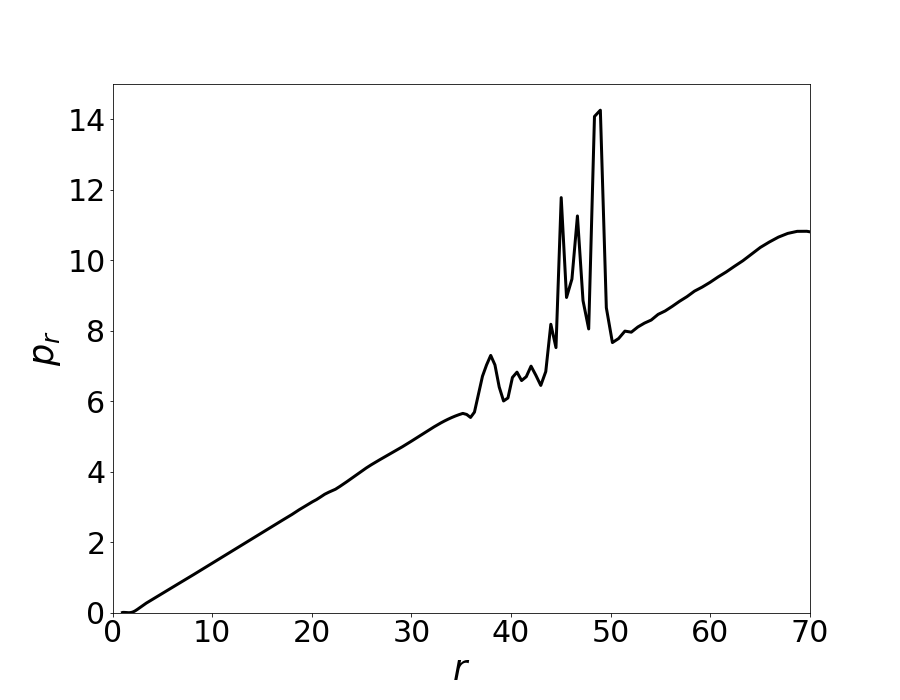}   
 \caption{Flux tube/\Alfven wave packet  launched on closed field lines  at  $60^{\circ}$ from zenith and radial distance  $a=2$,  "symmetric" injection,  at times  $t=2.01,2.04,3.2,3.8,4.1$ (launching at $t=2$,  $\Delta t = 0.06$. 
 Left and middle figures in top panel shows the zoomed-in version to capture the flux tube just after the launch. In the first plot the injected flux tube is a small patch, highlighted within a rectangle for visual clarity. We observe that the flux tube expands quickly within the \LC. The right grayscale figure in the middle panel shows the absolute value of effective \Bf\ $ ( B^2 -E^2)  \times r^2 \sin^2 \theta  $  at $t= 4.1$; it demonstrates that with the  the flux tube \Alfven pulse  there are no large variations of the effective \Bf.
  Bottom panel shows the magnitude of radial momentum  $p_r$ as a function of radial distance r for a slice at $60^{\circ}$ from the pole at $t=\{2.9,3.2,3.5\}$.  A forward-propagating pulse is clearly seen. We also repeated the calculations for angular slice at $120^{\circ}$ and the results were similar.  }
\label{xBphi_singletube_60Degrees}
\end{figure*}

\begin{figure*}
     \centering
       \includegraphics[width=0.49\textwidth]{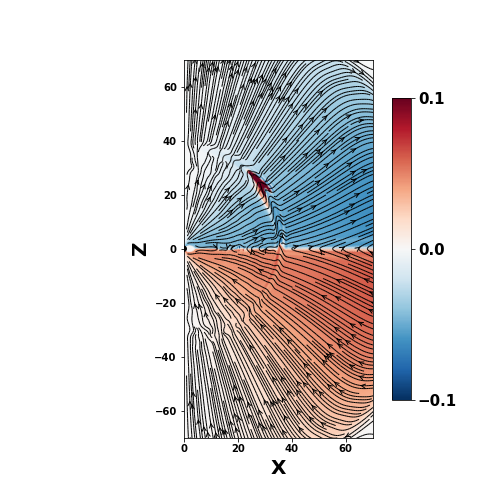}
         \includegraphics[width=0.49\textwidth]{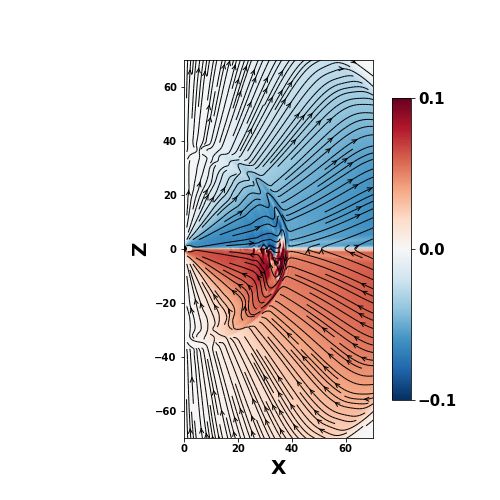}\\
          \includegraphics[width=0.49\textwidth]{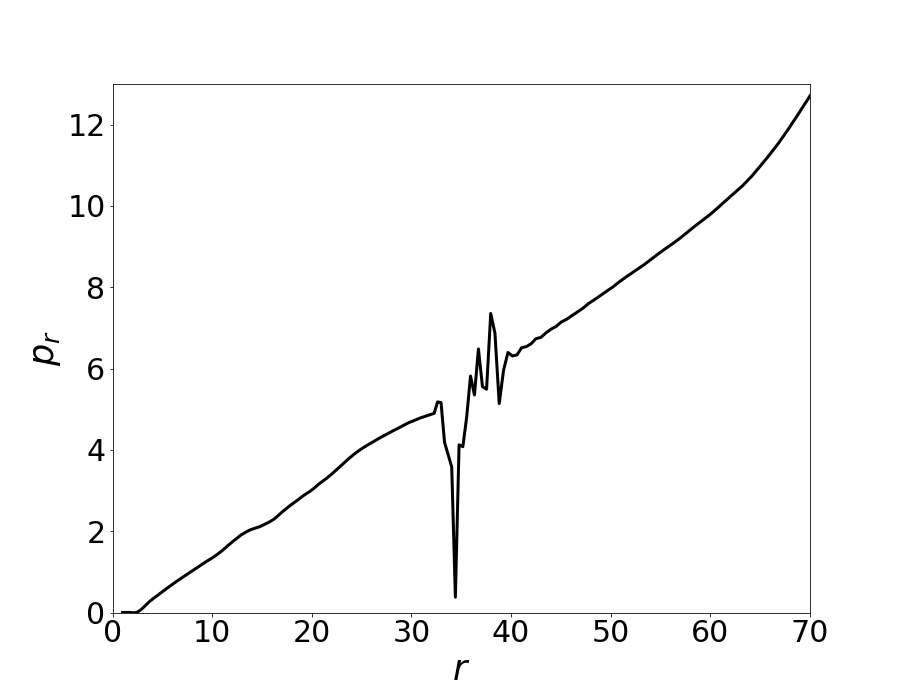}
     \includegraphics[width=0.49\textwidth]{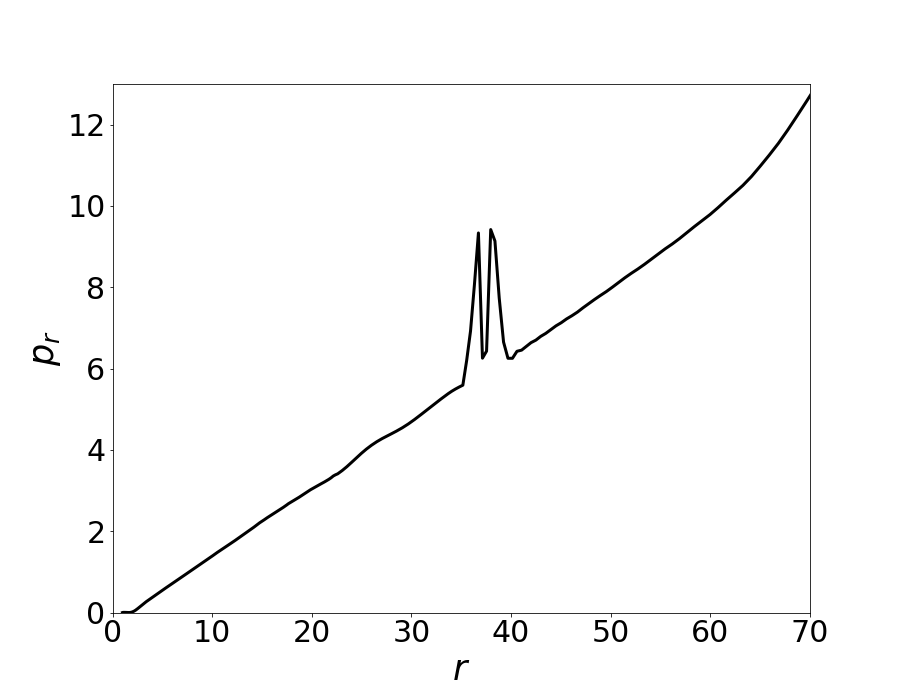}
 \caption{Comparison of injections on open field lines  $\theta=30^{\circ}$  (left column)  and closed field  $\theta=60^{\circ}$  (right column). Weak ``antisymmetric" injection. Notice that injection on open field lines produces a pulse that is backward propagating through the wind. For injection on closed field lines, the fast mode propagating with the \ms\ creates an \EM\ pulse that propagates in a forward direction through the wind independently on the initial  polarization of the pulse.
 }
\label{compare3060}
\end{figure*}

While the \Alfven components of the resulting pulse add differently to the wind flow (depending on the strength and polarization), see Fig. \ref{xBphi_singletube_60Degrees}, the X-mode component always produces a compression: a forward propagating wave, Fig. \ref{compare3060}.


Our basic results are plotted in Fig. \ref{xBphi_singletube_60Degrees} for injection on closed field lines ("symmetric" injection). We observe that the injected flux tube first expands within the \ms\ (top left two panels) and then propagates as an \Alfven pulse in the wind.

To further elucidate the underlying dynamics  in Fig. \ref{compare3060} we compare later behavior for ``antisymmetric" injection at two locations: $\theta = 30^{\circ}$ (left panels) and  $\theta = 60^{\circ}$ (right panel). In both cases the injection is ``weak'' - meaning that the injected toroidal flux is somewhat smaller than  (\ref{GJ_flux_analytics}). The two cases are clearly different: for ``antisymmetric" injection on open field lines {\it a backward propagating wave is launched} (in the panel \ref{compare3060}.a the radial momentum within the wave is {\it smaller} than that of the wind.

At the same time the similar injection but on closed field lines (right panels in  Fig. \ref{compare3060}) creates {\it forward} propagating pulse. The reason for the differences is the following. For ``antisymmetric"  injection on open field lines the resulting \Alfven pulse resembles the magnetospheric glitch, \S \ref{Jerked} -  regions with smaller toroidal field propagate slower. (We also verified that in case of ``strong  antisymmetric" injection, when the injected toroidal flux is larger than  (\ref{GJ_flux_analytics}), the resulting pulse is forward-propagating.

Qualitatively, using Michel's solution (\ref{michels_expressions}) a local toroidal \Bf\ corresponds to some local effective angular velocity. Reducing local toroidal \Bf\ (for weak antisymmetric injection) reduces the  effective angular velocity and the radial momentum. Since the radial momentum $p_r$ is a quadratic function of the field, strong antisymmetric injection (so that the total toroidal field is larger inside the pulse than in the surrounding wind) produces forward propagating pulse.

Injection on the closed field lines proceeds differently. For ``mild'' injection, the added toroidal field on the closed field lines corresponds to fast mode regardless of the polarization. The fast mode first propagates through the \ms\ and then creates {\it  compression}  of the field in the wind. The resulting pulse is always forward propagating. 

\subsection{Multiple injection events}
\label{multiple_flux_tube}

\begin{figure*}
     \centering
           \subfloat[\label{7a}]{\includegraphics[width=0.4\textwidth]{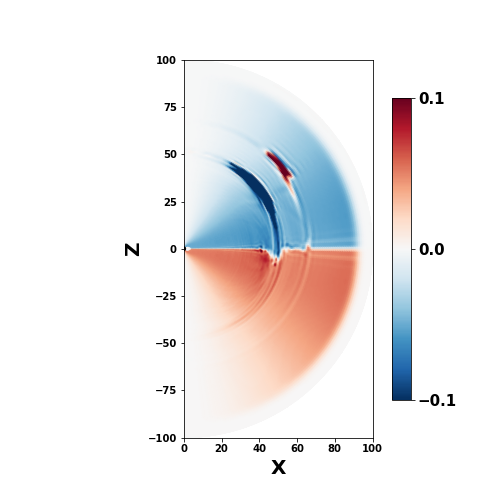}}
             \hspace{0 em}
             \subfloat[\label{7b}]{\includegraphics[width=0.4\textwidth]{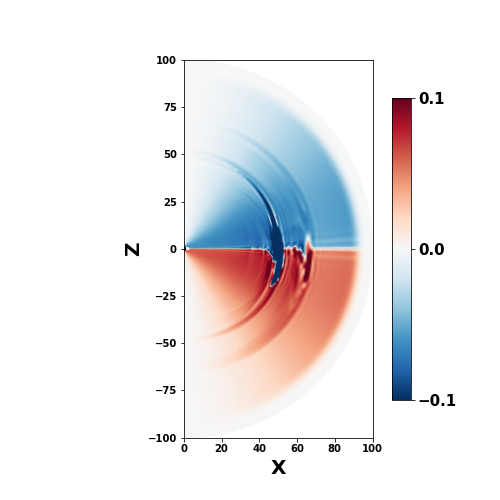}}\\
            \subfloat[\label{7c}]{\includegraphics[width=0.4\linewidth]{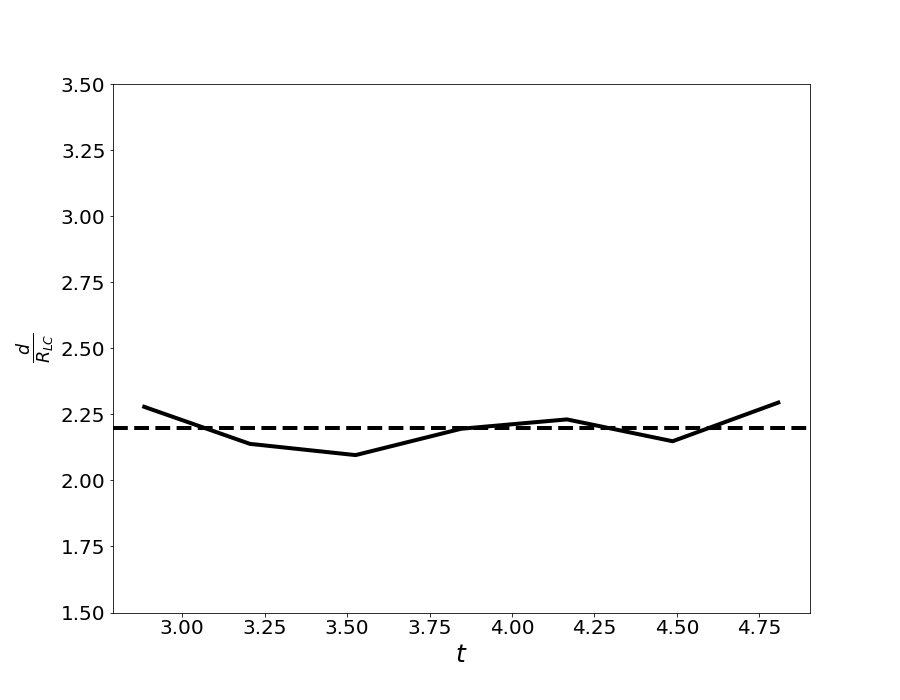}}
             \hspace{0 em}
            \subfloat[\label{7d}]{\includegraphics[width=0.4\linewidth]{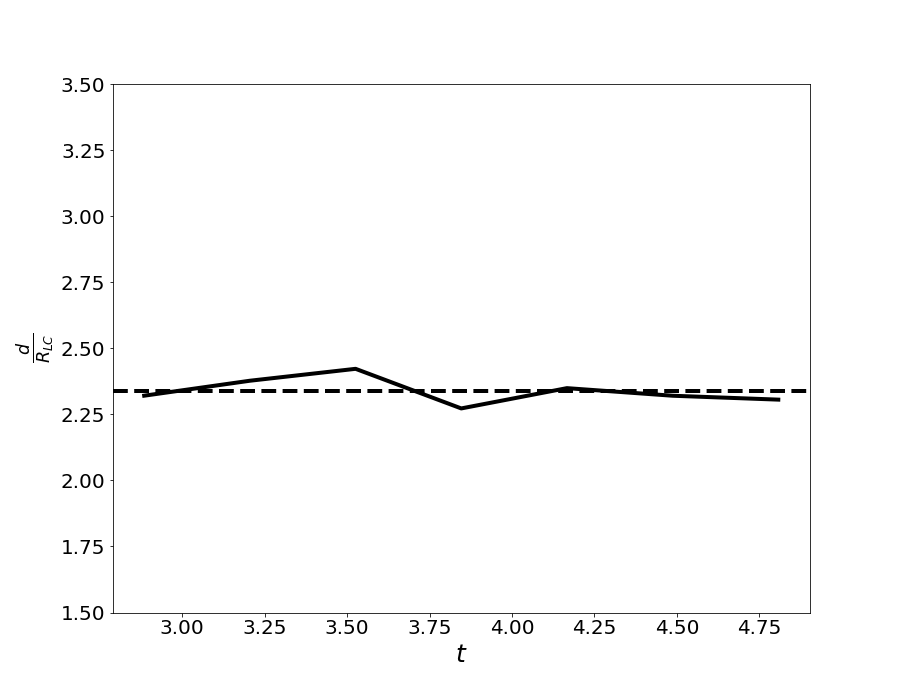}}
          \caption{Multiple injection events. Two tubes are injected inside the \LC\  at spherical radius $a=2$. Injection occurs at  $30^{\circ}$ (left column, on open field lines) and $60^{\circ}$  (right  column, on open field lines). The first tube is \emph{weak-antisymmetric}, the second tube was launched half rotational period after the first one.  Bottom row:  time evolution of the distance between the two flux tubes. Solid line represents the  separation between the tubes as a function of time whereas the dashed line shows the average separation. As observed, the separation between the injections remains (approximately) the same.}
\label{xBphi_twotubes_a=2}
\end{figure*}

We end this section by considering the scenario of multiple flux tubes. Here we add two flux tubes, the first one at $t=t_{insertion}$, and second one at $t=1.5 t_{insertion}$ with time expressed in terms of the rotational period of the star. The first tube is weaker and launched agains the wind (anti-symmetric scenario) while the second tube is stronger and launched along the wind. We consider injections on open field lines  $\theta=30^{\circ}$ and closed field  $\theta=60^{\circ}$. We show a snapshot of such multi flux tube system in Fig. \ref{7a} ,Fig. \ref{7b} 
The two tubes don't catch up, even if the second one is more powerful and the first is ``weak-antisymmetric'' launched on open field lines (hence propagating backwards through the wind).


This is clearly a result of relativistic kinematics, modified by the fact that the bulk flow is accelerating. In fact,  \Alfven waves propagating in the Michel's wind can be considered non-perturbatively, \cite{2011PhRvD..83l4035L}  and   \S \ref{Jerked}. Such waves can be parametrized by the local  spin $\Omega_{1,2} \neq \Omega_0$ ($ \Omega_0$ is  constant spin of the star). One then finds the location of the first  wave at time $t$ after leaving the \LC:
\be
R_1 = \frac{1}{2} \left(t+\frac{1}{\Omega _0}-\frac{\Omega _0}{\Omega _1^2}\right) + 
\sqrt{1+ \frac{\left(\Omega _1^2 (1+ t \Omega _0) -\Omega _0^2\right){}^2}{4 \Omega
   _0^2 \Omega _1^2}} \frac{1}{\Omega _1} \approx 
t+\frac{1}{\Omega _0}-\frac{\Omega _0}{\Omega _1^2}+ \frac{1}{t \Omega _1^2}
\label{R1} 
\ee
where $R_(0) = 1/\Omega_0$ (wave is launched at  time $t=0$ at the \LC); the latter relation is for $t \gg 1/\Omega_0$. Eq. (\ref{R1}) gives the location of the \Alfven pulse propagating through accelerating wind. 

If a second \Alfven pulse is launched after time $\Delta t$ with $\Omega_ 2 \geq \Omega_1$, the collision will occur approximately at 
\be
t_{coll} \Omega_0 \approx \left( 1+ \frac{{\Delta t} \Omega _1^2 \Omega _2^2}{\Omega _0 \left(\Omega _1^2-\Omega
   _2^2\right)} \right) ^{-1}  \approx \frac{1}{2} \Omega _0^2 \left(\frac{1}{\Omega _1^2}-\frac{1}{\Omega _2^2}\right)
   \ee
   where the last relation assumes that the two pulses are separated by one rotation. Typical collision times are long and not captured by our simulations. When the waves  eventually catch up  at $r\gg R_{LC}$, the interaction  will be resemble  interaction between two non-linear  packets of fast modes.

\section{Conclusion: whence to FRB}

In this work we continue, following \cite{2022MNRAS.509.2689L,2022ApJ...934..140B},   exploration of   the dynamics of the relativistic magnetized explosions:  how relativistic magnetically-driven explosions are produces by magnetar, and how they propagate through the preexisting magnetized wind. 
To search for answers we performed multiple 2D numerical simulations of a neutron star magnetosphere  and the winds. The simulations  focused on several different but related phenomenons:  production of magnetic flares via shearing of the foot-points of the \Bf\ lines,  and evolution of relativistic flux tube(s)/\Alfven pulses in magnetars' winds.

Two regimes of foot-points shearing we considered: (i) slow shear, so that the whole inflated magnetic arc structure is in a state of causal contact; (ii) fast shear, so that the corresponding dynamics  resembles large amplitude \Alfven waves injected into the \ms.

We stress again the importance of magnetic loading of magnetar flares \citep{2022MNRAS.509.2689L,2022ApJ...934..140B}: an injected  flux tube/plasmoid  looses a lot of energy trying to break out from the \ms.  For example,  we expect that a fraction of the injected  magnetic energy $E_{CME,0}$ will be emitted in X-rays. Yet the energy that gets deposited into the wind  even in the super-critical case is always much smaller at least by the small   numerical factor $\eta_E$ (the ratio of injected energy to the total  magnetospheric energy); for the flux tube scenario the decrease is even more dramatic, $\propto  \eta_E^2$, see Table \ref{scales11}.  For milder flares the wind adjusts to the perturbation right near the \LC, so no energy in deposited in the wind.

For subcritical injections, which do not experience detonation inside the \LC, the resulting CME is completely  ``cold turkey'': a structure in force balance and  advected passively with the wind. 
If the CME's energy exceeds the critical and detonation occurs,  then still only small fraction of the initial energy, at  most $\sim \eta_E$,  is transferred to the wind in the form of EM pulse.

We conclude that:
\begin{itemize}
\item For slow shear,  the \emph{Solar flare} paradigm:
 \begin{itemize}
 \item there are two possible stages of CME expansion within the \ms: for sufficiently large injection, a CME experiences internal detonation at some radius $r_{eq}$, when it starts expanding relativistically within the \ms\ and loses causal connection.
\item the magnetospheric dynamics depends both on the large scale structure and on the location of shearing foot points: to generate  rare powerful  events shear must occur on field lines that ``close in'' near the star; otherwise numerous weak events are generated.
\item Ejected magnetic blobs,  CMEs,  are frozen into the wind
\end{itemize}
\item For fast  shear, the \emph{Star quake} paradigm:
\begin{itemize}
\item  Shearing of foot-points leads to the generation of \Alfven wave; the  pressure of the \Alfven leads to opening of the \ms\ (no wave breaking). 
\item Resulting perturbations propagate in the wind as  shear \Alfven waves, with no breaking
\item multiple shear \Alfven waves are unlikely to collide within relativistically accelerating wind.
\end{itemize}
\item In both cases of slow and fast shear no considerable dissipation  occurs in the wind zone. In both cases  after the ejection the \ms\ first opens; afterwards the newly closed \ms\ is smaller, and recovers resistively.
\end{itemize}

Our results are complementary to those of \cite{2022ApJ...934..140B}, who investigated the dynamics of magnetic explosions with complicated, linked  magnetic internal stricture. 
 \cite{2022ApJ...934..140B} showed, using both relativistic  MHD and  force-free approaches, that  there is a clear regime of magnetic explosions - detonation. In this regime in MHD the expansion of a spheromak  becomes supersonic,  in the force-free case we see spheromak torn apart,   and also  becoming causally disconnected. 


Our results make a consistent picture:  powerful strongly magnetized ejected blobs/flux tube makes minimal distortion in the wind. They  either quickly reach force-balance with the wind, and propagate self-similarly, without producing shocks and/or dissipative structures, or propagate as highly weakened \EM\ disturbances.  This picture is in sharp contrast with the hydrodynamics, where over-pressurized regions create strong dissipative shock.

Our results  have implications for the generation of  FRBs. 
\begin{itemize}

\item \emph{FRBs as Coronal Mass Ejections -  large and small}. FRBs show large range of luminosities \citep{2020Natur.587...54C,2022arXiv220714316S}, which raises an obvious question: what's the control parameter that defines (both X-ray and radio) luminosity of magnetars'  bursts/flares?   The overall size involved in a  flare is one  obvious parameters. Another is the strength of the \Bf\ - both determine total energetics.

In the Solar flare paradigm of magnetar flares,  the 
 \Bf\ also enters via the rate of shearing the foot-point: the shearing rate is $\propto $ \Bf\ \citep{RG}; thus, qualitatively, the magnetar activity is a $B^3$ function of the \Bf\
 \citep{2015MNRAS.447.1407L}. 

In the present work we find that other, less clearly measured properties play a role:  (i) evolution of a CME within \ms\ proceeds in different regimes depending on the injected energy (possibly a detonation);
(ii)  location of the shear; (iii)  overall structure of the \ms. If shearing is done near the fields that extend far out from the star, then the twist is easily released in many small flares. (Along a given field line the twist concentrates near the regions of weakest \Bf, hence at the highest point in the \ms.)  In order to produce rare and powerful explosions the shearing should be done at the foot-points of field lines that close in, roughly speaking, within a stellar radius.

Qualitatively, a twist of a given \Bf\ line concentrates near the points where the guiding field is the weakest - at the furthest extent. It is there that the stability is determined. For field lines extending to large distances, the guiding field is small, so that the kink instability  is  easily initiated at small twists. The system then gets rid of the twist in many small events. 

Finally, the presence of the \LC\ effectively impedes the storage of the magnetic energy. If an inflated flux tube reaches the light cylinder, it opens up and releases the twist this limits the amount of  magnetic  energy  that can be stored. Thus, to produce strong flares the spin  period {\it should not} be too short.

\item \emph{Dynamics of  CMEs/\EM\ pulses in the preceding wind}.
Our results  on the wind dynamics are in some contradiction to the ``wind models'' of FRBs  \citep[\eg][]{2014MNRAS.442L...9L,2017ApJ...843L..26B,2019MNRAS.485.4091M,2022arXiv220911136T}. 
For slow shear, the Solar flare paradigm, energetically mild CME (non-detonating) produce minimal distortion of the wind:  topologically disconnected structures (``magnetic shells") come into force balance close to the \LC, and are then passively advected with the flow. In the super-critical detonating case, a highly weakened \EM\ pulse is launched into the wind.   
For fast shear, the Starquake paradigm, the energy is quickly deposited into the \ms\ in a form of \Alfven and X-modes that may also open the \ms. In doing so, the wave energy is deposited into the \ms, so is lost by the pulse. More powerful pulses open the \ms\ earlier and suffer larger energy losses.



In passing we note that the original shock model of \cite{1992ApJ...391...73G,1992ApJ...390..454H}, envisioned to explain {\it months-long} variability of Crab Nebula wisps, involves interaction of the relativistic wind with a heavy ejecta.  In that case the cyclotron instability occurs in  the termination shock of the wind, with only mildly relativistic post-shock flow. It is not applicable to generation of millisecond (and even shorter) radio pulses in FRBs. 

\end{itemize}


 \section{Acknowledgements}
This work had been supported by NASA grants 80NSSC17K0757 and 80NSSC20K0910, NSF grants 1903332 and 1908590. We would like to thank Spiro Antiochos, Jens Mahlmann, Bart Ripperda and Chris Thomson for comments and discussions. The work of the  organizers of the  ``Plenty of Room at the Bottom: Fast Radio Bursts in our Backyard''  workshop is acknowledged. 

\section{Data availability}
The data underlying this article will be shared on reasonable request to the corresponding author.


\newcommand{\newblock}{}
\bibliographystyle{mnras}
\bibliography{references,ft}

\appendix

\section{Numerical Method}
\label{NumericalMethod}
In this paper, we study the dynamics of sheared magnetospheres using time-dependent numerical simulations with the code PHAEDRA  \citep{2012MNRAS.423.1416P}.  The code solves  Maxwell's equations together with ideal constraints
\ba  &&
\mathbf{J} . \mathbf{E} =0 
\nn && 
 \mathbf{B^{2}} - \mathbf{E^{2}} \ge 0 
 \ea
in spherical, axisymmetric geometry.  

The simulation domain extends from $r_{in} = r_{\star}$, the \NS\ radius to the outer boundary $r_{out}$. For plasmoid ejection simulations, we set up the outer boundary at $r= 10 r_{\star}$ whereas for flux tube simulations the the outer boundary was set at $r= 100 r_{\star}$. We use smooth coordinate mapping for the radial grid, while the grid is equi-spaced in $\theta$ direction. The computational mesh consists of $280 \times 180$ cells in $(r, \theta)$ directions respectively. 


Our simulation region has two major boundaries: the inner boundary i.e. the surface of the star and the outer boundary defined by size of our simulation box. We assume axisymmetry as well as symmetry about the equatorial plane. The normal component of the magnetic field, $B_{r}$, and the tangential components of the electric field are continuous across the surface, and therefore are known. The required boundary conditions  at $r=r_{\star}$ are 
  
\ba  &&
B_{r} = B_{r} (\theta  )
\nn && 
 E_{\theta} = - \Omega B_{r} \sin \theta 
 \nn && 
  E_{\phi} = 0 
   \label{inner_boundary_boundary_condition}
\ea

For plasmoid ejection simulation, we introduce shearing after one rotation period by simply modifying the net angular velocity at the surface by $\Omega = \Omega + \omega_{s}$. For rotating cases, the rotational angular velocity of the star $\Omega$ is set as $0.2$ throughout the entire work.

\section{Dipole-plus-quadrupole  configurations}
\label{multipole}

In this case fairly simple analytical results can  guide us for the choice of shearing location.
The flux function for dipole and quadrupole, normalized to \Bf\ at the pole are
\ba&&
P_d = \frac{\sin^2 \theta}{r} \frac{ B_p R^3} {2} 
\nn &&
P_q = \frac{\sin (2  \theta) \sin \theta}{r^2} \frac{ B_q R^4} {4} 
\nn && 
\B_{d,q} = \nabla P_{d,q} \times \nabla \phi 
\ea
Total field is a linear sum 
\be
P_{tot} = P_d+ \mu_q  P_q 
\ee
where $\mu_q$ parametrizes the  relative strength of the dipole and quadrupole.  We use    $\mu_q=2$: this makes the \Bf\ at one pole 3 times larger than pure dipole, while at the other pole \Bf\ equal in value to the dipole value, but with the reverse sign. 

\begin{figure}
\includegraphics[width=.99\linewidth]{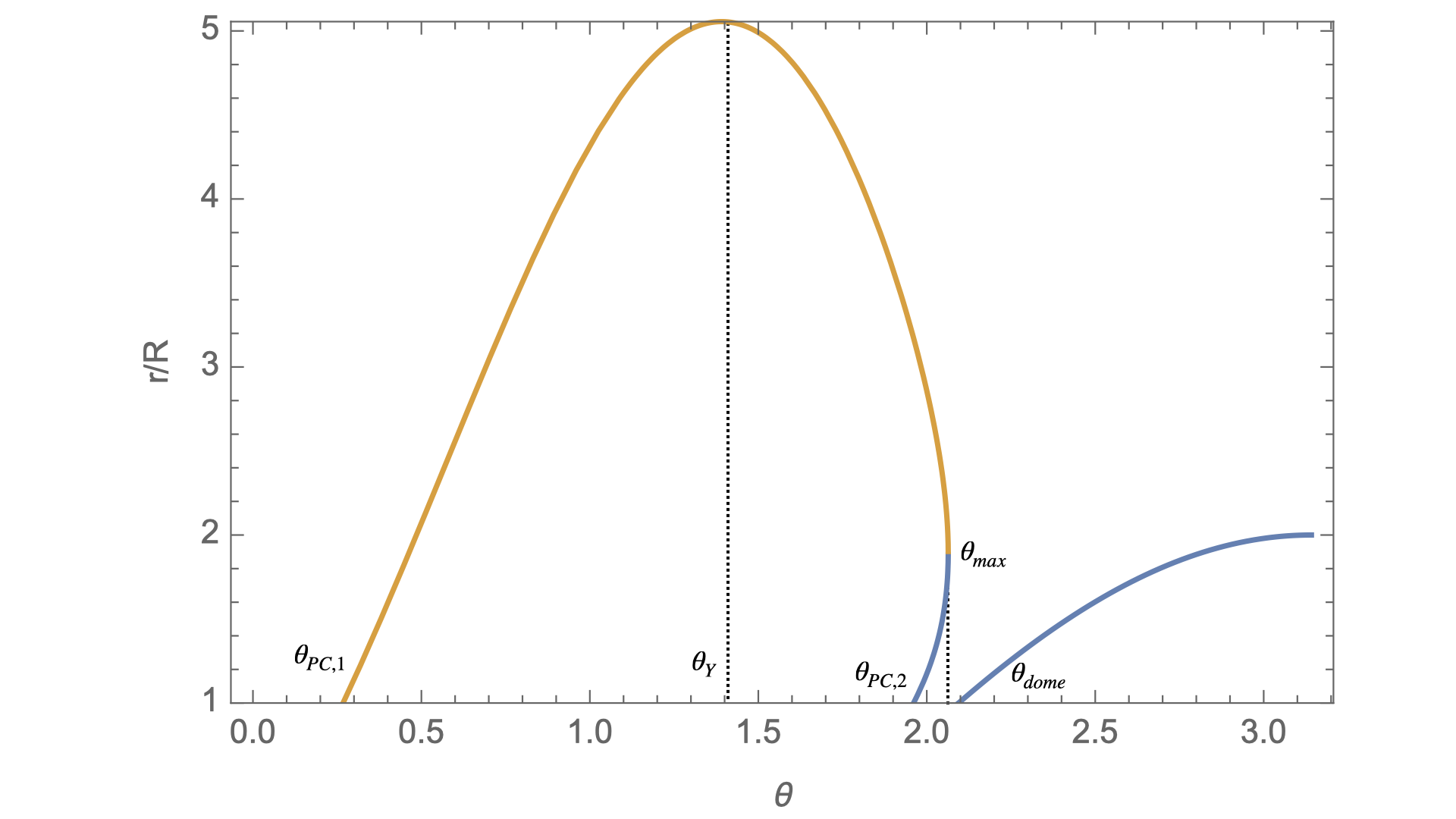}
\caption{Last closed field lines and the southern dome for the  dipole-plus-quadrupole  configuration with $\mu_q=2$.}
\label{multipoles001}
\end{figure}

We find  special points:
\begin{itemize}
\item Zero point at anti-pole  corresponds to $  \mu_{q}=1$
\item edge of the southern dome $\cos \theta_{dome} =-1/\eta$ ($\theta_{dome} = 2 \pi/3$ for $\mu_{q}=2$).
\item Upper polar cap. Furthest point at $R_{LC} = r \sin \theta_Y$,  ($\theta_Y $ is the angle of the Y-point 
\be
\eta = -\frac{6 \cot (\theta _Y)}{5 \cos (2 \theta_Y )+3} \frac{R_{LC}}{R}
\ee
For $R_{LC}/R= 5 $ and $\eta = 2$, $\theta_Y=   1.4480$ ($\theta_Y$ cannot be smaller than $ \cos (2 \theta_Y) = -3/5$). Y-point is at $r = R_{LC}/\sin \theta_Y \to 5.037$.
\item At the last open field line
\be
P_{tot}^{(0)} = \frac{\sin ^3\left(\theta _Y\right) \cos \left(2 \theta _Y\right)}{5 \cos \left(2 \theta
   _Y\right)+3} \frac{1}{R_{LC}}
   \ee
   Last closed field lines is given by
   \be
   r^{(0)} =  \frac{\sin ^2(\theta )\pm \sin (\theta ) \sqrt{\sin ^2(\theta )+8 \eta  P_{tot}^{(0)}  \cos (\theta )}}{4
   P_{tot}^{(0)} }
   \ee
   Maximal extent is when they are equal,
   \be
\cos \theta_{max} = 4 P_{tot}^{(0)} \mu_{q} - \sqrt{1 + (4 P_{tot}^{(0)}  \eta)^2}
\ee
Polar cap polar angles  are
\ba &&
\theta_{PC,1} = 0.267
\nn &&
\theta_{PC,2} = 1.96
\ea

\end{itemize}

\section{Rotating stars with no foot-point shearing}
 \label{noshear1}
  
  As a preliminary investigation, we considered rotating but unsheared configurations.   Rotation adds  a  characteristic scale to the problem viz., the radius of the light cylinder $R_{LC}$. The field lines opens to infinity beyond the light cylinder. We start with a non-rotating \NS\ and bring it to final rotational velocity $\Omega_{\star}$ and then allowed to relax to a steady equilibrium state.

\begin{figure}
\includegraphics[width=0.8\linewidth]{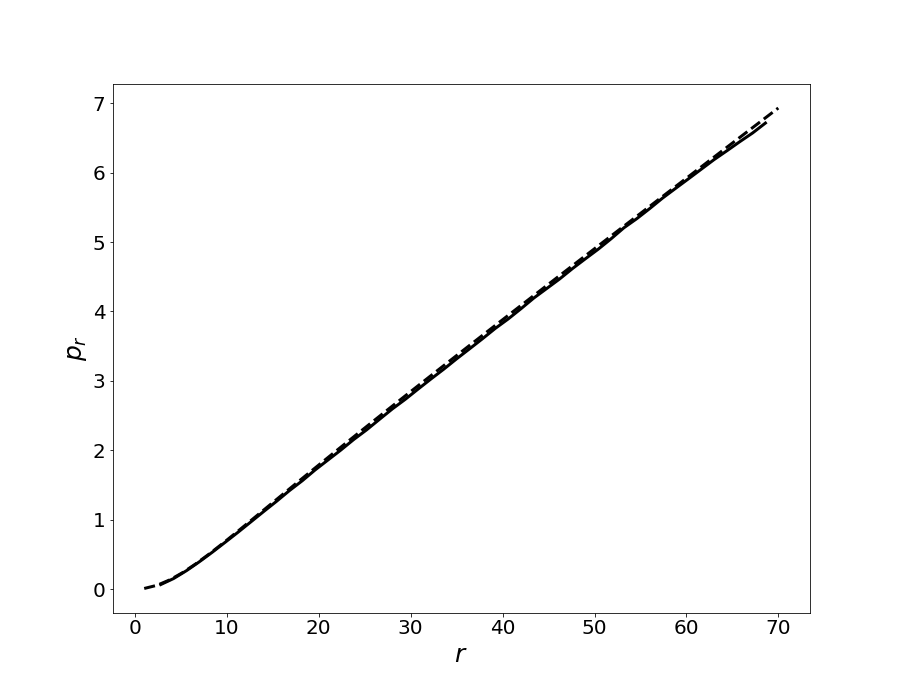}
    \caption{Radial momentum (solid) from analytical expression (Eq. (\ref{michels_expressions})) and (dashed) for a simple dipole with no flux tube from numerical simulation data at $\theta=30^{\circ}$ }
\label{pr_analytical_numerical}
    \end{figure}

\begin{figure*}
        \includegraphics[width=.49\linewidth]{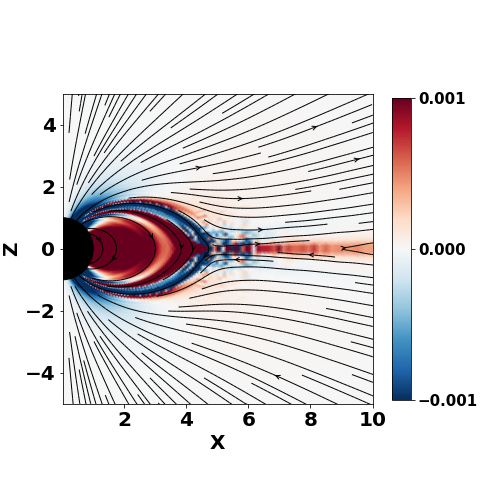}
        \includegraphics[width=.49\linewidth]{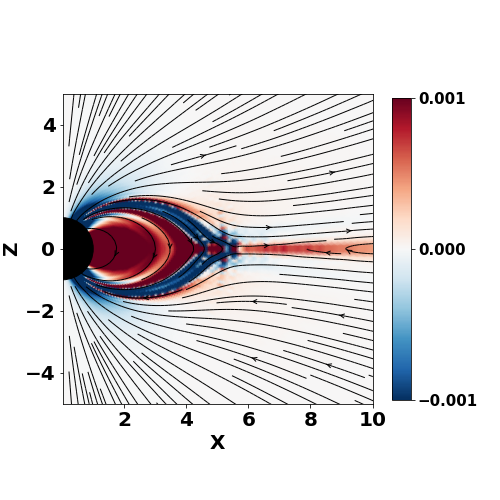}
     \caption{Toroidal current $( J_{\phi}) $ for an aligned rotating dipole ($\Omega=0.2$) with no shearing at two consecutive time slice clearly showing plasmoid ejection. }
    \label{simpledipole}
\end{figure*}

We first consider the case of an aligned \NS\ in purely dipolar field, and with \emph{no} shearing of the magnetic field lines.   We expect the solution to resemble that given by  \citep{1973ApJ...180L.133M} (Eq. (\ref{michels_expressions})), once the equilibrium has been achieved. 
We compared the radial momentum from simulations for a fixed $\theta$ with those generated from simulations and the values were in excellent agreement (Fig. \ref{pr_analytical_numerical}). 
We also  observe few weak plasmoid ejection events, consistent with \cite{2013ApJ...774...92P}, Fig. \ref{simpledipole}.


\end{document}